\documentclass[12pt, preprint]{aastex}
\usepackage{graphics}
\begin{document}

\title{Astronomical Spectroscopy}
\author{Philip Massey}
\affil{Lowell Observatory, 1400 W Mars Hill Road, Flagstaff, AZ 86001, USA; phil.massey@lowell.edu}
\and
\author{Margaret M. Hanson}
\affil{Department of Physics, University of Cincinnati, PO Box 210011, Cincinnati, OH 45221-0011}

\begin{abstract}

Spectroscopy is one of the most important tools that an astronomer has for studying the universe.  
This chapter begins by discussing the basics, including the different types of optical
spectrographs, with extension to the ultraviolet and the near-infrared.
Emphasis is given to the fundamentals of how spectrographs
are used, and the trade-offs involved in designing an observational experiment.
It then covers
observing and reduction techniques, noting that some of the standard practices of flat-fielding
often actually degrade the quality of the data rather than improve it. Although the
focus is on point sources, spatially resolved spectroscopy of extended sources is also briefly discussed.   Discussion of differential extinction, the impact of crowding,
multi-object techniques,  optimal extractions,  flat-fielding considerations, and determining
radial
velocities and velocity dispersions provide the spectroscopist with the fundamentals
needed to obtain the best data.  Finally the chapter combines the previous material by providing some examples of real-life observing experiences
with several typical instruments.   
\end{abstract}

\clearpage

\section{Introduction}

\begin{verse}

{\it
They're light years away, man, and that's pretty far \\
(lightspeed's the limit, the big speed limit) \\
But there's plenty we can learn from the light of a star \\
(split it with a prism, there's little lines in it) }\\
--Doppler Shifting, Alan Smale (AstroCappella\footnote{http://www.astrocappella.com/})
\end{verse}

Spectroscopy is one of the fundamental tools at an astronomer's disposal, allowing one to determine
the chemical compositions, physical properties, and radial velocities of astronomical
sources. Spectroscopy is the means used to measure the dark matter content of galaxies, the masses of two stars in orbit about each other, the mass of a cluster of galaxies, the rate of expansion
of the Universe,  or discover
an exoplanet around other stars, all using the Doppler shift.  It makes it possible for the astronomer
to determine the physical conditions in distant stars and nebulae, including the chemical composition and temperatures,
by quantitative analysis of the strengths of spectral features, 
thus constraining models of chemical enrichment in galaxies and the evolution of the universe.  As one well-known astronomer
put it, ``You can't do astrophysics just by taking pictures through 
little colored pieces of glass," contrasting
the power of astronomical spectroscopy with that of broad-band imaging.

Everyone who has seen a rainbow has seen the light of the sun dispersed into a spectrum, but it was
Isaac Newton (1643-1727) who first showed that sunlight could be dispersed into a continuous series
of colors using a prism.   Joseph von Fraunhofer (1787-1826) extended this work by discovering and
characterizing the dark bands evident in the sun's spectrum when sufficiently dispersed.  The
explanation of these dark bands was not understood until the work of Gustav Kirchhoff (1824-1887) and 
Robert Bunsen (1811-1899), who proposed that they were due to the selective
absorption of a continuous
spectrum produced by the hot interior of the sun by cooler gases at the surface. 
The spectra of stars were first observed visually by
Fraunhofer and Angelo Secchi (1818-1878), either of whom may be  credited with
having founded the science of astronomical spectroscopy.  

The current chapter will emphasize observing and reduction techniques primarily for optical 
spectroscopy obtained
with charge coupled devices (CCDs) and the techniques needed for near-infrared (NIR)
spectroscopy obtained
with their more finicky arrays.  Spectroscopy in the ultraviolet (UV) will also be briefly
discussed.
Very different techniques
are required for gamma-ray, x-ray, and radio spectroscopy, and these topics will not be included here.
Similarly the emphasis here will be primarily on stellar (point) sources, but with some discussion of how
to extend these techniques to extended sources.

The subject of astronomical spectroscopy has received a rich treatment in the literature. The volume
on {\it Astronomical Techniques} in the original {\it Stars and Stellar Systems} series
contains a number of seminal treatments of spectroscopy.  In particular, the introduction to 
spectrographs by Bowen (1962) remains useful even 50 years
later, as the fundamental physics remains the same even though photographic plates have given way
to CCDs as detectors.  The book on diffraction gratings by Loewen \& Popov (1997) is also
a valuable resource.  Grey (1976)  and Schroeder (1974) provide
very accessible descriptions of astronomical spectrographs, while the ``how to" guide by Wagner (1992)
has also proven to be very useful.  Similarly the monograph by Walker (1987) delves into the
field of astronomical spectroscopy in a more comprehensive manner than is possible in a single chapter,
and is recommended.

\section{An Introduction to Astronomical Spectrographs}

This section will concentrate on the hardware aspect of astronomical spectroscopy.  The basics are discussed first.
 The following subsections then describe specific types of astronomical spectrographs, citing examples  in current 
 operation.
 
\subsection{The Basics}
\label{Sec:Basics}
\begin{figure}[htp]
\plotone{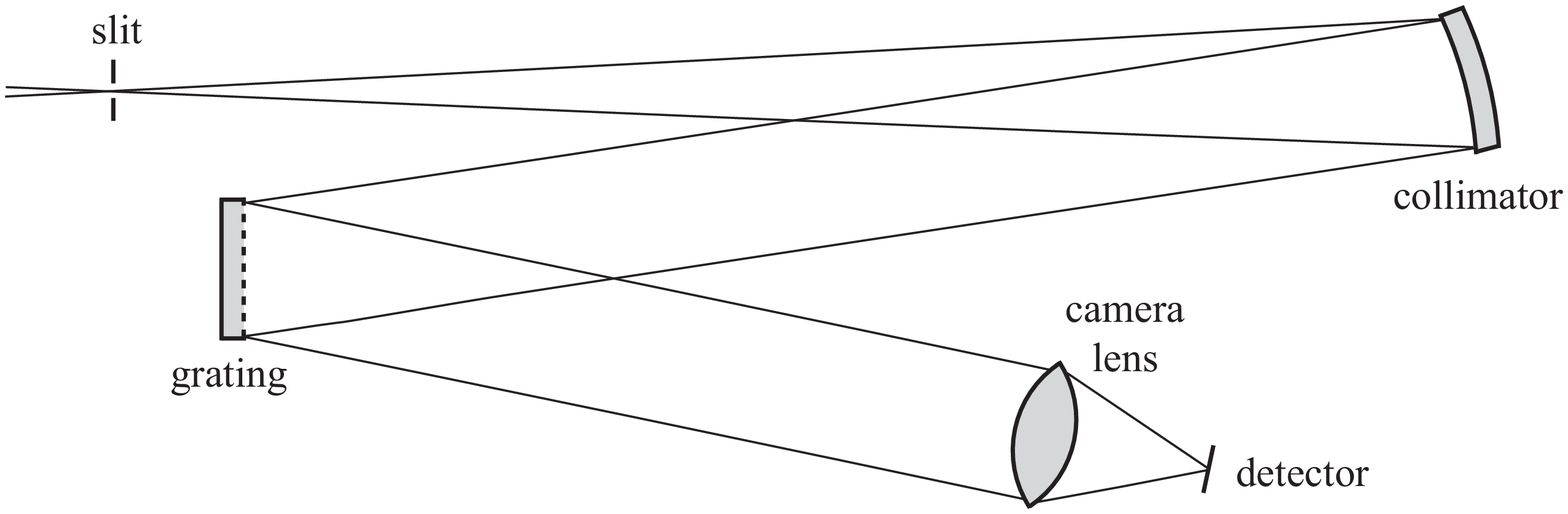}
\caption{\label{fig:spect} The essential components of an astronomical spectrograph.}
\end{figure}

When the first author was an undergraduate, his astronomical techniques professor, one Maarten Schmidt, drew a schematic
diagram of a spectrograph on the blackboard, and said that all astronomical spectrographs contained these 
essential elements: a slit on to which the light from the telescope would be focused; a collimator, which would take the
diverging light beam and turn it into parallel light; a disperser (usually a reflection grating); and a camera that would
then focus the spectrum onto the detector.  In the subsequent 35 
years of doing astronomical
spectroscopy for a living,  the first author has yet to encounter a spectrograph
that didn't meet this description, at least in functionality.
 In a multi-object fiber spectrometer, such as Hectospec on the MMT (Fabricant et al.\ 2005),
the slit is replaced with a series of fibers.  In the case of an echelle, such as MagE on the Clay 6.5-m telescope (Marshall et al.\ 2008),
 prisms are inserted into the beam after
the diffraction grating to provide cross-dispersion.  In the case of an objective-prism spectroscopy, the star itself
acts as a slit ``and the Universe for a collimator" (Newall 1910; see also Bidelman 1966). 
Nevertheless, this heuristic picture provides the reference for such
variations, and a version is reproduced here  in Figure~\ref{fig:spect} in the hopes that it will prove equally
useful to the reader.

The slit sits in the focal plane, and usually has an adjustable width {\it w}.   The image of
the star (or galaxy or other object of interest) is focused onto
the slit.  The diverging beam continues to the collimator, which has focal length $L_{\rm coll}$.  The {\it f-ratio} of the collimator
(its focal length divided by its diameter)
must match that of the telescope beam, and hence its diameter  has to be larger the further away it is from the
slit, as the light from a point source should just fill the collimator.  The collimator is usually an off-axis paraboloid, so that it both
turns the light parallel and redirects the light towards the disperser.

In most astronomical spectrographs the disperser is a grating, and is ruled with a certain number of grooves per mm, usually
of order 100-1000.  
If one were to place one's eye near where the camera is shown
in Figure~\ref{fig:spect} the wavelength $\lambda$ of light seen would depend upon exactly what angle {\it i} the grating was set at relative to the
incoming beam (the angle of incidence), and the angle $\theta$ the eye made with the normal to the grating (the angle of diffraction).   How much one has to move one's head by in order
to change wavelengths by a certain amount is called the dispersion, and generally speaking
the greater the projected number of grooves/mm (i.e., as seen along the light path), the
higher the dispersion, all other things being equal.  
The relationship governing all of this is called 
{\it the grating equation} and is given as 
\begin{equation}
\label{equ-grating}
m\lambda=\sigma (\sin i+\sin \theta).
\end{equation}  

In the grating equation,  {\it m} is an integer representing
the {\it order} in which the grating is being used.  Without moving one's head, and in the absence of any order blocking filters, one could  see 8000\AA\ light
from first order and 4000\AA\ light from second order at the same time\footnote{This is because of the
basics of interference: if the extra path length is any integer multiple of a given wavelength,
constructive interference occurs.}. An eye would also have to see further into the red and blue
than human eyes can manage, but CCDs typically have sensitivity extending from 3000-10000\AA, so this is a real issue, and is solved
by inserting a {\it blocking filter} into the beam that excludes unwanted orders, usually right after the light has passed through the slit.

The angular spread  (or dispersion\footnote{Although we derive the true dispersion here,  the 
characteristics of a grating used in a particular spectrograph usually describe this quantity
 in terms of
the ``reciprocal dispersion", i.e., a certain number of \AA\ per mm or \AA\  per pixel.  Confusingly, some refer to this
 as the dispersion rather than the reciprocal dispersion.}) of a given order $m$ with wavelength can be found by differentiating the grating equation:
\begin{equation}
\label{equ-diff}
d\theta/d\lambda = m/(\sigma \cos \theta) 
\end{equation}
for a given angle of incidence $i$.
Note, though, from Equation~\ref{equ-grating} that $m/\sigma=(\sin i + \sin \theta)/\lambda$, so
\begin{equation}
\label{equ-diff2}
d\theta/d\lambda =(\sin i + \sin \theta)/(\lambda \cos\theta)
\end{equation}
In the Littrow condition ($i=\theta$), the angular dispersion $d\theta/d\lambda$ is given by:
\begin{equation}
\label{equ-dispersion}
d\theta/d\lambda=(2/\lambda) \tan \theta.
\end{equation}

Consider a conventional grating spectrograph.
These must be used in low order ($m$ is typically 1 or 2) to avoid overlapping wavelengths from different orders, as
discussed further below.  These spectrographs are designed to be used with a small angle of incidence, i.e., the light
comes into and leaves the grating almost normal to the grating) and the only way of achieving high dispersion is by using
a large number of groves per mm (i.e., $\sigma$ is small in Equation~\ref{equ-diff}). (A practical
limit is roughly 1800 grooves per mm, as beyond this polarization effects limit the efficiency
of the grating.)  Note from the above that  $m/\sigma=2\sin\theta/\lambda$ in the Littrow condition.  So, if the angle of incidence
is very low, $\tan \theta \sim \sin \theta \sim \theta$, and the angular dispersion $d \theta/d\lambda \sim m/\sigma$.
If $m$ must be small to avoid overlapping orders, then the only way of increasing the dispersion is to decrease $\sigma$; i.e., use a larger
number of grooves per mm.  Alternatively, if the 
angle of incidence is very high, one can achieve high dispersion with a low number of groves per mm by operating in a high
order.  This is indeed how echelle spectrographs are designed to work, with typically $\tan \theta \sim$ 2 or greater. 
A typical echelle grating might have $\sim 80$ grooves/mm, so, $\sigma\sim25\lambda$ or so for visible light. The order
 $m$ must be of
order 50.
Echelle spectrographs can get away with this because
they cross-disperse the light (as discussed more below) and thus do not have to be operated in a particular low order to avoid
overlap. 

Gratings have a {\it blaze angle} that results in their having maximum efficiency for a particular value of m$\lambda$.   Think of the grating as having little triangular facets, so that if one
is looking at the grating perpendicular to the facets, each will act like a tiny mirror.  It is easy to
to envision the efficiency being greater in this geometry.
When speaking
of the corresponding {\it blaze wavelength,} $m=1$ is assumed.  When the blaze wavelength
is centered, the angle $\theta$ above is this blaze angle. The blaze wavelength is typically computed for the Littrow configuration, but that is seldom the case for astronomical spectrographs, so the effective
blaze wavelength is usually a bit different.  

As one moves away from the blaze wavelength $\lambda_b$,
gratings fall to 50\% of their peak efficiency at a wavelength 
\begin{equation}
\label{equ-blue}
\lambda=\lambda_b/m- \lambda_b/3m^2 
\end{equation}
on the blue side and 
\begin{equation}
\label{equ-red}
\lambda=\lambda_b/m+\lambda_b/2m^2
\end{equation}
on the red side\footnote{The actual efficiency is very complicated to calculate, as it depends upon blaze angle, polarization, and 
diffraction angle. See Miller \& Friedman (2003) and references therein for more discussion. Equations~\ref{equ-blue} and \ref{equ-red} are a modified version of the  ``2/3-3/2 rule" used
to describe the cut-off of a first-order grating as 2/3$\lambda_b$ and 3/2$\lambda_b$; see
Al-Azzawi (2007).}.
Thus the efficiency falls off faster to the blue than to the red, and the useful wavelength range is smaller for higher orders.
Each
spectrograph usually offers a variety of gratings from which to choose.  The selected grating can then be tilted, adjusting
 the central wavelength.

The light then enters the camera, which has a focal length of $L_{\rm cam}$.  The camera takes the dispersed light, and focuses it on the CCD, which is assumed to have 
a pixel size $p$, usually 15$\mu$m or so.  The camera design often dominates in the overall efficiency of most
spectrographs.

Consider the trade-off involved in designing a spectrograph.  On the one hand, one would
like to use a wide enough slit to include most of the light of a point source, i.e., be comparable
or larger than the seeing disk.  But the wider the slit, the poorer the spectral resolution,
if all other components are held constant. 
Spectrographs are designed so that when the slit width is some reasonable match to the
seeing (1-arsec, say) then the projected slit width on the detector corresponds to at least
2.0 pixels in order to satisfy the tenet of the Nyquist-Shannon sampling theorem.
The magnification factor
of the spectrograph is the ratio of the
focal lengths of the camera and the collimator, i.e., $L_{\rm cam}/L_{\rm coll} $.  This is a good approximation if all of the angles in the
spectrograph are small, but if the collimator-to-camera angle is greater than about 15 degrees one should include a factor of $r$,
the ``grating anamorphic demagnification", where $r=\cos(t+\phi/2)/cos(t-\phi/2)$, where $t$ is the grating tilt and $\phi$ is collimator-camera angle (Schweizer 1979)\footnote{Note that some observing manuals give the reciprocal of $r$.  As defined here, $r\leq1$.}.
 Thus the projected size of the slit  on the
detector will be $ W r L_{\rm cam} / L_{\rm coll},$ where $W$ is the slit width.
This projected size should be equal to at least 2 pixels, and preferably 3 pixels.

The spectral resolution is  characterized as $R=\lambda/\Delta \lambda$,
where $\Delta \lambda$ is the resolution element, the difference in wavelength between two equally strong (intrinsically skinny) spectral lines that can be resolved, corresponding to the projected slit width in wavelength units. 
Values of a few thousand are considered ``moderate resolution", while values of several tens of thousands are 
described as ``high resolution".  For comparison, broad-band filter  imaging has a resolution in the single digits,
while most interference-filter imaging has an $R\sim 100$.

The {\it free spectral range} $\delta \lambda$ is the difference between two wavelengths $\lambda_m$ and $\lambda_{(m+1)}$ in successive orders for a given angle $\theta$:
\begin{equation}
\label{equ-fsr}
\delta \lambda=\lambda_m-\lambda_{m+1} = \lambda_{m+1}/m.
\end{equation}
 For
conventional spectrographs that work in low order ($m$=1-3) the free spectral range is large, and blocking filters are needed to restrict the observation to a particular
order.  For echelle spectrographs, $m$ is large ($m \ge 5$) and the free spectral range is small, and 
the orders must be cross-dispersed to prevent overlap. 

Real spectrographs do differ in some regards from the simple heuristic description here.
For example, the collimator for a conventional long-slit spectrograph
must have a diameter that is larger
than would be needed just for the on-axis beam for a point source, because it has to
efficiently accept the light from each end of the slit as well as the center.  One would
like the exit pupil 
of the telescope
imaged onto the grating, so that small inconsistencies in guiding etc will minimize how
much the beam ``walks about" on the grating.
An off-axis paraboloid can do this rather well, but only if the geometry of the rest 
of the system matches it
rather well.

\subsubsection{Selecting a Blocking Filter}

\begin{figure}[hpt]
\plotone{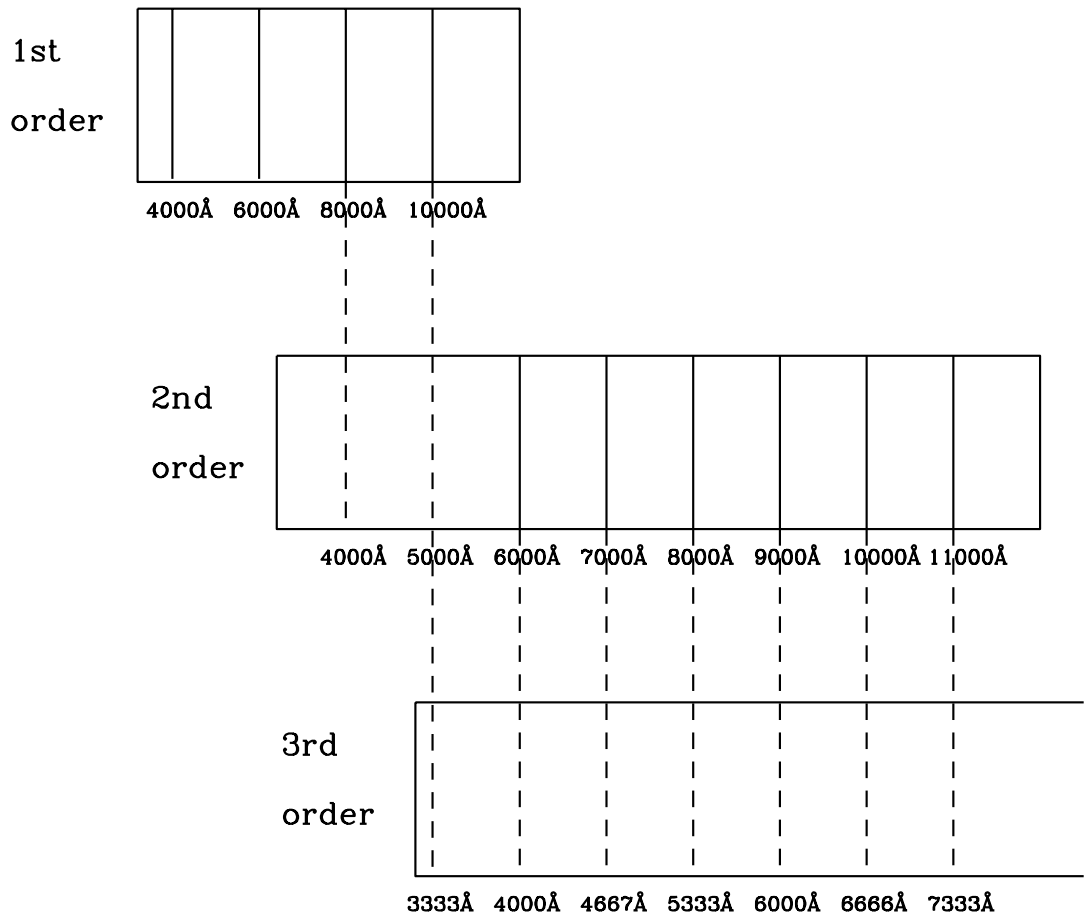}
\caption{\label{fig:orders} The overlap of various orders is shown. }
\end{figure}

There is often confusion over the use of order separation filters.  Figure~\ref{fig:orders} shows the potential problem.
Imagine trying to observe from 6000\AA\ to 8000\AA\ in first order.   At this particular angle,
one will encounter 
overlapping light from 3000\AA\ to 4000\AA\ in second order, and, in principle, 2000\AA\ to 2666\AA\ in third order, etc.  

\begin{figure} [htp]
\epsscale{0.5}
\plotone{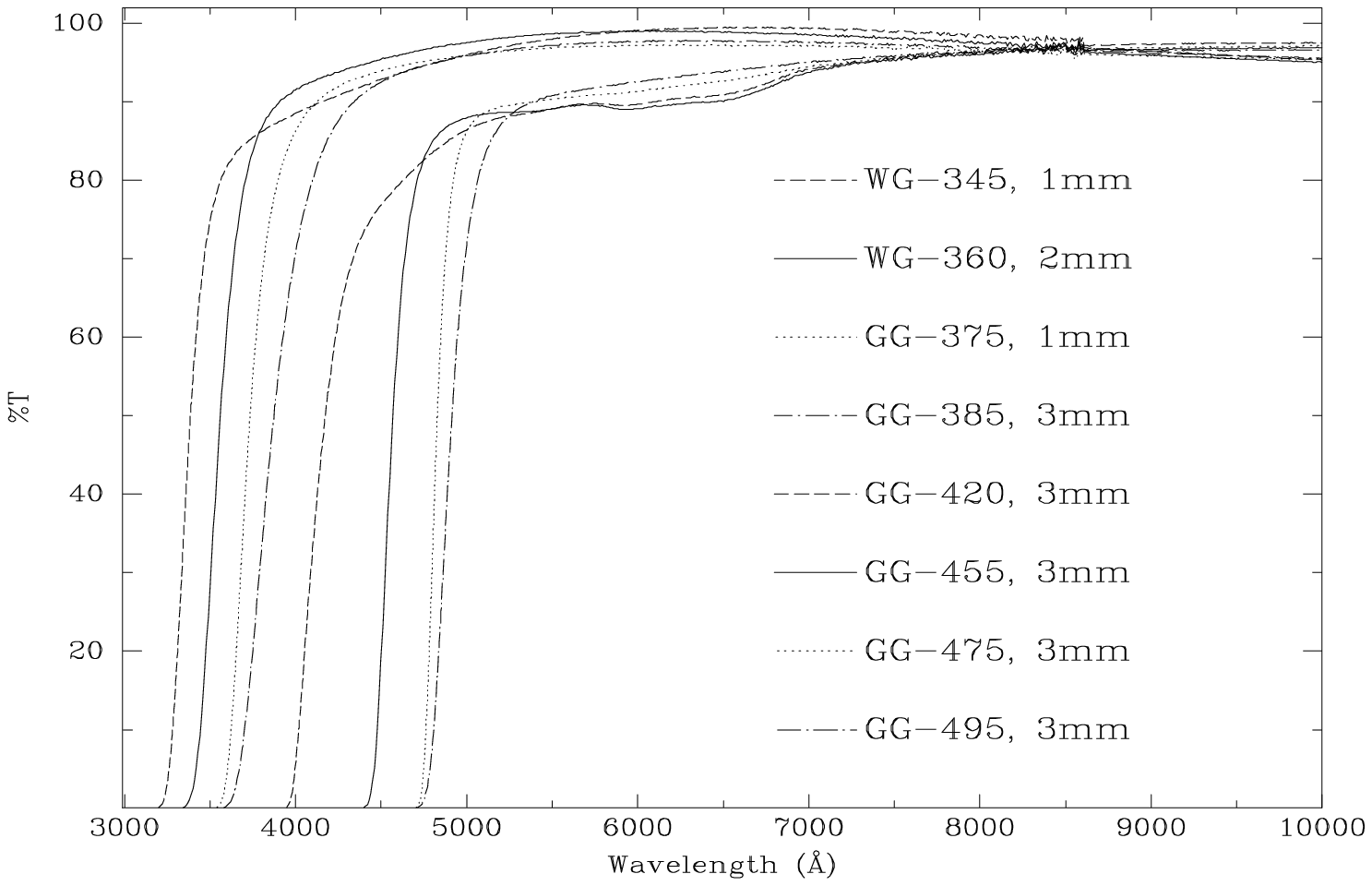}
\plotone{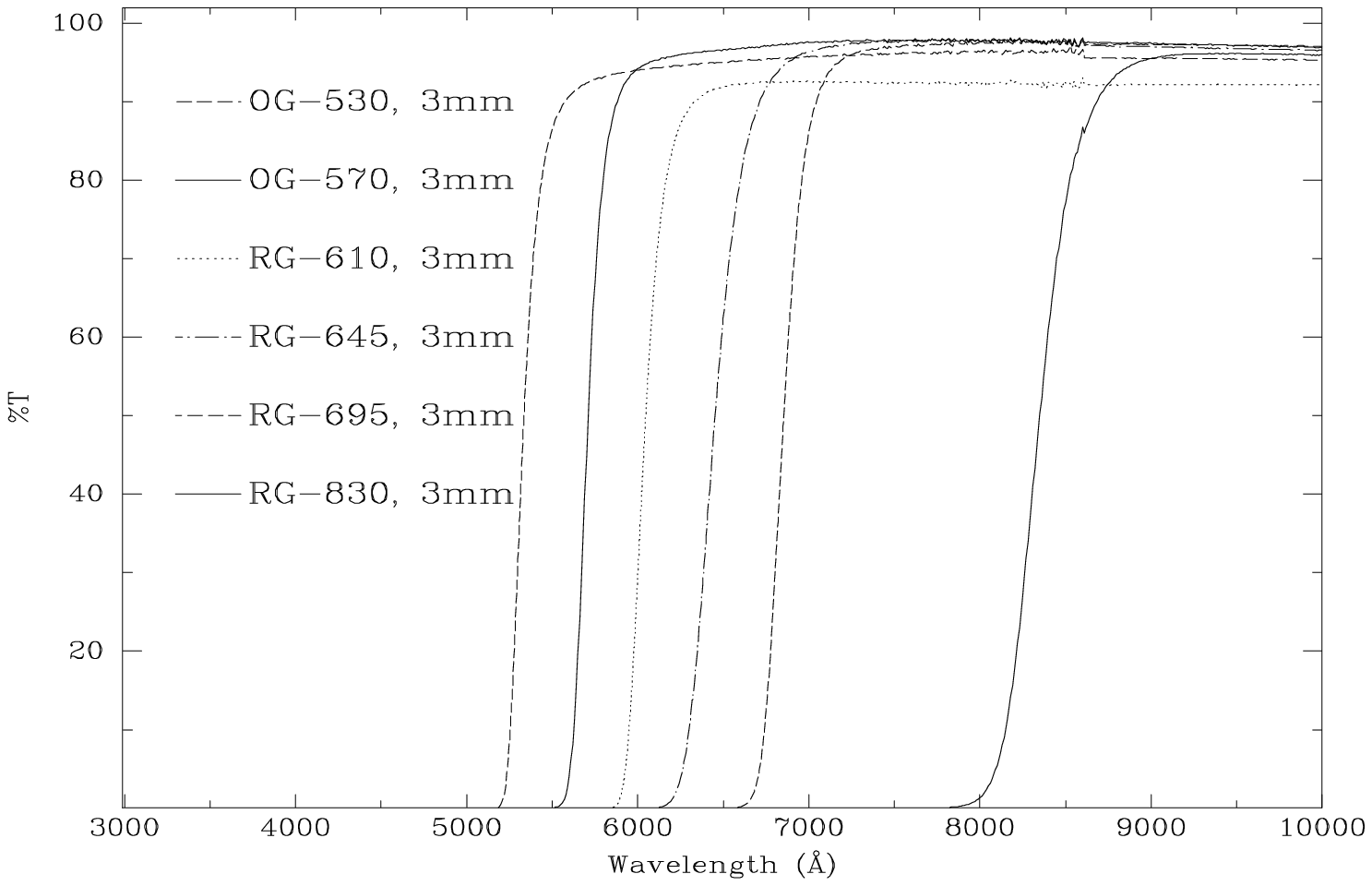}
\plotone{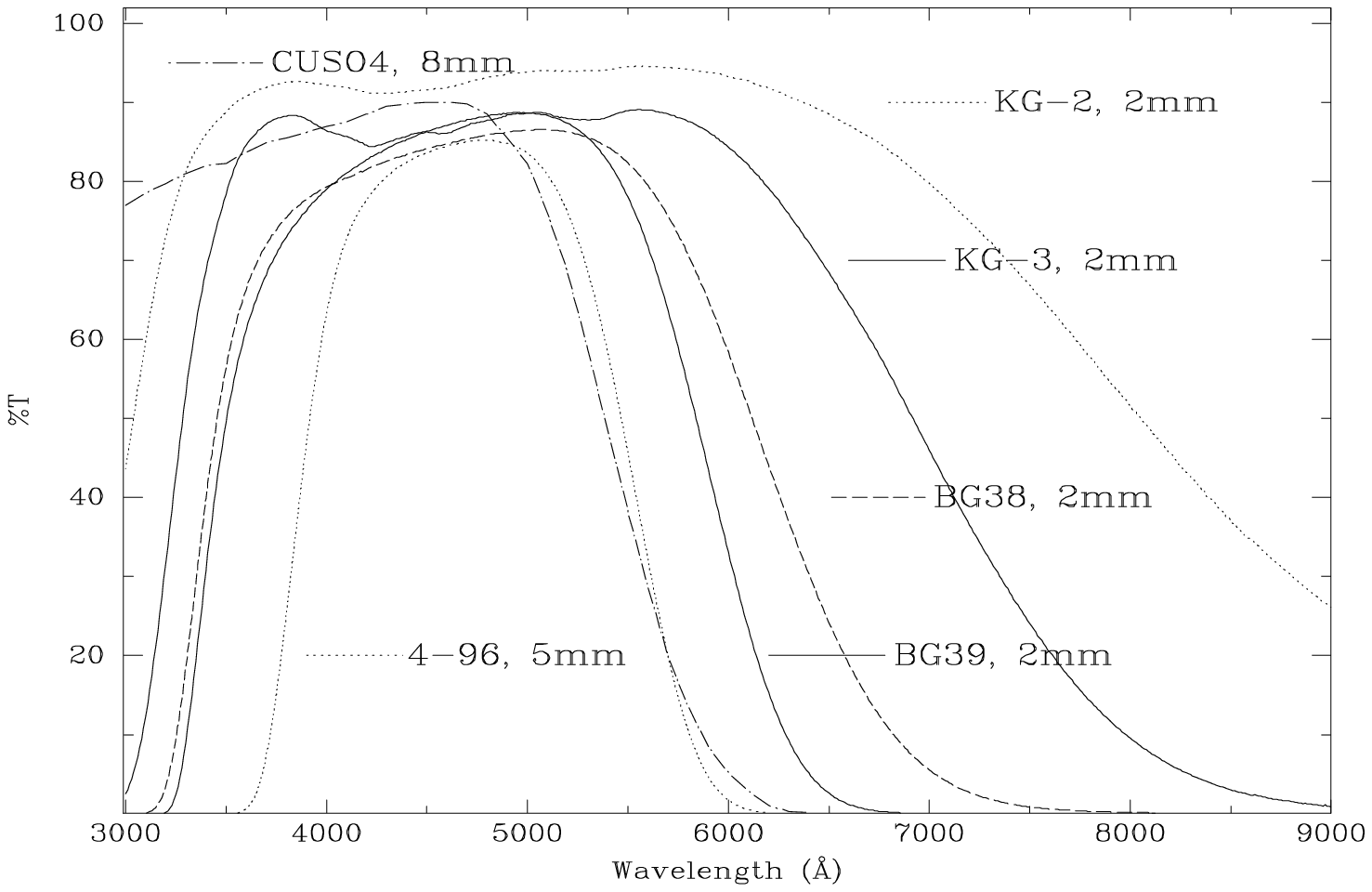}
\caption{\label{fig:filters}Examples of the transmission curves of order blocking filters, taken from Massey et al.\  (2000).}
\end{figure}

Since the atmosphere transmits very little light below 3000\AA, there is no need to worry about third or higher orders.  However,
light from 3000-4000\AA\ {\it does} have to be filtered out. There are usually a wide variety of blue cut-off
filters to choose among; these cut off the light in the blue but pass all the light longer than a particular wavelength.  In this example,
any cut-off filter that passed 6000\AA\ and higher would be fine.  The transmission curves of some typical order blocking
filters are shown in Figure~\ref{fig:filters}.  The reader will see that there are a number of
good choices, and that either a GG455, GG475, GG495, OG530, or an 
OG570 filter could be used.  The GG420 might work, but it looks as if it is still passing {\it some} light at 4000\AA, so why take the chance?

What if instead one wanted to observe from 4000\AA\ to 5000\AA\ in second order?   Then there is an issue
about first order red light
contaminating the spectrum, from 8000\AA\ on.  Third order light {\it might} not be a problem---at most there
would be a little 3333\AA\ light at 5000\AA,
but one could trust the source to be faint there and for the atmosphere to take its toll.  So, a good choice for a blocking filter would seem
be a CuSO$_4$ filter.  However, 
one should be relatively cautious though in counting on the atmosphere to exclude light. Even though many astronomers would argue
that the atmosphere doesn't transmit ``much" in the near-UV,  it is worth noting that actual extinction at 3200\AA\ is typically
only about 1 magnitude per airmass, and is  0.7 mag/airmass at 3400\AA.
So, in this example if one were using a very blue spectrophotometric standard to flux calibrate
the data, one could only count on its second-order flux at 3333\AA\ being attenuated by a factor of 2 (from  the atmosphere) and
another factor of 1.5  (from the higher dispersion of third order).   One might be better off using a BG-39 filter (Figure~\ref{fig:filters}).

One can certainly find situations for which no available filter will do the job.  If  instead one had wanted to observe from 4000\AA\ to 6000\AA\ in 
second order, one would have such a problem: not only does the astronomer now have to worry about $>$8000\AA\ light from first order, but also about $<$4000\AA\ light from 3rd order.  And there simply is no good glass blocking filter that transmits well from 4000-6000\AA\ but
also blocks below 4000\AA\ and long-wards of 8000\AA.  One could buy a special (interference) filter that did this but these tend to be
rather expensive and may not transmit as well as a long pass filter.  The only good solution in this
situation is to observe in first order with a suitably blazed grating.

\subsubsection{Choosing a grating}
\label{Sec-grating}
What drives the choice of one grating over another?   There usually needs to be some minimal spectral resolution, 
and some minimal
wavelength coverage.  For a given detector these two may be in conflict; i.e., if there are only 2000 pixels and a minimum
(3-pixel) resolution of 2\AA\ is needed, then no more than about 1300\AA\ can be covered in a single exposure. 
The larger the number of lines per mm,
the higher the dispersion (and hence resolution) for a given order.  Usually the observer also has in mind a specific wavelength region,
e.g.,  4000\AA\ to 5000\AA.    There may still be various choices to be made.  For instance, a 1200 line/mm grating blazed at 4000\AA\ and
a 600 line/mm grating blazed at 8000\AA\ may be (almost) equally good for such a project, as the 600 line/mm could be used in second order
and will then have the same dispersion and effective blaze as the 1200 line grating.  The primary difference is that the efficiency will fall
off much faster for the 600 line/mm grating used in second order.  As stated above (Equations~\ref{equ-blue} and \ref{equ-red}), gratings
fall off to 50\% of their peak efficiency at roughly $\lambda_b/m-\lambda_b/3^2$ and
$\lambda_b/m+\lambda_b/2m^2$ where $\lambda_b$ is the first-order blaze wavelength
and $m$ is the order. 
So, the 4000\AA\ blazed 1200 line/mm grating used in first order will fall to 50\% by roughly 6000\AA. 
 However, the
8000\AA\ blazed 
600 line/mm grating used in second order will fall to 50\% by 5000\AA.  Thus most likely the first order grating would be a better choice, although one should check the efficiency curves
 for the specific gratings (if such are available) to make sure one is making the right choice.
Furthermore, it would be easy to block unwanted light if one were operating in second order in this example, but generally it is a lot easier
to perform blocking when one is operating in first order, as described above.

\subsection{Conventional Long-Slit Spectrographs}
\label{Sec-longslit}

Most of what has been discussed so far corresponds to a conventional long-slit spectrograph, the simplest type of astronomical spectrograph,
and in some ways the most versatile.  The spectrograph can be used to take spectra of a bright star or a faint quasar, and the long-slit
offers the capability of excellent sky subtraction.  
Alternatively the long-slit can be used to obtain spatially resolved spectra of extended sources, such as galaxies (enabling kinematic,
abundance, and population studies) or HII regions.  They are usually easy to use, with straightforward acquisition using a TV imaging the
slit, although in some cases (e.g., IMACS on Magellan, discussed below in \S~\ref{Sec-multi}) the situation is more complicated.

Table~1 provides characteristics for a number of commonly used long-slit spectrographs.  Note that the resolutions are given
for a 1-arcsec wide slit.

\leftskip=-30pt
\begin{tabular} [h] {l l l l l l}
\multicolumn{6}{c}{Table 1. Some Long Slit Optical Spectrographs} \\ \hline \hline
\multicolumn{1}{c}{Instrument}
&\multicolumn{1}{c}{Telescope}
&\multicolumn{1}{c}{Slit}
&\multicolumn{1}{c}{Slit scale}
&\multicolumn{1}{c}{$R$}
&\multicolumn{1}{l}{Comments} \\  
&&\multicolumn{1}{c}{length} & \multicolumn{1}{c}{(arcsec/pixel)}\\ \hline
LRIS & Keck I & 2.9' & 0.14 & 500-3000 & Also multi-slits\\
GMOS&Gemini-N,S & 5.5' & 0.07 & 300-3000 & Also multi-slit masks\\
IMACS&Magellan I    &  27' &  0.20    &500-1200 & f/2.5 camera,  also multi-slit masks\\
           &           & 15' &   0.11    &300-5000 & f/4 camera, also multi-slit masks \\
Goodman & SOAR 4.2-m &  3.9' &   0.15 & 700-3000 &also multi-slit masks \\
RCSpec & KPNO 4-m & 5.4' & 0.7 &  300-3000 \\     
RCSpec & CTIO 4-m  & 5.4'  & 0.5  &  300-3000 \\
STIS      & HST           & 0.9'  & 0.05  &  500-17500  \\
GoldCam & KPNO 2.1-m & 5.2' & 0.8 & 500-4000 \\
RCSpec   & CTIO 1.5-m  &  7.5' & 1.3 & 300-3000\\ 
\hline \hline
\end{tabular}

\leftskip=0pt

\subsubsection{An Example: The Kitt Peak RC Spectrograph}
\label{Sec-RCSpec}

\begin{figure}[htp]
\epsscale{1.05}
\plotone{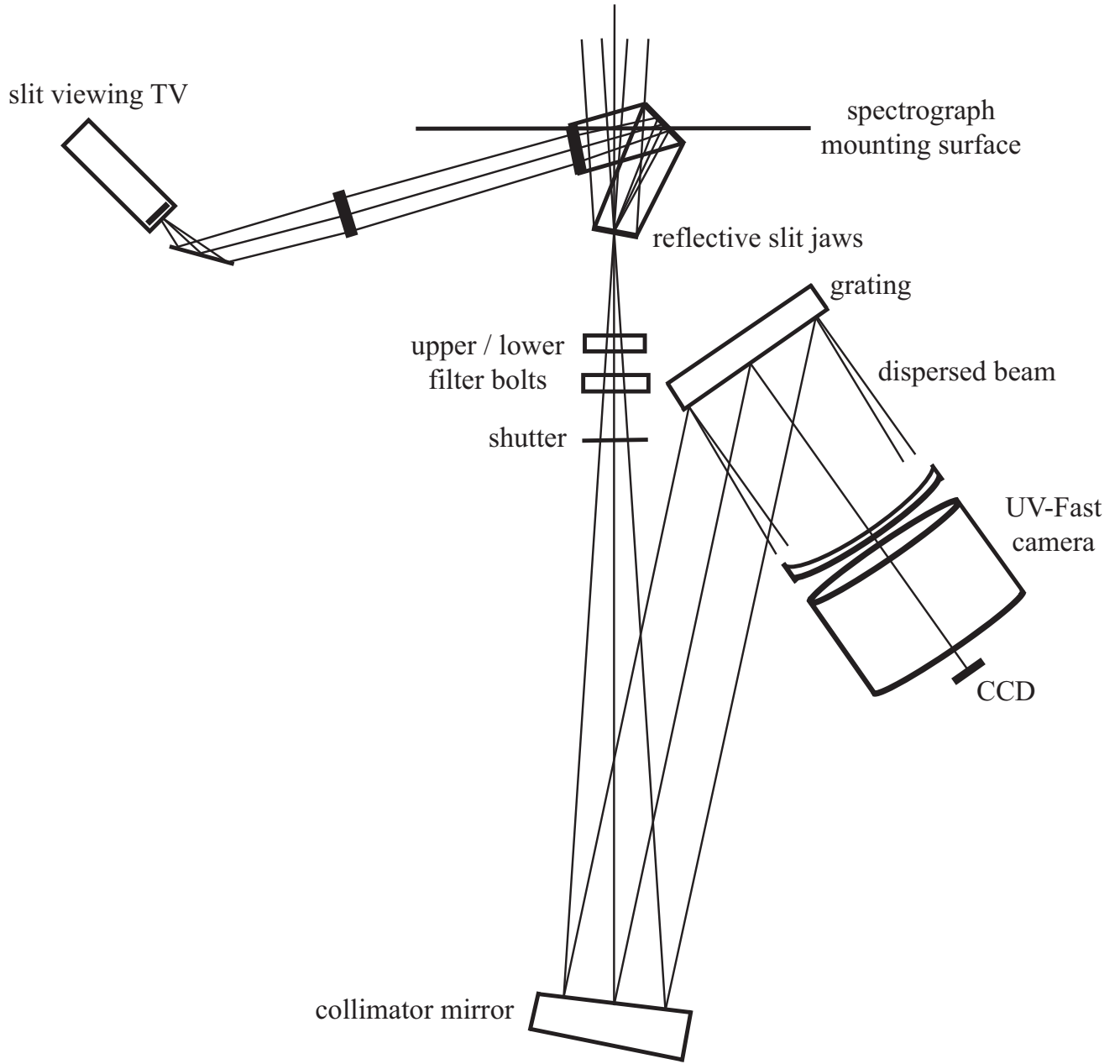}
\caption{\label{fig:rc} The optical layout of the Kitt Peak 4-meter RC Spectrograph.  This figure is based upon an illustration from
the Kitt Peak instrument manual by James DeVeny.}
\end{figure}

Among the classic workhorse instruments of the Kitt Peak and Cerro Tololo Observatories have been the Ritchey-Chretien (RC) 
spectrographs on the Mayall and Blanco 4-meter telescopes.
Originally designed in the era of photographic plates, these instruments were subsequently outfitted with CCD cameras.
The optical diagram for the Kitt Peak version is shown in Figure~\ref{fig:rc}. 
It is easy to relate this to the heuristic schematic of Figure~\ref{fig:spect}.  There are a few
additional features that make using the spectrograph practical.  First, there is a TV mounted to view the slit jaws, which are highly
reflective on the front side.  This makes it easy to position
an object  onto the slit.   The two filter bolts allow inserting either neutral density filters or order blocking filters into
the beam.  A shutter within the spectrograph controls the exposure length.   The f/7.6 beam is turned into collimated (parallel) light by the
collimator mirror before striking the grating.  The dispersed light then enters the camera, which images the spectrum onto the CCD.  

The ``UV fast camera" used with the CCD has a focal length that provides an appropriate
magnification factor.  
The  magnification of the spectrograph $r L_{\rm cam}/L_{\rm coll}$ is $0.23r$,
with $r$ varying from 0.6 to 0.95, depending upon the grating.   The CCD has 24$\mu$m
pixels and thus for 2.0 pixel resolution one can open the slit to 250$\mu$m, corresponding to
1.6 arcsec,   a good match to less than perfect seeing.

The spectrograph has 12 available gratings to choose among, and  their properties are
given in Table~2.

\clearpage

\begin{tabular} [h] {l c c c c c c c}
\multicolumn{8}{c}{Table 2. 4-m RC Spectrograph Gratings} \\ \hline \hline
\multicolumn{6}{c}{  }
&\multicolumn{1}{c}{Reciprocal}\\
\multicolumn{1}{c}{Name}
 & \multicolumn{1}{c}{l/mm}
 & \multicolumn{1}{c}{order}
 & \multicolumn{1}{c}{Blaze}
 & \multicolumn{2}{c}{Coverage(\AA)}
 & \multicolumn{1}{c}{Dispersion}
 & \multicolumn{1}{c}{Resolution$^a$} \\  \cline{5-6}
\multicolumn{3}{c}{}
 & \multicolumn{1}{c}{(\AA)}
 & \multicolumn{1}{c}{1500 pixels}
 & \multicolumn{1}{c}{1700 pixels}
 & \multicolumn{1}{c}{(\AA/pixel)}
 & \multicolumn{1}{c}{(\AA)} \\ \hline
BL 250 & 158 &1& 4000 & \multicolumn{2}{c}{1 octave$^b$} & 5.52 & 13.8\\
BL 400 & 158 &1& 7000 & \multicolumn{2}{c}{1 octave$^b$} & 5.52 & 13.8\\
       &     &2& 3500 & \multicolumn{2}{c}{$<$4100$^c$} & 2.76 & 6.9\\
KPC-10A & 316 &1& 4000 &  4100 & 4700 & 2.75 & 6.9\\
BL 181  & 316 &1& 7500 &  4100 & 4700 & 2.78 & 7.0\\
       &     &2& 3750 & \multicolumn{2}{c}{$<$2000$^c$} & 1.39 & 3.5\\
KPC-17B & 527 &1& 5540 &  2500 & 2850 &1.68 & 4.2\\
BL 420  & 600 &1& 7500 &  2300 & 2600 &1.52 & 3.8\\
       &     &2& 3750 &  1150 & 1300 & 0.76 & 1.9\\
KPC-007 &632 &1& 5200 &  2100 & 2350 & 1.39 & 3.5\\
KPC-22B & 632 &1& 8500 &  2150 & 2450 & 1.44 & 3.6\\
        &    &2& 4250 &  1050 & 1200 & 0.72 & 1.8\\
BL 450  & 632 &2&5500 &  1050 & 1200 & 0.70 & 1.8\\
        &     &3&3666 &   690 & 780  & 0.46 & 1.2\\
KPC-18C& 790 &1& 9500 &  1700 & 1900 & 1.14 & 2.9\\
       &     &2& 4750 &   850 & 970 & 0.57 & 1.4\\
KPC-24 & 860 &1&10800 &  1600 & 1820 & 1.07 & 2.7\\
       &     &2& 5400 &   800 & 900 & 0.53 & 1.3 \\
BL 380  &1200 &1& 9000 &  1100 &1250 & 0.74 & 1.9\\
       &     &2& 4500 &   550 & 630 &0.37 & 0.9\\
\hline \hline
\end{tabular}

\smallskip
\noindent
Notes: (a) Based on 2.5 pixels FHWM
corresponding to 300 $\mu$m slit (2 arcsec) with no anamorphic factor.
  (b) Spectral coverage limited
by overlapping orders. (c) Spectral coverage limited by grating
efficiency and atmospheric cut-off.

How does one choose from among all of these gratings? Imagine that a particular
project required obtaining 
radial velocities at the Ca II triplet ($\lambda \lambda 8498, 8542, 8662$) as well
as  MK classification spectra (3800-5000\AA) of the same objects.  For the
radial velocities, suppose that  3-5 km s$^{-1}$ accuracy
was needed, a pretty sensible limit to be achieved with a spectrograph mounted on the
backend of a telescope and the inherent flexure that comes with this.  
At the wavelength of the Ca II lines, 5 km s$^{-1}$ corresponds to how
many angstroms?   A velocity $v$ will just be $c \Delta \lambda/\lambda$ 
according to the Doppler formula.  Thus for an uncertainty of 5 km s$^{-1}$
one would like to locate the center of  a spectral line to 0.14\AA.  
In general it is easy to centroid to 1/10th of a pixel,
and so one needs a reciprocal dispersion smaller than about 1.4\AA/pixel.
 It is hard to observe in the red in 2nd order so one probably wants to look at gratings
 blazed at the red.  One could do well with KPC-22B (1st order blaze at 8500\AA) with
a  reciprocal dispersion of 1.44\AA\ per pixel.  One would have to employ some sort of blocking filter
to block the blue 2nd order light, with the 
choice dictated by exactly how the Ca II triplet was centered within the 2450\AA\ wavelength
coverage that the grating would provide. The blue
spectrum can then be obtained by just changing the blocking filter to block 1st order red while
allowing in 2nd order blue. The blue  
1200\AA\ coverage would be just right for covering the MK classification region from
3800-5000\AA.  By just changing the blocking filter, one would then obtain coverage
in the red from  7600\AA\ to 1$\mu$m,  with the Ca II lines relatively well centered.  
An OG-530 blocking filter would be a good choice for the 1st order red observations.  For
the 2nd order blue, 
either the BG-39 or CuSO$_4$ blocking filters would be a good choice
 as either would filter out light
with a wavelength of $>$7600, as shown in Figure~\ref{fig:filters}\footnote{The BG-38 also looks like it would do a good job, but careful inspection of the actual transmission curve reveals that it has a significant red leak at wavelengths $>$9000\AA.  It's a good idea to check the actual numbers.}.
The advantage to this set up would be that by just moving the filter bolt from one
position to another one could observe in either wavelength region.

\subsection{Echelle Spectrographs}

In the above sections the issue of order separation for conventional slit spectrographs have been discussed extensively.   Such spectrographs
image a single order at a given time. On a large two-dimensional array most of the area is ``wasted" with the spectrum of the night sky,
unless one is observing an extended object, or unless the slit spectrograph is used with a multi-object slit mask, as described below.

\begin{figure}[htp]
\epsscale{0.8}
\plotone{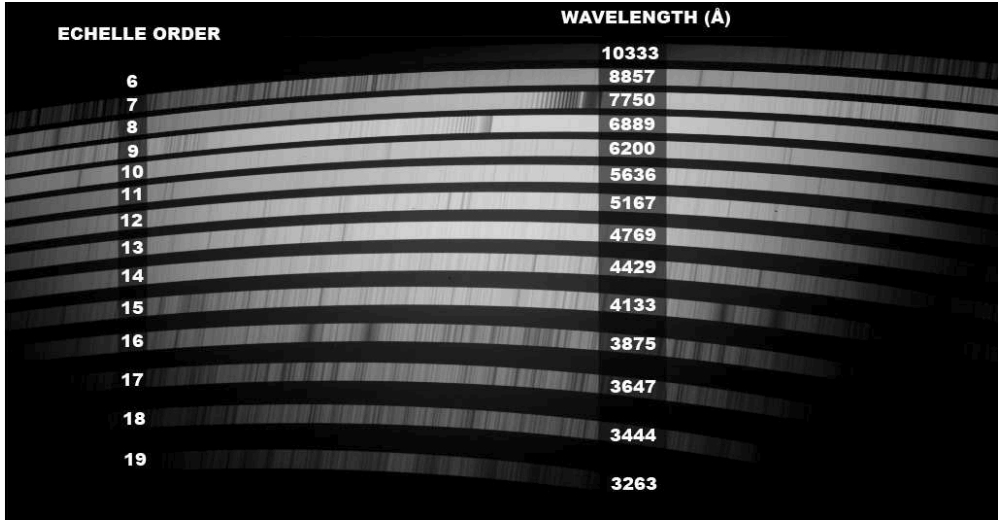}
\caption{\label{fig:MagE} The spectral format of MagE on its detector.  The various orders are shown, along with the approximate
central wavelength.}
\end{figure}

Echelle spectrographs use a second dispersing element (either a grating or a prism) to {\it cross disperse} the various orders, spreading
them across the detector.  An example is shown in Figure~\ref{fig:MagE}.  The trade off with designing echelles and selecting
a cross-dispersing grating is to balance greater wavelength coverage, which would have adjacent orders crammed close together,  with the
desire to have a ``long" slit to assure good sky subtraction, which would have adjacent orders more highly separated\footnote{Note that some ``conventional" near-IR spectrographs are cross dispersed in order to take advantage of the fact that the JHK bands are coincidently
centered one with the other in orders 5, 4, and 3  respectively (i.e., 1.25$\mu$m, 1.65$\mu$m, and 2.2$\mu$m).}.

Echelles are designed to work in higher orders (typically $m\geq 5$)
and both $i$ and $\theta$ in the grating equation (\S~\ref{Sec:Basics})
are large\footnote{Throughout this section the term "echelle" is used to include the so-called echellette.  Echellette gratings
have smaller blaze angles ($\tan \theta \le 0.5$) and are used in lower orders ($m=$5-20) than
classical echelles ($\tan \theta \ge 2$, $m=$20-100.)  However, both are cross-dispersed and provide higher dispersions
than conventional grating spectrographs.}.  At the detector one obtains multiple orders side-by-side.  Recall from 
above that the wavelength difference $\delta \lambda$ between successive orders at a given angle 
(the free spectral range) will
scale inversely with the order number (Equation~\ref{equ-fsr}).   Thus for low order numbers (large central wavelengths) 
the free spectral range will be larger.

The angular spread $\delta \theta$ of a single order will be $\delta \lambda d\theta/d\lambda$.   Combining this with the equation
for the angular dispersion (Equation~\ref{equ-dispersion}) then yields:
$$\lambda/\sigma \cos  \theta = \delta \lambda (2/\lambda) \tan \theta,$$ and hence the wavelength covered in a  single
order will be 
\begin{equation}
\label{equ-perorder}
\delta \lambda= \lambda^2/(2\sigma \sin \theta).
\end{equation}
The angular spread of a single order will be
\begin{equation}
\label{equ-perorder2}
\Delta \theta = \lambda /(\sigma \cos \theta).
\end{equation}

Thus the number of angstroms covered in a single order will increase by the {\it square} of the wavelength (Equation~\ref{equ-perorder}), while
the length of each order increases only {\it linearly} with each order (Equation~\ref{equ-perorder2}). This is apparent from  
Figure~\ref{fig:MagE},
as the shorter wavelengths (higher orders) span less of the chip.  
At lower orders the wavelength coverage actually exceeds the length
of the chip.   Note that the
same spectral feature may be found on adjacent orders, but usually the blaze function
is so steep that good signal is obtained for a feature in one particular order.  This can
 be seen for the very strong H and K Ca II lines apparent near the center of order 16 and to
the far right in order 15 in Figure~\ref{fig:MagE}.

If a grating is used as the cross disperser, then the separation between orders
should increase for lower order numbers (larger wavelengths) as gratings provide
fairly linear dispersion and the free spectral range is larger for lower order numbers.
(There is more of a difference in the wavelengths between adjacent orders and hence
the orders will be more spread out by a cross-dispersing grating.)  However, Figure~\ref{fig:MagE} 
shows that just the opposite is true for MagE:  the  separation between
adjacent orders actually
decreases towards lower order numbers.  Why?  MagE uses prisms for cross-dispersing,
and (unlike a grating) the dispersion of a prism is greater in the blue than in the red.
In the case of MagE the decrease in dispersion towards larger wavelength (lower orders)
for the cross-dispersing prisms more than compensates for the increasing separation
in wavelength between adjacent orders at longer wavelengths.

Some echelle spectrographs are listed in Table~3.  HIRES, UVES, and the KPNO 4-m echelle have a variety of gratings and cross-dispersers
available; most of the others provide a fixed format but give nearly full wavelength coverage in the optical in a single exposure.

\begin{tabular} [h] {l l r l l}
\multicolumn{5}{c}{Table 3. Some Echelle Spectrographs} \\ \hline \hline
\multicolumn{1}{c}{Instrument}
&\multicolumn{1}{c}{Telescope}
&\multicolumn{1}{c}{$R$ (1 arcsec slit)}
&\multicolumn{1}{c}{Coverage(\AA)}
&\multicolumn{1}{l}{Comments} \\   \hline
HIRES & Keck I & 39,000 & Variable &   \\
ESI      & Keck II &  4,000  & 3900-11000& Fixed format\\
UVES  & VLT-UT2 & 40,000 & Variable & Two arms \\
MAESTRO & MMT & 28,000 & 3185-9850 & Fixed format \\
MIKE & Magellan II  & 25,000 & 3350-9500 & Two arms \\
MagE & Magellan II  & 4,100 & $<$3200-9850 &Fixed format \\
Echelle & KPNO 4-m & $\sim$30,000 & Variable & \\

\hline \hline
\end{tabular}

\leftskip=0pt

\subsubsection{An Example: MagE}

\begin{figure} [htp]
\epsscale{1.0}
\plotone{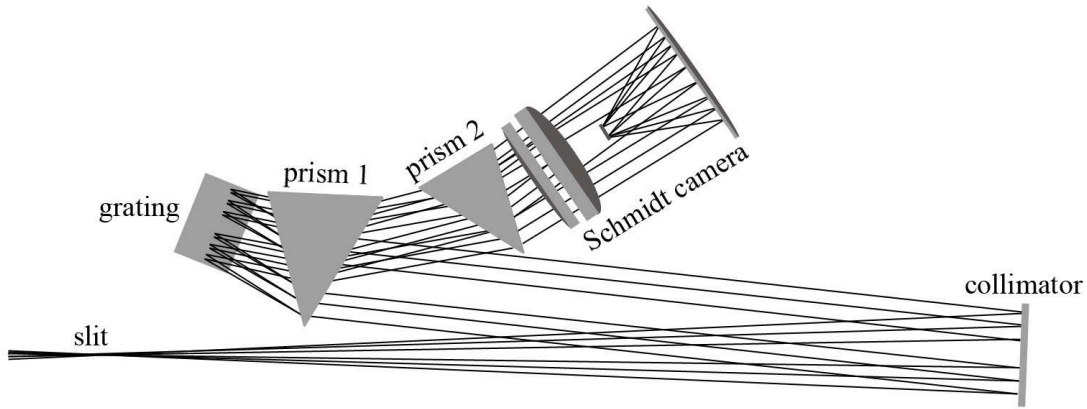}
\caption{\label{fig:mageopt} The optical layout of MagE.  Based upon Marshall et al.\ (2008).}
\end{figure}

The Magellan Echellette (MagE) was deployed on the Clay (Magellan II) telescope in November 2007, and provides full wavelength coverage
from 3200\AA\ to 10,000\AA\ in a single exposure, with a resolution $R$ of 4,100 with a 1 arsecond slit.  The instrument is described in detail by
Marshall et al.\ (2008).  The optical layout is shown in Figure~\ref{fig:mageopt}.  Light from the telescope is focused onto a slit,
and the diverging beam is then collimated by a mirror.  Cross dispersion is provided by two prisms, the first of which
is used in double pass mode, while the second has a single pass.  The echelle grating has 175 lines/mm and is used in a quasi-Littrow configuration.
The Echelle Spectrograph and Imager (ESI) used on Keck II has a similar design (Sheinis et al.\ 2002).  MagE has a fixed format
and uses orders 6 to 20, with central wavelengths of 9700\AA\ to 3125\AA, respectively.  

The spectrograph is remarkable for its extremely high throughput and ease of operation.  The spectrograph was optimized for use in the
blue, and the measured efficiency of the instrument alone is $>$30\% at 4100\AA.  (Including the telescope the efficiency is about 20\%.)
Even at the shortest wavelengths (3200\AA\ and below) the overall efficiency is 10\%.  The greatest challenge in using the instrument is
the difficulties of flat-fielding over that large a wavelength range.  This is typically done using a combination of in- and out-of-focus
Xe lamps to provide sufficient flux in the near ultra-violet, and quartz lamps to provide good counts in the red.  Some users have found
that the chip is sufficiently uniform that they do better by not flat-fielding the data at all; in the case of very high signal-to-noise one
can dither along the slit. (This is discussed in general in \S~\ref{Sec-flats}.) The slit length of MagE is 10 arcsec, allowing good sky subtraction for stellar sources, and still providing clean
separation between orders even at long wavelengths (Figure~\ref{fig:MagE}).

It is clear from an inspection of Figure~\ref{fig:MagE} that there are significant challenges to the data reduction: the orders are curved
on the detector (due to the anamorphic distortions of the prisms)
and in addition the spectral features are also tilted, with a tilt that varies along each order.   One
spectroscopic pundit has likened echelles to space-saving storage travel bags: a lot of things are packed together very efficiently, but extracting the particular sweater one wants can be a real challenge. 

\subsubsection{Coude Spectrographs}

Older telescopes have equatorial mounts, as it was not practical to utilize an
 altitude-azimuth (alt-az) design until modern computers were
available.  Although alt-az telescopes allow for a more compact design (and hence
a significant cost savings in construction of the telescope enclosure), the equatorial systems
provided the opportunity for a coude focus.  By adding three additional mirrors, one could
direct the light down the stationary polar axis of an equatorial system.  From there the light
could enter a large ``coude room", holding a room-sized spectrograph that would be
extremely stable.  Coude spectrographs are still in use at Kitt Peak National Observatory
(fed by an auxiliary 0.9-m telescope), McDonald Observatory (on the 2.7-m telescope), and at the
Dominion Astrophysical Observatory (on a 1.2-m telescope), among other places.  Although such spectrographs
occupy an entire room, the basic idea was the same, and these instruments afford very
high stability and high dispersion.  To some extent, these functions are now provided by high resolution instruments mounted on the Nasmyth foci of large alt-az telescopes, although these platforms
provide relatively cramped quarters to achieve the same sort of stability and dispersions
offered by the classical coude spectrographs.


\subsection{Multi-object Spectrometers}

There are many instances where an astronomer would like to observe multiple objects in the same  field of view, such as studies
of the stellar content of a nearby, resolved galaxy, the members of a star cluster, or individual galaxies in a group.  If the density of
objects is relatively high (tens of objects per square arcminute) and the field of view small (several arcmins) then one often will use a {\it slit mask} containing not one but
dozens or even hundreds of slits.  If instead the density of objects is relatively 
low (less than 10 per square arcminute)
but the field of view required is large (many arcmins) one can employ a multi-object fiber positioner feeding a bench-mounted
spectrograph. Each kind of device is discussed below.

\subsubsection{Multi-slit Spectrographs}
\label{Sec-multi}

\begin{figure} [htp]
\epsscale{0.35}
\plotone{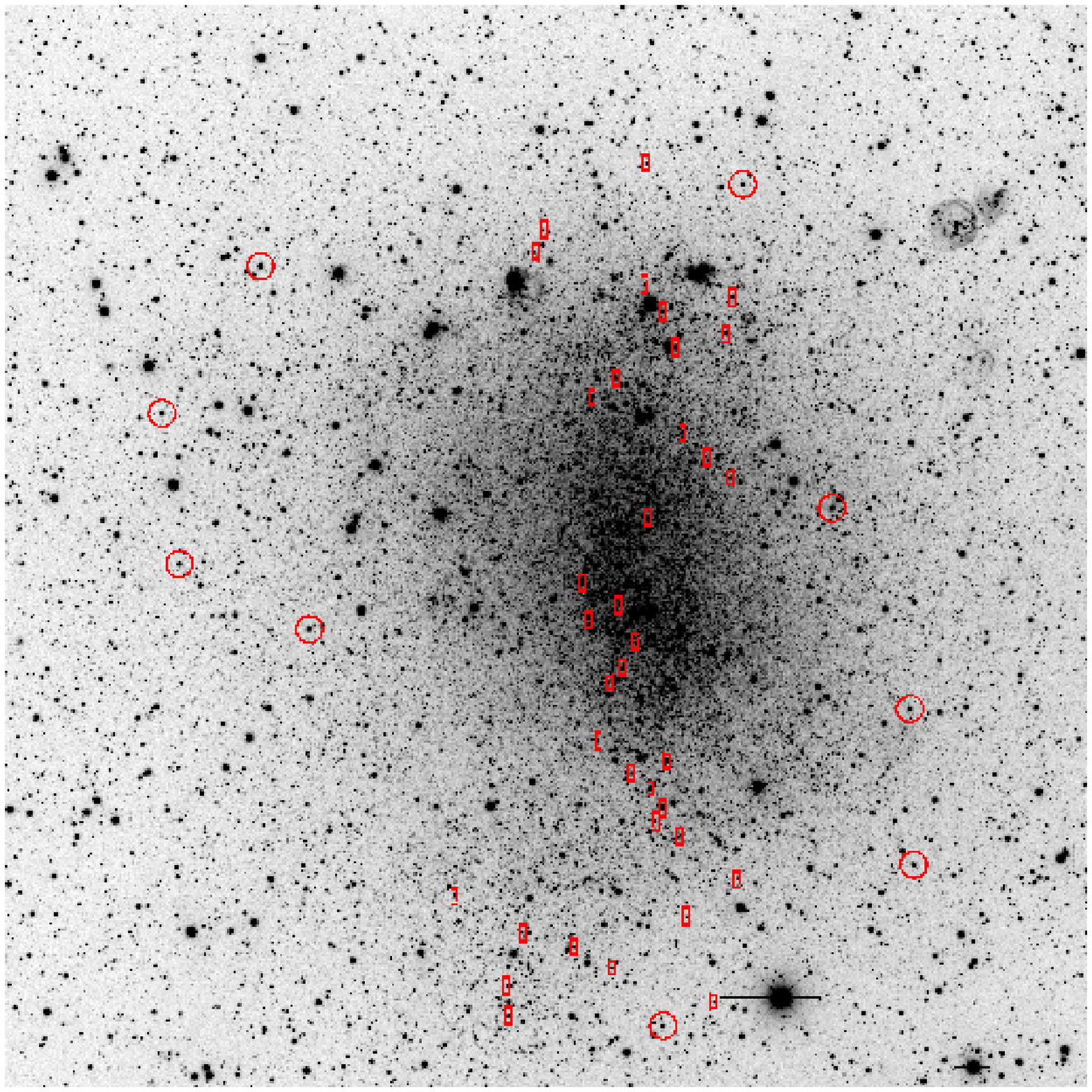}
\epsscale{0.48}
\plotone{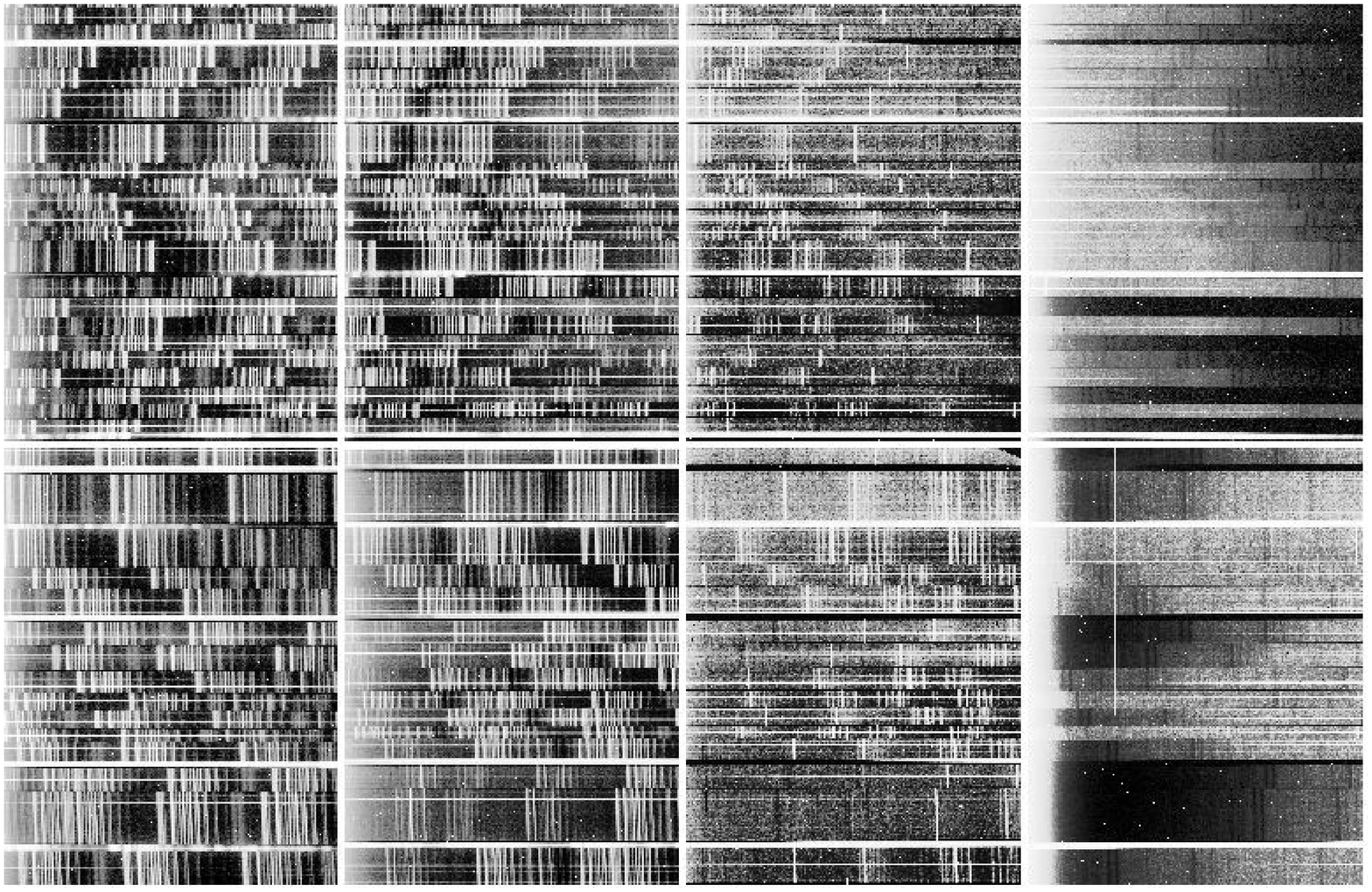}
\epsscale{0.5}
\plotone{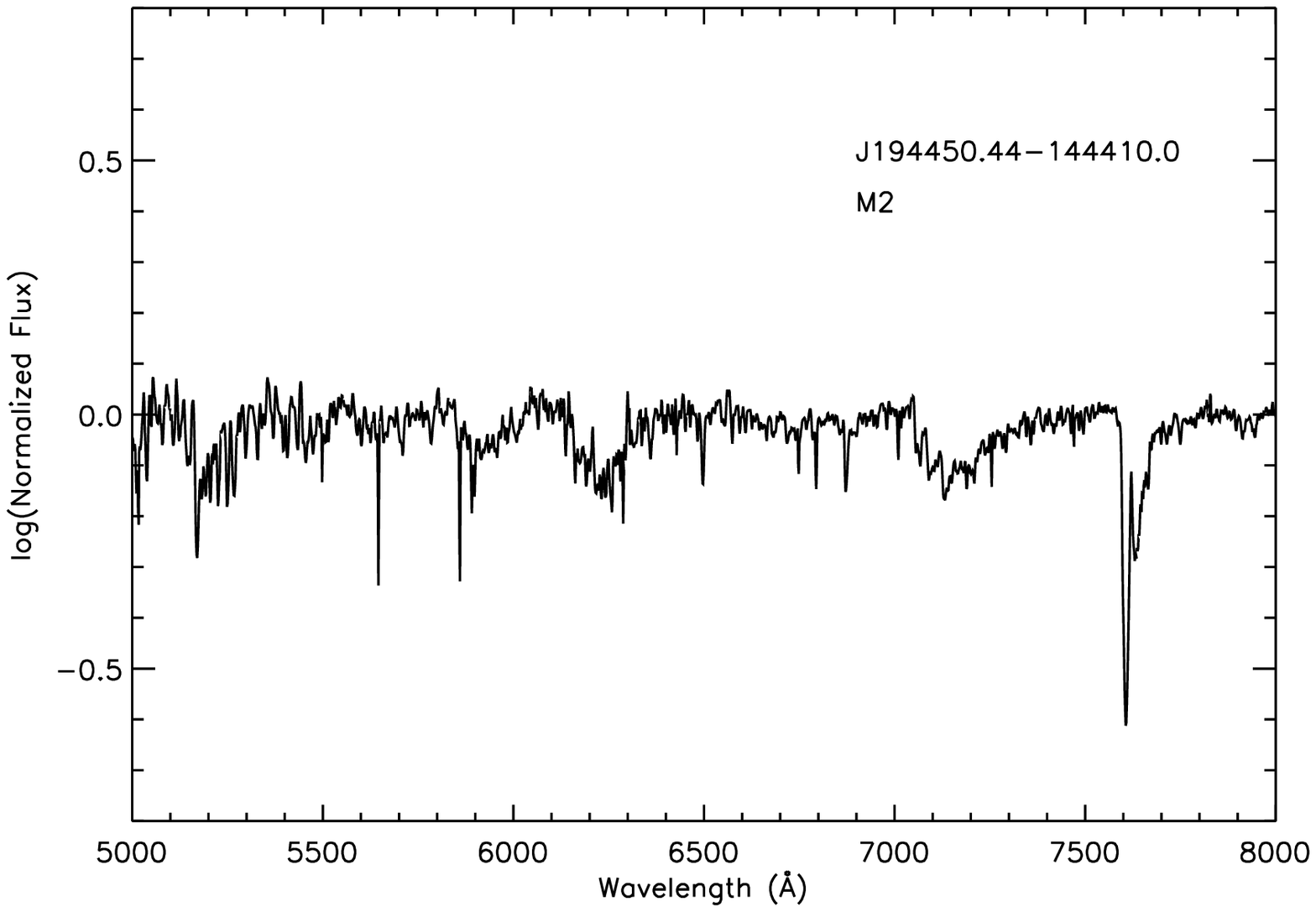}
\caption{\label{fig:mask}Multi-object mask of red supergiant candidates in NGC 6822.  The upper left figure shows an image of the Local Group galaxy NGC 6822, taken from the Local Group Galaxies Survey (Massey et al.\ 2007).  The red circles correspond to ``alignment" stars, and the small rectangles indicate the position of red supergiants candidates to be observed.  The slit mask consists of a large metal plate machined
with these
holes and slits. The upper right figure is the mosaic of the 8 chips of IMACS.  The vertical lines are night-sky emission lines, while the 
spectra of individual stars are horizontal narrow lines.  A sample of one such reduced spectrum, of an M2 I star, is shown in the lower
figure. These data were obtained by Emily Levesque, who kindly provided parts of this figure.}
\end{figure}

Several of the ``long slit" spectrographs described in \S~\ref{Sec-longslit} were really designed to be used with multi-slit masks.  
These masks allow
one to observe many objects at a time by having small slitlets machined into a mask 
at specific locations.  The design of these masks can be quite challenging, as the slits cannot overlap spatially on the mask.  An example 
is shown in Figure~\ref{fig:mask}.  Note that in addition to slitlet masks, there are also small alignment holes centered on modestly
bright stars, in order to allow the rotation angle of the instrument and the position of the telescope to be set exactly.

In practice, the slitlet masks need to be at least  5 arcsec in length in order to allow sky subtraction on either side of a point source.  Allowing
for some small gap between the slitlets, one can then take the field of view and divide by 
a typical slitlet length to estimate the maximum number of slitlets an instrument would
accommodate. 
Table 1 shows that an instrument such as
GMOS on the Gemini telescopes has a maximum (single) slit length of 5.5 arcmin, or 330 arcsec.  Thus
at most, one might be able to cram in 50 slitlets, were the objects of interest properly aligned on
the sky to permit this.  An instrument with a 
larger field of view, such as IMACS (described below) really excels in this game, as over a hundred slitlets can be machined onto a single mask.

Multi-slit masks offer a large multiplexing advantage, but there are some disadvantages as well.  First, the masks typically need to be machined
weeks ahead of time, so there is really no flexibility at the telescope  other than to change
exposure times.  Second, the setup time for such masks is non-negligible, 
usually of order 15 or 20 minutes.  This is not an issue when exposure times are long, but
can be a problem if the objects are bright and the exposure times short.
 Third, and perhaps most significantly, the
wavelength coverage will vary from slitlet to slitlet, depending upon  location within the field.  As shown in the example of
Figure~\ref{fig:mask}, 
the mask field has been rotated so that the slits extend north and south, and indeed the body
of the galaxy is mostly located north and south, minimizing this problem.  
The alignment holes are located well to the east and west, but one does
not care about their wavelength coverage.  In general, though, 
if one displaces a slit off center by X arcsec,
then the central wavelength of the spectrum associated with that slit is going to shift by Dr(X/p), where p is the scale
on the detector in terms of arcsec per pixel, r is the anamorphic demagnification factor
 associated with this particular grating and tilt ($\leq$1), and D is the dispersion in \AA\ per pixel. 

Consider the case of the IMACS multi-object spectrograph.  Its basic parameters
are included in Table 1, and the instrument is described in more
detail below. The field of view with the f/4 camera is 15 arcmins $\times$15 arcmins.  
A slit on the edge of the field of view will be displaced by 7.5 arcmin,  or 450 arcsec.
With a scale of 0.11 arcsec/pixel this corresponds to an offset of 4090 pixels ($X/p=4090$).   With a 1200 line/mm grating centered at 4500\AA\ for a slit
on-axis, the wavelength coverage is 3700-5300\AA\ with a dispersion $D=0.2$\AA/pixel.  The anamorphic demagnification is 0.77.  
So, for a slit on the edge the wavelengths are shifted by 630\AA, and the spectrum is centered at 5130\AA\ and covers 4330\AA\ to 5930\AA.
On the other edge the wavelengths will be shifted by -630\AA, and will cover 3070\AA\ to 4670\AA.  The only wavelengths in common
to slits covering the entire range in X is thus 4330\AA-4670\AA, only 340\AA!

\paragraph{Example: IMACS}
\label{Sec-IMACS}

The Inamori-Magellan Areal Camera \& Spectrograph (IMACS) is a innovative slit spectrograph attached to the Nasmyth focus of the
Baade (Magellan I) 6.5-m telescope (Dressler et al.\ 2006).  The instrument can be used either for imaging or for spectroscopy.  Designed primarily for multi-object spectroscopy, the instrument is sometimes used with a mask cut with a single long (26-inch length!)   slit.
There are two cameras, and either is available to the observer at any time: an f/4 camera with a 15.4 arcmin coverage, or an f/2.5 camera with a 27.5 arcmin coverage.

The f/4 camera is usable with any of 7 gratings, of which 3 may be mounted at any time, and which provide resolutions of 300-5000 with a
1-arcsec wide slit.  The delivered image quality is often better than that (0.6 arcsec fwhm is not unusual) and so one can use a 
narrower slit resulting in higher spectral resolution.
  The spectrograph is really designed to take advantage of such good seeing, 
as a 1-arcsec wide slit projects to 9 unbinned pixels.  Thus binning is commonly used.
The f/2.5 camera is used with a grism\footnote{A ``grism" is a prism attached to a diffraction grating.  The diffraction grating provides the
dispersive power, while the (weak) prism is used to displace the first-order spectrum back to the straight-on position.  The idea was
introduced by Bowen \& Vaughn (1973), and used successfully by Art Hoag at the prime focus of the Kitt Peak 4-m telescope
(Hoag 1976).}, providing a longer spatial coverage but lower dispersion. 
Up to two grisms can be inserted for use during a night.

\begin{figure}[htp]
\epsscale{0.7}
\plotone{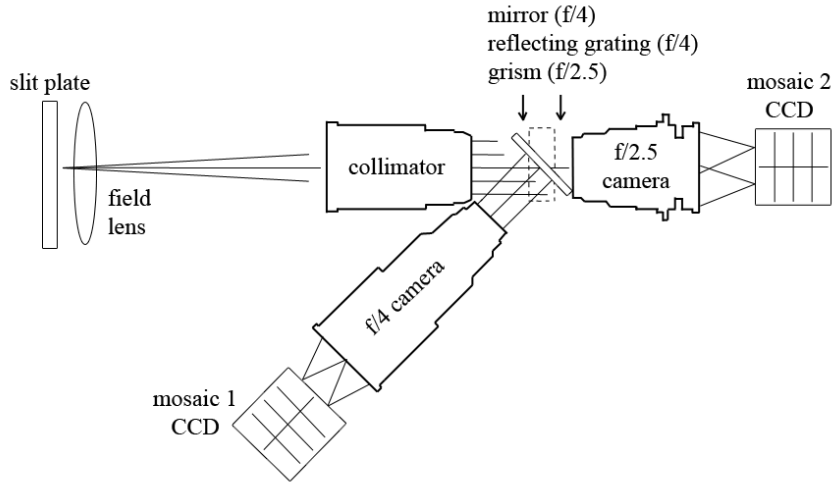}
\caption{\label{fig:IMACS} Optical layout of Magellan's IMACS.  This is based upon an illustration in the IMACS user manual.}
\end{figure}

The optical design of the spectrograph is shown in Figure~\ref{fig:IMACS}.  Light from the f/11 focus of the Baade Magellan telescope
focuses onto the slit plate, enters a field lens, and is collimated by transmission optics.  The light is then either directed into the f/4 or f/2.5
camera.  To direct the light into the f/4 camera, either a mirror is inserted into the beam (for imaging)  or a diffraction grating is inserted
(for spectroscopy).  If the f/2.5
camera is used instead, either the light enters directly (in imaging mode) or a transmission ``grism" is inserted.  Each camera has its
own mosaic of eight CCDs, providing 8192x8192 pixels.  The f/4 camera provides a smaller field of view but higher dispersion 
and plate scale; see Table~1.   Pre-drilled ``long-slit" masks are available in a variety of slit widths. Up to six masks can be inserted for a night's observing, and
selected by the instrument's software. 

\subsubsection{Fiber-fed Bench-Mounted Spectrographs}

As an alternative to multi-slit masks, a spectrograph can be fed by multiple optical fibers. The fibers can be arranged in the focal
plane so that light from the objects of interest enter the fibers, while at the spectrograph end the fibers are arranged in a line,
with the ends acting like the slit in the model of the basic spectrograph (Figure~\ref{fig:spect}).   Fibers were first commonly used
for multi-object spectroscopy in the 1980s, prior even to the advent of CCDs; for example, the Boller and Chivens spectrograph
on the Las Campanas du Pont 100-inch telescope was used with a plug-board fiber system when the detector was an intensified
Reticon system.   Plug-boards are like multi-slit masks in that there are a number of holes pre-drilled at specific locations in which
the fibers are then ``plugged".    For most modern fiber systems, the fibers are positioned robotically in the focal plane, although
the Sloan Digital Sky Survey used a plug-board system.  A major advantage of a fiber system is that the spectrograph can be mounted
on a laboratory air-supported optical bench in a clean room, and thus not suffer flexure as the telescope is moved.  This can result in high stability, needed
for precision radial velocities.  The fibers themselves provide additional ``scrambling" of the light, also significantly improving the
radial velocity precision, as otherwise the exact placement of a star on the slit may bias the measured velocities.   

There are three
down sides to fiber systems.  First, the fibers themselves tend to have significant losses of light at the slit end; i.e., not all of the light
falling on the entrance end of the fiber actually enters the fiber and makes it down to the spectrograph.  These losses can be as high as
a factor of 3 or more compared to a conventional slit spectrograph. Second, although typical fibers are 200-300$\mu$m in diameter,
and project to a few arcsec on the sky, each fiber must be surrounded by a protective sheath, resulting in a minimal spacing between
fibers of 20-40 arcsec.  Third, and most importantly, sky subtraction is never ``local".  Instead,
fibers are assigned to blank sky locations just like objects, and the accuracy of the sky subtraction is dependent on how accurately one can remove the fiber-to-fiber transmission variations by flat-fielding.

\leftskip=-30pt
\begin{tabular} [h] {l l c c c c c c c}
\multicolumn{5}{c}{Table 4. Some Fiber Spectrographs} \\ \hline \hline
\multicolumn{1}{c}{Instrument}
&\multicolumn{1}{c}{Telescope}
&\multicolumn{1}{c}{\# }
&\multicolumn{2}{c}{Fiber size}
&\multicolumn{1}{c}{Closest}
&\multicolumn{1}{c}{FOV}
&\multicolumn{1}{c}{Setup}
&\multicolumn{1}{c}{$R$} \\  \cline{4-5}
& & fibers& ($\mu$m) & (") & \multicolumn{1}{c}{spacing (arsec)} & (') & (mins) \\
\hline
Hectospec  & MMT 6.5-m &300               &250&1.5& 20       &  60   & 5   &1000-2500 \\
Hectochelle & MMT 6.5-m &240          &250& 1.5 &20  &  60    &5  & 30,000 \\
MIKE            & Clay 6.5-m & 256          & 175     &1.4& 14.5   & 23 & 40 & 15,000-19,000\\
AAOMega    &   AAT  4-m   & 392         &140 &2.1  &35 & 120 &      65 &1300-8000 \\
Hydra-S         & CTIO 4-m  & 138          &300 &  2.0 & 25 &  40 &20      &1000-2000 \\
Hydra (blue)    &     WIYN 3.5-m &83   &310  & 3.1  &37 &  60  &20  &1000-25000 \\
Hydra (red)      &     WIYN 3.5-m &90  &200 & 2.0 & 37   & 50 &20   & 1000-40000 \\
\hline
\end{tabular}

\leftskip=0pt

\paragraph{An Example: Hectospec}

Hectospec is a 300-fiber spectrometer on the MMT 6.5-m telescope on Mt Hopkins. 
The instrument is described in detail by Fabricant et al.\ (2005).  
The focal surface consists of a 0.6-m diameter stainless steel plate onto which
the magnetic fiber buttons are placed by two positioning robots (Figure~\ref{fig:hecto}).
The positioning speed of the robots is unique among such instruments and 
is achieved without sacrificing positioning accuracy (25$\mu$m, or 0.15 arcsec).
The field of view is a degree across.  The fibers subtend 1.5 arsec on the sky, and can be
positioned to within 20 arcsec of each other.  
Light from the fibers is then fed into a bench mounted spectrograph, which uses either
a 270 line/mm grating ($R\sim1000$) or a 600 line/mm grating ($R\sim 2500$).  
The same fiber bundle can be used with a separate spectrograph, known as Hectochelle.

\begin{figure}[htp]
\caption{\label{fig:hecto} View of the focal plane of Hectospec.  From Fabricant et al.\ (2005).
Reproduced by permission. THIS FIGURE WAS REMOVED FOR THE ASTROPH POSTING BUT WILL APPEAR IN THE SPRINGER EDITION.}
\end{figure}

Another unique aspect of Hectospec is the ``cooperative" queue manner in which the data are obtained, made possible in part because
multi-object spectroscopy with fibers is not very flexible and configurations are done well in advance of going to the telescope.
  Observers are awarded a certain number of nights and scheduled on the telescope in the classical way.   The astronomers design their  fiber configuration files in advance; 
these files contain the necessary
positioning information for the instrument, as well as exposure times, grating setups, etc.
All of the observations however become part of a collective pool.  The astronomer goes to
the telescope on the scheduled night, but a ``queue manager" decides on a night-by-night
basis which fields should be observed and when.  The observer has discretion to select
alternative fields and vary exposure times depending upon weather conditions, seeing, etc.
The advantages of this over classical observing is that weather losses are spread
amongst all of the programs in the scheduling period (4 months).  The advantages over
normal queue scheduled observations is that the astronomer is actually present for some
of his/her own observations, and there is no additional cost involved in hiring queue
observers.

\subsection{Extension to the UV and NIR}

The general principles involved in the design of optical spectrographs extend to those used to observe in the ultraviolet (UV) and
near infrared (NIR), with some modifications.  CCDs have high efficiency in the visible region, but poor sensitivity
at shorter ($<$3000\AA) and longer ($>1\mu$m) wavelengths.   At very short wavelengths (x-rays, gamma-rays) and very long
wavelengths (mid-IR through radio and mm) special techniques are needed for spectroscopy, and are beyond the scope of the
present chapter.

Here we provide examples of two non-optical instruments, one whose domain is the ultraviolet (1150-3200\AA) and one whose
domain is in the near infrared (1-2$\mu$m).

\subsubsection{The Near Ultraviolet}
\label{Sec-STIS}
For many years, astronomical ultraviolet spectroscopy was the purview of the privileged few, mainly instrument Principle Investigators (PIs)
who
flew their instruments on high-altitude balloons or short-lived rocket experiments.  The Copernicus (Orbiting Astronomical Observatory 3)
was a longer-lived mission (1972-1981), but the observations were still PI-driven.  This all changed drastically due to the 
{\it International Ultraviolet Explorer (IUE)} satellite, which operated from 1978-1996.  Suddenly any astronomer could apply for time and obtain
fully reduced spectra in the ultraviolet.   {\it IUE}'s primary was only 45 cm in diameter, and there was considerable demand for the
community to have UV spectroscopic capability on the much larger (2.4-m) {\it Hubble Space Telescope (HST)}.

The Space Telescope Imaging Spectrograph (STIS) is the spectroscopic work-horse of {\it HST}, providing spectroscopy from the far-UV through the
far-red part of the spectrum.   Although a CCD is used for the optical and far-red, another type of detector (multi-anode microchannel
array, or MAMA) is used for the UV.  Yet, the demands are similar enough for optical and UV spectroscopy that the rest of the
spectrograph is in common to both the UV and optical.  The instrument is described in detail by Woodgate et al.\ (1998), and the
optical design is shown in Figure~\ref{fig:STIS}.

\begin{figure}[htp]
\caption{\label{fig:STIS} Optical design of STIS from Woodgate et al.\ (1998). Reproduced by permission. THIS FIGURE WAS REMOVED FOR THE ASTROPH POSTING BUT WILL APPEAR IN THE SPRINGER EDITION.}
\end{figure}

In the UV, STIS provides resolutions of $\sim$1000 to 10,000 with first-order gratings.  With the echelle gratings, resolution
as high as 114,000 can be achieved.  No blocking filters are needed as the MAMA detectors are insensitive to longer wavelengths.
From the point of view of the astronomer who is well versed in optical spectroscopy, the use of STIS for UV spectroscopy seems transparent.

With the success of the Servicing Mission 4 in May 2009, the Cosmic Origins Spectrograph (COS) was added to {\it HST's} suite of
instruments.  COS provides higher through put than {\it STIS} (by factors of 10 to 30) in the far-UV, from 1100-1800\AA.
In the near-UV (1700-3200\AA) STIS continues to win out for many applications.

\subsubsection{Near Infrared Spectroscopy and OSIRIS}
\label{Sec-irspectrographs}

Spectroscopy in the near-infrared (NIR) is complicated by the fact that the sky is much brighter
than most sources, plus the need to remove the strong telluric bands in the spectra.
In general, this is handled by moving a star along the slit on successive, short exposures
(dithering), and subtracting adjacent frames, such that the sky obtained in the first exposure
is subtracted from the source in the second exposure, and the sky in the second exposure
is subtracted from the source in the first exposure.  Nearly featureless stars are observed
at identical airmasses to that of the program object in order to remove the strong telluric
absorption bands. 
These issues will be discussed further in \S~\ref{Sec-irred} and \S~\ref{Sec-ObsNIR} below.

The differences in the basics of infrared arrays compared to optical CCDs also
affect how NIR astronomers go about their business.
CCDs came into use in optical astronomy in the 1980s because of their very high efficiency ($\ge $50\%, relative to photographic plates of a few percent) and high linearity (i.e., the counts above bias are proportional to the number of photons falling on their surface over a large dynamic range).  CCDs work by exposing a thin wafer of silicon to light and to collect the resulting freed charge carriers under electrodes. By manipulating the voltages on those electrodes, the charge packets can be carried to a corner of the detector array where a single amplifier can read them out successively. (The architecture may also be used to feed multiple output amplifiers.) This allows for the creation of a single, homogenous silicon structure for an optical array (see Mackay 1986 for a review).Ê For this and other reasons, optical CCDs are easily fabricated to remarkably large formats, several thousand pixels to a side. 

Things are not so easy in the infrared.Ê The band gap (binding energy of the electron) in silicon is simply too great to be dislodged by an infrared photon.Ê For detection between 1 and 5 $\mu$m, either Mercury-Cadmium-Telluride (HgCdTe) or Indium-Antimonide (InSb) are typically used, while the read out circuitry still remains silicon-based. From 5 to 28 $\mu$m, silicon-based (extrinsic photoconductivity) detector technology is used, but they continue to use similar approaches to array construction as in the near-infrared. The two layers are joined electrically and mechanically with an array of Indium bumps. (Failures of this Indium bond lead to dead pixels in the array.)  Such a two-layered device is called a {\it hybrid} array (Beckett 1995).Ê For the silicon integrated circuitry, a CCD device could be (and was originally) used, but the very cold temperatures required for the photon detection portion of the array produced high read noise.Ê Instead, an entirely new structure that provides a dedicated readout amplifier for each pixel was developed (see Rieke 2007 for more details).Ê These direct read-out  arrays are the standard for infrared instruments and allow for enormous flexibility in how one reads the array.Ê For instance, the array can be set to read the charge on a specific, individual pixel without even removing the accumulated charge (non-destructive read).Ê Meanwhile, reading through a CCD removes the accumulated charge on virtually every pixel on the array.

Infrared hybrid arrays have some disadvantages, too.  Having the two components (detection and readout) made of different  materials limits the size of the array that can be produced.  This is due to the challenge of matching each detector to its readout circuitry to high precision when flatly pressed together over  millions of unit cells.  Even more challenging is the stress that develops from differential thermal contraction when the hybrid array is chilled down to very cold operating temperatures.   
However, improvements in technology now make it possible to fabricate 2K $\times$ 2K hybrid arrays,  and it is expected that 4K $\times$ 4K will eventually be possible.  Historically, well depths have been lower in the infrared arrays, though hybrid arrays can now be run with a higher gain.  This allows for well depths approaching that available to CCD arrays (hundreds of thousands of electrons per pixel).  All infrared hybrid arrays have a small degree of nonlinearity, of order a few percent, due to a slow reduction in response as signals increase (Rieke 2007).  In contrast, CCDs are typically linear to a few tenths of a percent over five orders of magnitude.   Finally, the infrared hybrid arrays are far more expensive to build than CCDs.  This is because of the extra processing steps required in fabrication and their much smaller commercial market compared to CCDs.

The Ohio State Infrared Imager/Spectrometer (OSIRIS) provides an example of such an instrument, and how the field has evolved over the past two decades.
 OSIRIS is a multi-mode infrared imager and spectrometer designed and built by The Ohio State University (Atwood et al.\ 1992, Depoy et al.\ 1993).  Despite being originally built in 1990, it is still in operation today, most recently spending several successful years at the Cerro Tololo Inter-American Observatory  Blanco 4-m telescope.  Presently, OSIRIS sits at the Nasmyth focus on the 4.1-m Southern Astrophysical Research (SOAR) Telescope on Cerro Pach\'on.  

When built twenty years ago, the OSIRIS  instrument was designed to illuminate the best and largest infrared-sensitive arrays available at the time, the 256 x 256 pixel NICMOS3 HgCdTe arrays, with 27 $\mu$m pixels.   This small array has long since been upgraded as infrared detector technology has improved.  The current array on OSIRIS is now 1024 x 1024 in size, with 18.5 $\mu$m pixels (NICMOS4, still HgCdTe).  As no design modifications could be afforded to accommodate this upgrade, the larger array now used is not entirely illuminated  due to vignetting in the optical path.  This is seen as a fall off in illumination near the outer corners of the array. 

OSIRIS provides two cameras, $f/2.8$ for lower resolution work ($R \sim 1200$ with a 3.2-arcmin long slit) and $f/7$ for higher-resolution work ($R \sim 3000$, with a 1.2-arcminute long slit).  One then uses broad-band filters in the $J$ (1.25 $\mu$m), $H$ (1.65 $\mu$m) or $K$ (2.20 $\mu$m)
bands, to select the desired order, 5th, 4th and 3rd, respectively.   The instrument grating tilt is set to simultaneously select the central regions of these three primary transmission bands of the atmosphere.  However, one can change the tilt to optimize observations at wavelengths near the edges of these bands.  OSIRIS does have a cross-dispersed mode, achieved by introducing a cross-dispersing grism in the filter wheel.  A final filter, which effectively blocks light outside of the $J$, $H$, and $K$ bands, is needed for this mode.  The cross-dispersed mode allows observing at low resolution ($R \sim 1200$) in all three bands simultaneously, albeit it with a relatively short slit (27-arcsecs).

Source acquisition in the infrared is not so straightforward. While many near-infrared
objects have optical counterparts, many others do not or show rather different morphology
or central positions offsets between the optical and infrared. This means acquisition and alignment
must be done in the infrared, too. OSIRIS, like most modern infrared spectrometers, can image its own slit onto the science detector when the grating is not deployed in the light path. This greatly facilitates placing objects on the slit (some NIR spectrometers have a dedicated slit viewing imager so that objects may be seen through the slit during an actual exposure).
This quick-look imaging configuration is available with an imaging mask too, and deploys a flat mirror in place of the grating (without changing the grating tilt) thereby displaying an infrared image of the full field or slit with the current atmospheric filter. This change in configuration only takes a few seconds and allows one to align the target on the slit then quickly return the
grating to begin observations. Even so, the mirror/grating flip mechanism will only repeat to a fraction of a pixel when being moved to change between acquisition and spectroscopy modes. The most accurate observations may then require new flat fields and or lamp spectra be taken before returning to imaging (acquisition) mode.

For precise imaging observations,
OSIRIS can be run in "full" imaging mode which includes placing a cold mask in the light path to block out-of-beam back ground emission for the telescope primary and secondary. Deployment of the mask can take several
minutes. This true imaging mode is important in the K-band where background emission becomes significant beyond 2$\mu$m due to the warm telescope and sky.

There are fantastic new capabilities for NIR spectroscopy about to become available as
modern multi-object spectrometers come on-line on large telescopes (LUCIFER on the Large Binocular Telescope,  MOSFIRE on Keck, FLAMINGOS-2 on Gemini-South, and MMIRS on the Clay Magellan telescope).   As with
the optical, utilizing the multi-object capabilities of these instruments effectively requires
proportionately greater observer preparation, with a significant increase in the complexity of
obtaining the observations and performing the reductions.  Such multi-object NIR observations are
not yet routine, and as such details are not given here.  One should perhaps master the ``simple" NIR
case first before tackling these more complicated situations.

\subsection{Spatially Resolved Spectroscopy of Extended Sources: Fabry-Perots and Integral Field Spectroscopy}

The instruments described above allow the astronomer to observe single or multiple point sources at a time.  If instead one wanted to
obtain spatially resolved spectroscopy of a galaxy or other extended source, one could place a long slit over the object at a
particular location and obtain a one-dimensional, spatially resolved spectrum.  If one wanted to map out the velocity structure of an
HII region or galaxy, or measure how various spectral features changed across the face of the object, one would have to take multiple
spectra with the slit rotated or moved across the object to build up a three dimensional image ``data cube": two-dimensional spatial location plus
wavelength.  Doing this is sometimes practical with a long slit: one might take spectra of a galaxy at four different position angles,
aligning the slit along the major axis, the minor axis, and the two intermediate positions.  These four spectra would probably give a pretty
good indication of the kinematics of the galaxy.  But, if the velocity field or ionization structure is complex, one would really
want to build up a more complete data cube.   Doing so by stepping the slit by its width over the face of an extended object would be one
way, but clearly very costly in terms of  telescope time.

An alternative would be to use a Fabry-Perot interferometer, basically a tunable filter.  A series of images through
a narrow-band filter is taken, with the filter tuned to a slightly different wavelength for each exposure.  The
resulting data are spatially resolved, with the spectral resolution dependent upon the step size between adjacent wavelength settings.
(A value of 30 km s$^{-1}$ is not atypical for a step size; i.e., a resolution of 10,000.)
The wavelength changes slowly as a function of radial position within
the focal plane, and thus a ``phase-corrected" image cube is constructed which yields both an intensity map (such being a direct image)
and radial velocity map for a particular spectral line (for instance,  H$\alpha$).  

This works fine in the special case where one is interested in only a few spectral features in an extended object.  Otherwise, the issue
of scanning spatially has simply been replaced with the need to scan spectrally. 

Alternative approaches broadly fall under the heading of {\it integral field spectroscopy}, which simply means obtaining the
full data cube in a single exposure.  There are three methods of achieving this, following Allington-Smith et al.\ (1998).

{\it Lenslet arrays:} One method of obtaining integral field spectroscopy is to place a microlens array (MLA) at the focal plane.  
The MLA produces a series of images of the telescope pupil, which enter
the spectrograph and are dispersed.  By tilting the MLA, one can arrange it so that the spectra do not overlap with one another.

{\it Fiber bundles:}  An array of optical fibers is placed in the focal plane, and the fibers then transmit the light to the spectrograph,
where they are arranged in a line, acting as a slit.  This is very similar to the use of multi-object fiber spectroscopy, except that the
ends of the fibers in the focal plane are always in the same configuration, with the fibers bundled as close together as possible.
There are of course gaps between the fibers, resulting in incomplete spatial coverage without dithering.  

It is common to use both lenslets and fibers together, as for instance is done with the integral field unit of the FLAMES 
spectrograph  on the VLT.

{\it Image slicers:} A series of mirrors can be used to break up the focal plane into congruent ``slices" and arrange these slices
along the slit, analogous to a classic Bowen image slicer\footnote{When observing an astronomical object, a narrow slit is needed to maintain good spectral resolution, as
detailed in \S~\ref{Sec:Basics}.  Yet, the size of the image may be much larger than the size of the slit, resulting in a significant
loss of light, known as ``slit losses".   Bowen (1938) first described a novel device for reducing slit losses by changing the shape of the incoming image to match that of a long, narrow slit: a cylindrical lens is used to slice up the image into a series of strips with a width
equal to that of the slit, and then to arrange them end to end along the slit.   This needs to be accomplished without altering the focal ratio of the incoming beam.  Richardson et al.\ (1971) describes a variation of the same principle, while Pierce (1965) provides detailed construction notes for
such a device.  
The heyday of image slicers was in the photographic era, where sky subtraction was impractical and most astronomical spectroscopy
was not background limited.  Nevertheless, they have not completely fallen into disuse.  For instance, GISMO (Gladders
Image-Slicing Multislit Option) is an image slicing device available for use with multi-slit plates on Magellan's IMACS (\S~\ref{Sec-IMACS})
to re-image the central 3.5 arcmin $\times$ 3.2 arcmin of the focal plane into the full field of view of the instrument, allowing an 8-fold
increase in the spatial density of slits.}.  One advantage of the image slicer technique for integral-field spectroscopy is that spatial information
within the slice is preserved.

\section{Observing and Reduction Techniques}

This section will begin with a basic outline of how spectroscopic CCD data are reduced, and
then extend the treatment to the reduction of NIR data.  
Some occasionally overlooked 
details will be discussed.  The section will
conclude by placing these together by describing a few sample observing runs.   
It may seem a little backwards to start with the
data reduction rather than with the observing.  But, only by 
understanding how to reduce data can one really understand how to best take data.

The basic premise throughout  this section is that one should neither observe nor reduce data by rote.  
Simply subtracting biases because all of one's colleagues subtract
biases is an inadequate reason for doing so.  One needs to examine the particular data to see if doing so helps or harms.
Similarly, unless one is prepared to do a little math, one might do more harm than good by flat-fielding.  
Software reduction packages, such as {\it IRAF}\footnote{The Image Reduction and Analysis  Facility is distributed by
the National Optical Astronomy Observatory, which is operated by the Association of Universities for Research in Astronomy,
under cooperative agreement with the National Science Foundation.} or {\it ESO-MIDAS}\footnote{The European Southern Observatory
Munich Image Data Analysis System} are extremely useful tools---in the right hands. But, one should never let the software provide a guide to
reducing data.  Rather, the
astronomer should do the guiding.  One should strive to understand the steps involved at the level
that one could (in principle) reproduce the results with a hand calculator!

In addition, there are very specialized data reduction pipelines, such as {\it HST's} CALSTIS and IMACS's COSMOS, which may
or may not do what is needed for a specific application.   The only way to tell is to try them, and compare the results with what one obtains
by more standard means.  See \S~\ref{Sec-optimal} for an example where the former does not do very well.

\subsection{Basic Optical Reductions}
\label{Sec-Basicred}

The simplest reduction example involves obtaining a spectrum of a star obtained from a long slit spectrograph.
What calibration data does one need, and why?

\begin{itemize}
\item {\bf The data frames themselves} doubtless contain an overscan strip along the side, or possibly along the top.  As a CCD is
read out, a ``bias" is added to the data to assure that no values go below zero.  Typically this bias is several hundred or even 
several thousand electrons (e$^-$)\footnote{We use ``counts" to mean what gets recorded in the image; sometimes these are also
known as analog-to-digital units (ADUs).  We use ``electrons" (e$^-$) to mean the things that behave like Poisson statistics, with the
noise going as the square root of the number of electrons.  The gain $g$ is the number of e$^-$ per count.}.  The exact value is likely to  change slightly from exposure to exposure, due to slight
temperature variations of the CCD.  The overscan value allows one to remove this offset.  In some cases the overscan should be used
to remove a one dimensional bias structure, as demonstrated below.

\item {\bf Bias frames} allow one to remove any residual bias {\it structure}.  Most modern CCDs, used with modern controllers, have very
little (if any) bias structure, i.e., the bias levels are spatially uniform across the chip. So, it's not clear that one needs to use bias frames.  If one takes enough bias frames (9 or more) and
averages them correctly, one probably does very little damage to the data by using them.  Still, in cases where read-noise dominants your program spectrum,
subtracting a bias could increase the noise of the final spectrum.

\item {\bf Bad pixel mask} data allows one to interpolate over bad columns and other non-linear pixels. These can be identified by comparing the average of a series of exposures of high
counts with the average of a series of exposures of low counts.

\item {\bf Dark frames} are exposures of comparable length to the program objects but obtained with the shutter closed.  In the olden
days, some CCDs generated significant dark current due to glowing amplifiers and the like.  Dark frames obtained with modern CCDs
do little more than reveal light leaks if taken during daytime.  Still, it is generally harmless to obtain them. 

\item {\bf Featureless flats} (usually ``dome flats") 
allow one to correct the pixel-to-pixel variations within the detector.  There have to be sufficient
counts to not significantly degrade the signal-to-noise ratio of the final spectrum.  This will be discussed in more detail in \S~\ref{Sec-flat1}.

\item {\bf Twilight flats} are useful for removing any residual spatial illumination mismatch between the featureless flat and the object
exposure.  This will be discussed in more detail in \S~\ref{Sec-flat2}.

\item {\bf Comparison arcs} are needed to apply a wavelength scale to the data.  These are usually short exposures of a combination
of discharge tubes containing helium, neon, and argon (HeNeAr), or thorium and argon (ThAr), with the choices dictated by what lamps are available
and the resolution.  HeNeAr lamps are relatively sparse in lines and useful at low to moderate dispersion;  ThAr
lamps are rich in lines and useful at high dispersion, but have few unblended lines
at low or moderate dispersions.

\item {\bf Spectrophotometric standard stars} are stars with smooth spectra with calibrated fluxes
used to determine the instrument response.  Examples of such stars can
be found in Oke (1990), Stone (1977, 1996), Stone \& Baldwin (1983), Massey et al.\ (1988, 1990), and particularly Hamuy et al.\ (1994).
Exposures should be long enough to obtain at least 10,000 e$^-$ integrated over 50\AA\ bandpass, i.e., at least 200 e$^-$ per \AA\ integrated
over the spatial profile.

\end{itemize}

The basic reduction steps are described here.  For convenience to the reader, reference is made to the
relevant IRAF tasks.  Nevertheless, the goal is to explain the steps, not
identify what buttons need to be pushed.

\begin{enumerate}
\item {\bf Fit and subtract overscan.}  Typically the overscan occupies 20-40 columns on the right side (high column numbers) of the chip.
The simplest thing  one can do is to average the results of these columns on a row-by-row basis, and then fit these with a low-order
function, possibly a constant.  

An example of such a nicely behaved 
overscan in shown in the left panel of Figure~\ref{fig:overscan}.
In other cases, there is a clear gradient in the bias section that should be removed by a higher-order fit.  The IMACS chips, for instance, have an overscan at both the top and the right, and it is clear
from inspection of the bias frames that the top overscan tracks a turn-up on the left side.  Such an overscan is shown in
the middle panel of Figure~\ref{fig:overscan}.   Occasionally the overscan allows one to correct for some problem that occurred
during the read-out of the chip.  For instance, the GoldCam spectrometer chip suffers 10-20 e$^-$ striping along rows
if the controller electronics
get a bit too hot. (This situation is sometimes referred to as ``banding".) A similar problem may occur if there is radio transmission from
a nearby walkie-talkie during readout.  However, these stripes extend into the overscan, which can be used to eliminate them by 
subtracting the overscan values in a row-by-row manner rather than by subtracting a low- or high-order function.
Such banding is shown in the right panel of Figure~\ref{fig:overscan}. In IRAF, 
overscan subtraction is a task  performed in the first pass
through {\it ccdproc}.

\begin{figure} [htp]
\epsscale{0.32}

\plotone{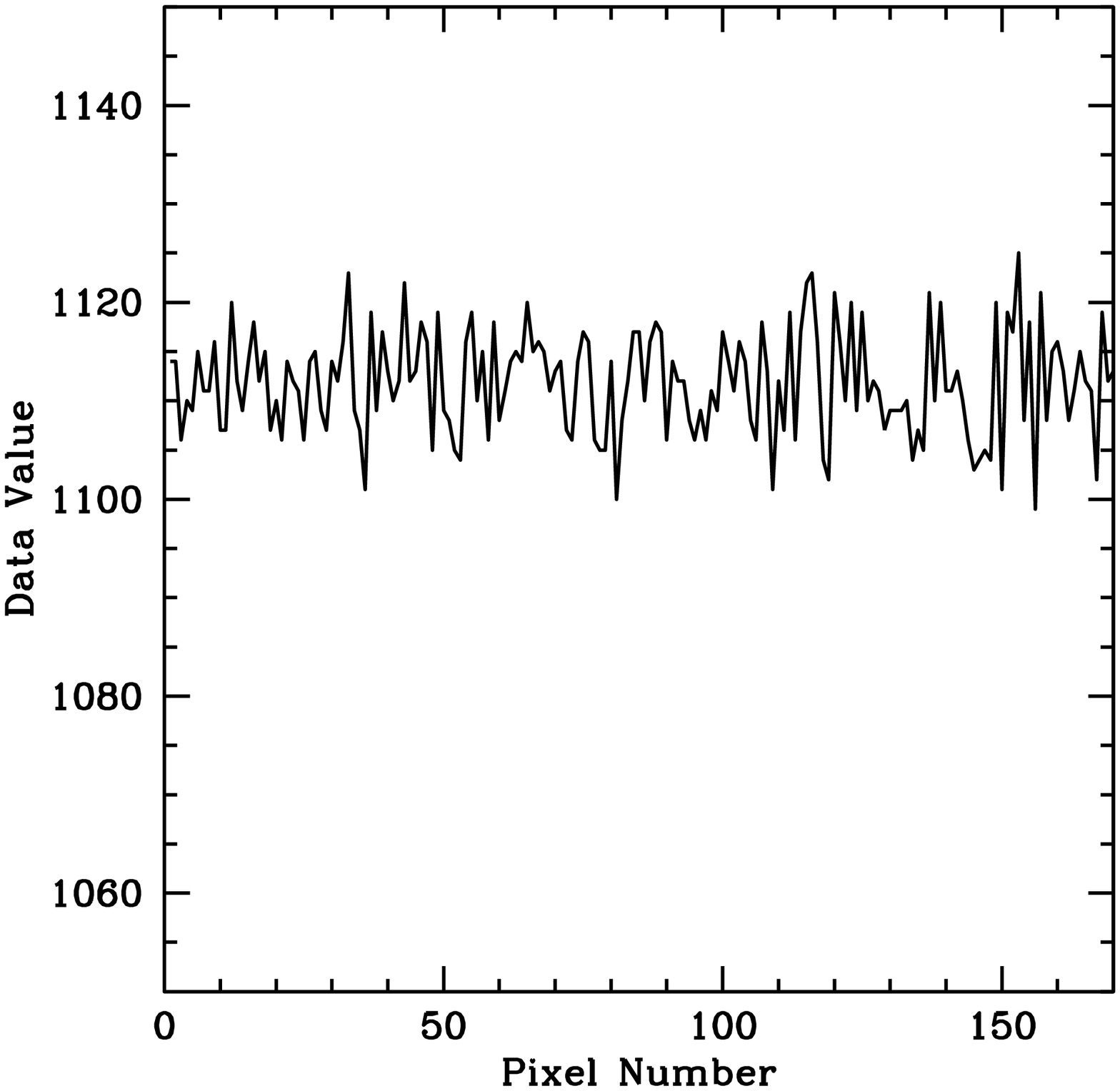}
\plotone{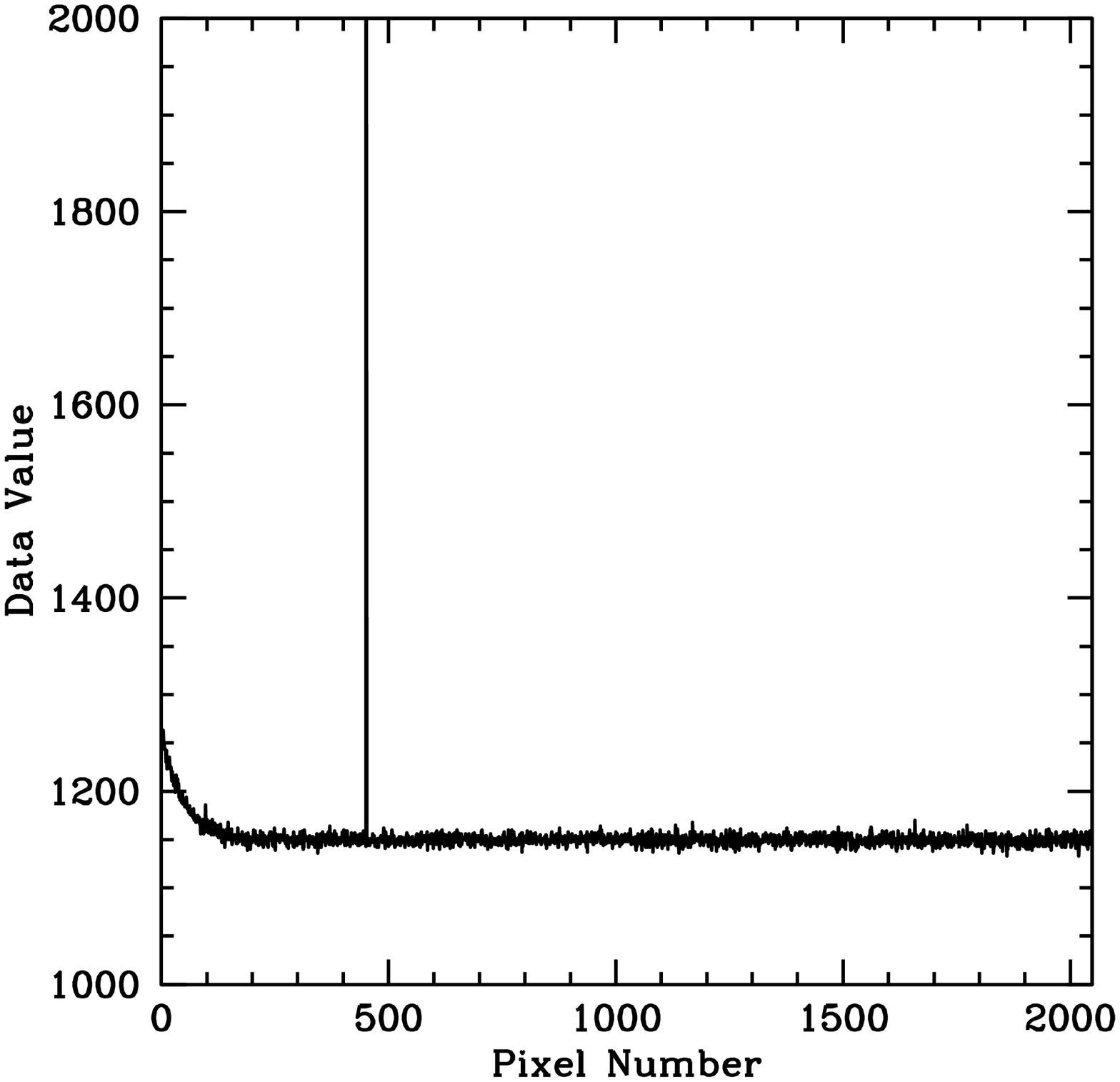}
\plotone{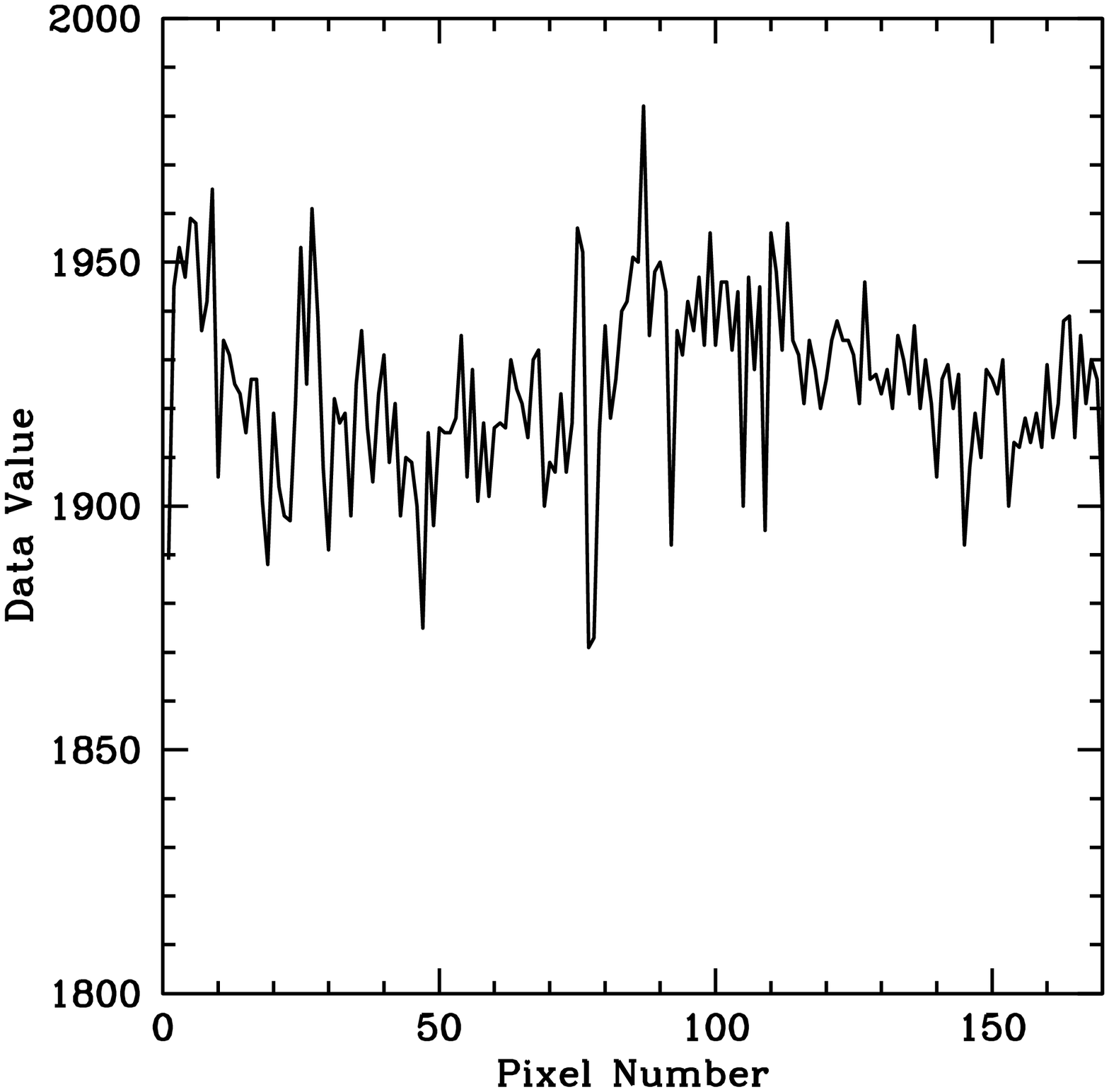}
\caption{\label{fig:overscan}  Three examples of overscan structure.  The overscan values have been averaged over the 20-40 widths of the overscan region, and are shown plotted against the read-axis.  The overscan on the left is from the GoldCam spectrometer on the Kitt Peak
2.1-m and is best fit by a constant value.  The overscan in the middle is from chip 3 of IMACS, and shows a pronounced upturn on the left,
and is best fit with a low-order function.  (Note too the bad column around pixel 420.)  The overscan on the right is also from GoldCam,
but during an electronics problem that introduced a number of bands with low values that were also in the data region.  Subtracting the
average values in the overscan row-by-row removed them to a good approximation.}
\end{figure}

\item {\bf Trim.}  Most spectroscopic data do not extend over the entire face of the chip, and the blank areas will
need to be trimmed off.  
Often the first or last couple of rows and columns are not well behaved and should be trimmed as well. 
At the very least, the overscan
regions need to be removed. In IRAF this task is performed in the same step as the overscan subtraction by {\it ccdproc}.

\item {\bf Dark subtraction.} In the highly unlikely event that there is significant dark current the dark frame should be subtracted
from all of the images after scaling by the relevant exposure times. 

\item {\bf Interpolate over bad pixels.} This step is mostly cosmetic, but is necessary if one plans to flux-calibrate. 
Most CCDs have a
few bad columns and non-linear pixels.  To identify these one can use two series of exposures of featureless flats: one of high
intensity level (but be careful not to saturate!) and one of low intensity.  Enough of these 
low-intensity flats need to be taken in order for the average to have a reasonable signal-to-noise ratio.   Typically 3 exposures of an internal
quartz lamp (say) might be taken, along with
 50 frames of similar exposure time with a 100$\times$ attenuating neutral density filter.  The first series may contain
10,000 counts/pixel, while the latter only 100 counts/pixel.   Averaging each series and dividing the results identifies the non-linear
pixels.   One can construct a bad pixel ``mask" (an image that is only a few bits deep) to use to decide whether to interpolate 
along rows (in order to handle bad columns) or over rows and columns (to handle isolated bad pixels), etc. 
 The relevant tasks
in IRAF are {\it ccdmask} which generates a bad pixel mask, followed by using the mask in {\it ccdproc}.

\item {\bf Construct (and use?) a master bias frame.}  All the bias frames (which have been treated by steps 1-2 as well) are combined
and examined to see if there is any spatial structure.  If there is, then it is worth subtracting this master bias frame from all the
program frames.  The relevant IRAF tasks are {\it zerocombine} and {\it ccdproc}.

\item \label{step-flat} {\bf Construct and apply a master normalized  featureless flat.}  
All of the dome flat exposures  are combined using a bad pixel
rejection algorithm.  The IRAF task for this combination is {\it flatcombine}.  The question then is how to best normalize the flat?  

Whether the final step in the reduction process is to flux-calibrate the spectrum, or simply to normalize
the spectrum to the continuum, the effort is going to be much easier if
one makes the correct choice as to how to best normalize the flat.  Physically the question comes down to whether or not the shape
of the featureless flat in the wavelength direction is dominated by the same effects as the astronomical object one is trying to reduce,
or if instead it is dominated by the calibration process.  For instance, all flats are likely to show some bumps and wiggles.  Are the bumps
and wiggles due to the grating?  If so, one is better off fitting the flat with a constant value (or a low order function) and using that as the
flat's normalization.  That way the same bumps and wiggles are present in both the program data as in the flat, and will divide out.
  If instead some of
the bumps and wiggles are due to filters in front of the flat-field lamp, or the extreme difference in color temperature between the
lamps and the object of interest, or due to wavelength-dependent variations in the reflectivity of the paint used for the flat-field
screen, then one is better off fitting the bumps and wiggles by using a very high order function, and using the flat only to remove pixel-to-pixel
variations.   It usually isn't obvious which will be better {\it a priori}, and really the only thing to do is to 
select an object whose spectrum is expected to be smooth (such as a spectrophotometric standard)
and reduce it through the entire process using each kind of flat.  It will be easy to tell in the end.
The IRAF task for handing the normalization of the flat is {\it response}, and the flat field division is handled by {\it ccdproc}.

An example is shown in Figure~\ref{fig:flat}. The flat-field 
 has a very strong gradient in the wavelength direction, and a bit of a bend around pixel 825 (about 4020\AA).  Will one do better
by just normalizing this by a constant value, ascribing the effects to the grating and spectrograph?   Or should one normalize the flat
with a higher order
function based on the assumption that these features are due to the flat-field system and very red lamps?  (Blue is on the right and
red is on the left.) The only way to answer the question is to reduce some data both ways
and compare the results, as in Figure~\ref{fig:flat}. The flat fit by a constant value does a better job removing both the gradient
and the bump.  (The spectrum needs to be magnified to see that the bump at 4020\AA\ has been removed.) 
 Thus the flat
was a good reflection of what was going on in the spectrograph, and is not due to issues with the calibration system.

Once one determines the correct normalized flat, it needs to be divided into all of the data.  The IRAF task for handling this is again {\it ccdproc}.

\begin{figure} [htp]
\epsscale{0.30}

\plotone{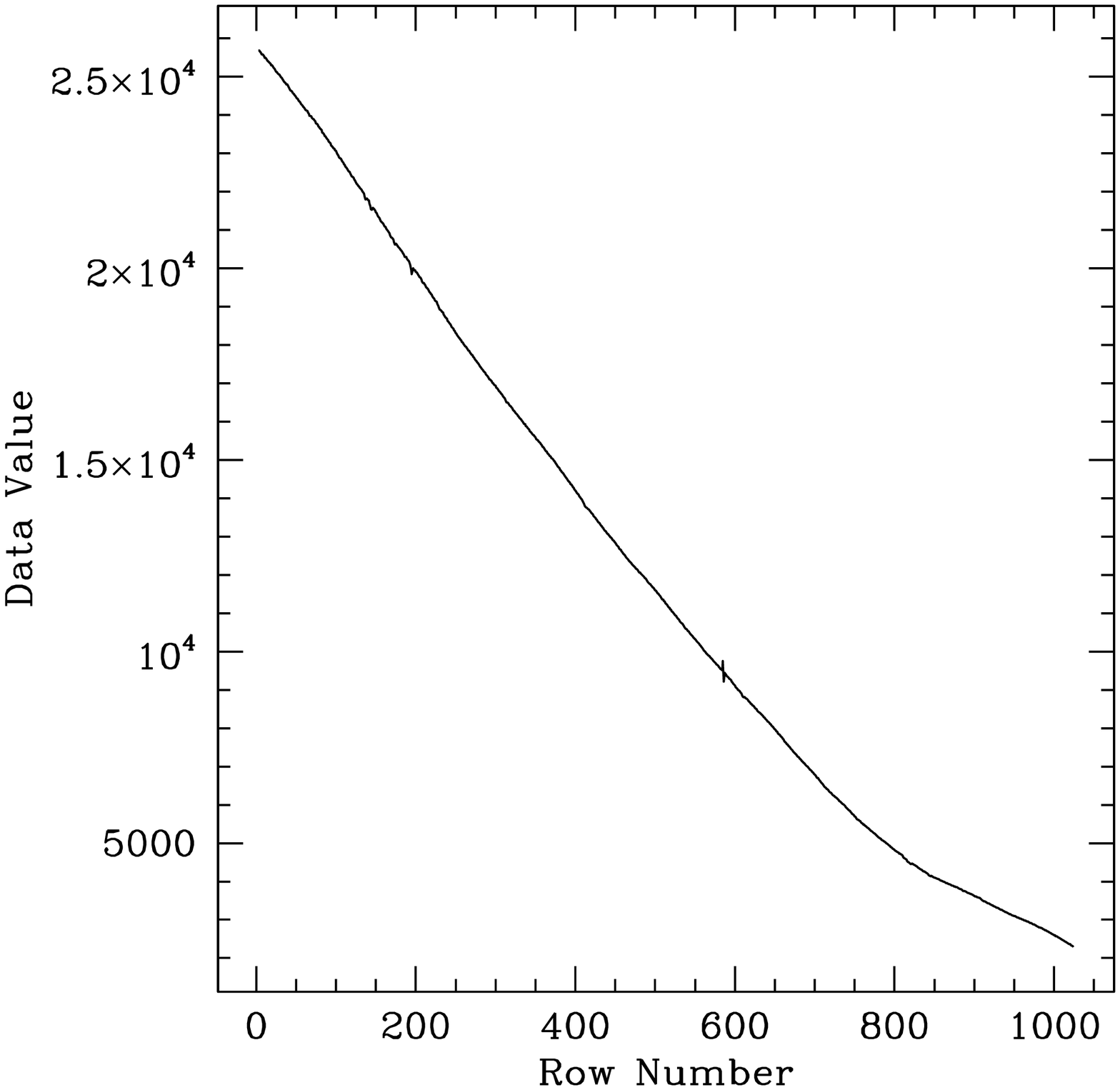}
\plotone{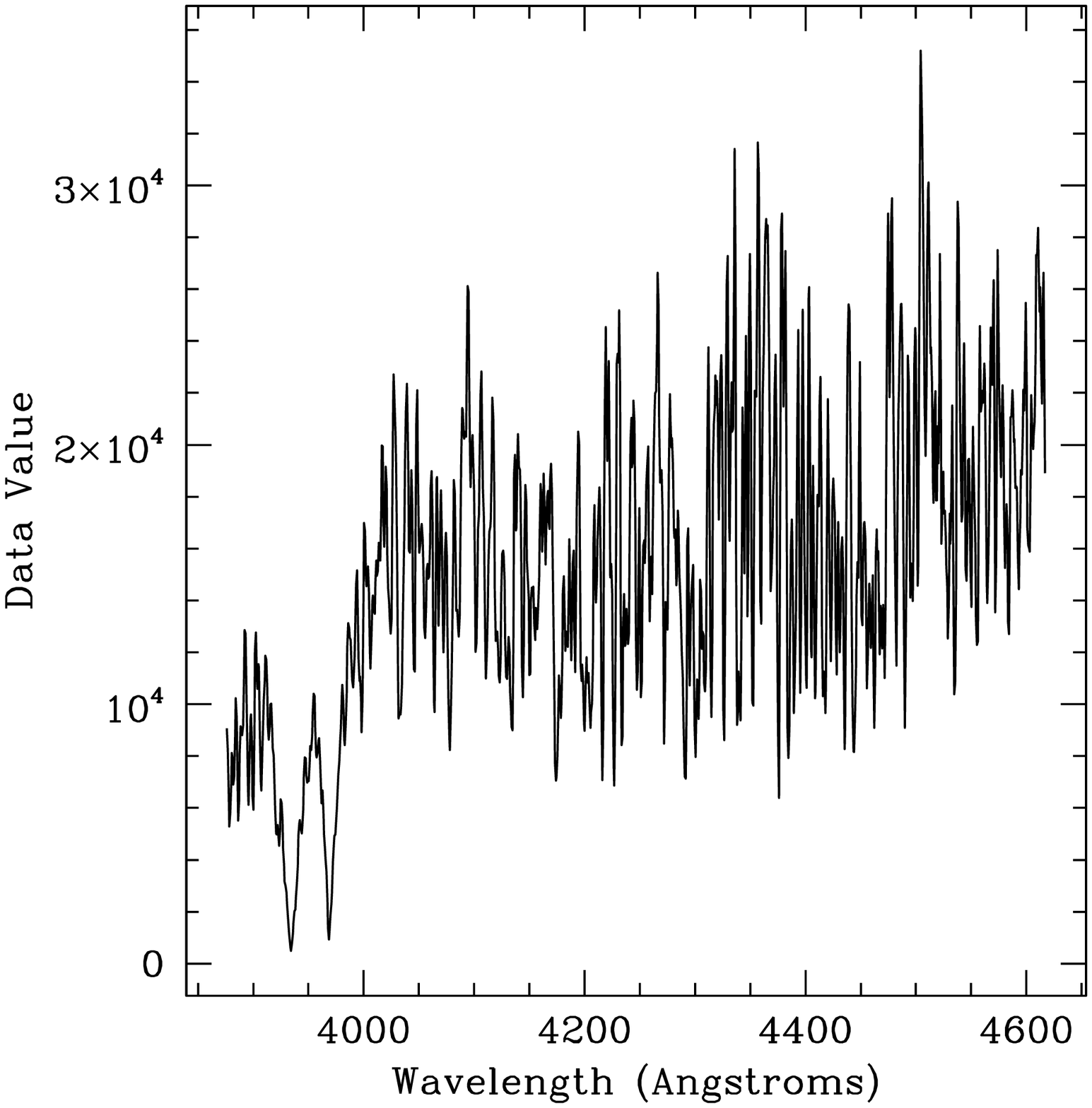}
\plotone{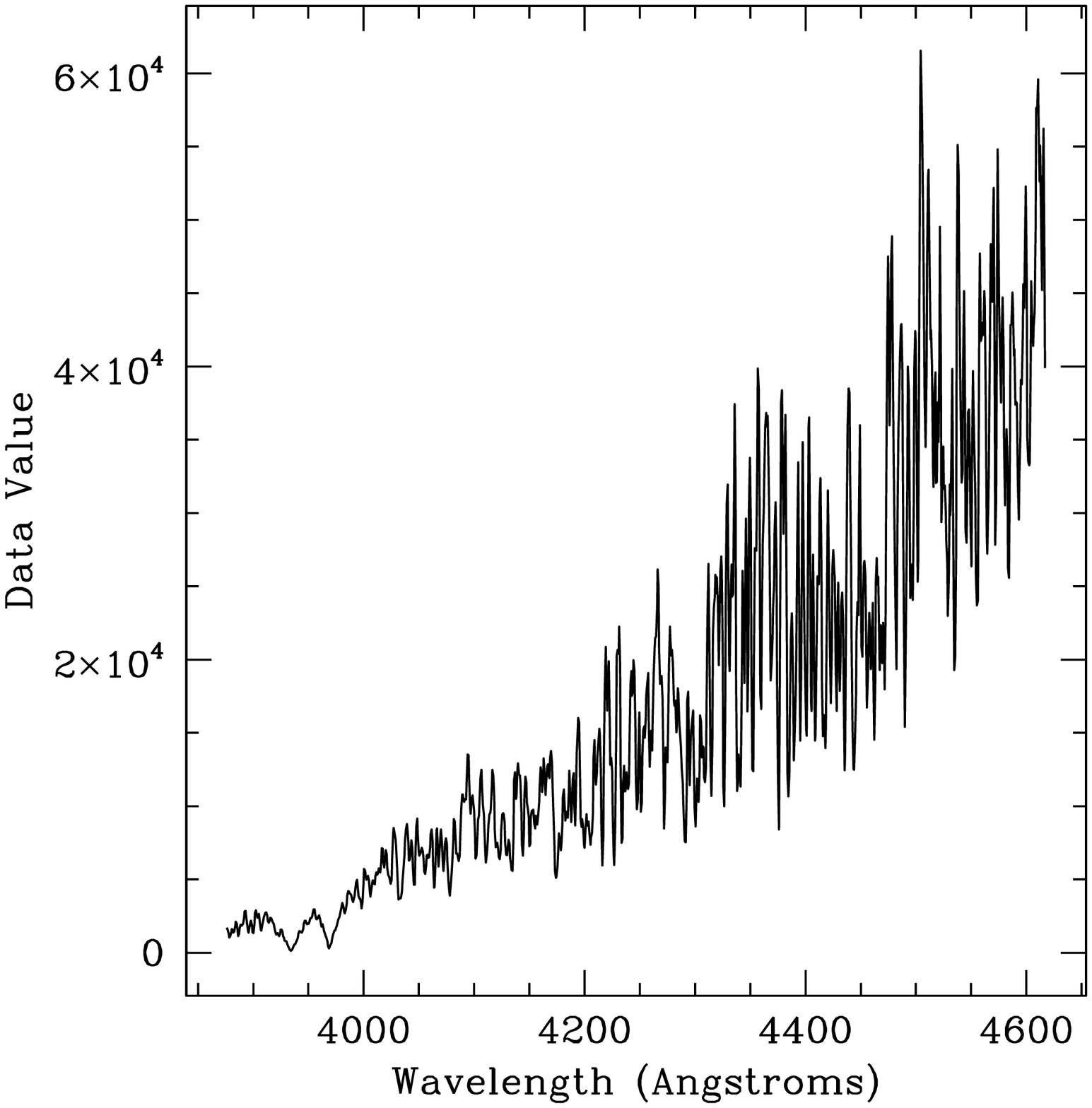}
\caption{\label{fig:flat}  Normalization of Flats.  {\it Left.} The flat field counts varies by a factor of of 10 or more with wavelength, and there
is a discontinuity in the derivative around pixel 820. {\it Middle.} Normalizing the flat by a constant yields a smooth stellar spectrum.
{\it Right.} Normalizing the flat by a higher order function results in an artificially large gradient in the counts.}

\end{figure}

\item {\bf Construct and Use Illumination Function Correction.}  The non-uniformity in the spatial direction is referred
to as the ``slit illumination function" and is a combination of any unevenness in the slit jaws, the vignetting within the focal plane
or within the spectrograph, etc.  Most of these issues should have been removed by dividing by the flat field in step~\ref{step-flat}.
Is this good enough?  If the remaining non-uniformity (relative to the night sky) is just a few percent, and has a linear gradient,
this is probably fine for sky subtraction from a point source as one can linearly interpolate from the sky values on either side
of the object.  However, if the remaining non-uniformity is peaked or more complex,
or if the goal is to measure the surface brightness of an extended source, then getting the sky really and truly flat is crucial,
and worth some effort.

The cheapest solution (in terms of telescope time) is to use exposures of the bright twilight sky.  
With the telescope tracking, one obtains 3-5 unsaturated exposures,
moving the telescope slightly (perpendicular to the slit) between exposures.  To use these, one can combine the exposures (scaling
each by the average counts), averaging the spatial profile over all wavelengths, and then fitting a smooth function to the spatial profile.
The result is an image that is uniform along each row of the dispersion axis, and hopefully matches the night sky perfectly in
the spatial direction.  Sadly this is usually not the case, as during 
bright twilight there is light bouncing off of all sorts of surfaces in the dome. Thus one might need to make a further correction
using exposures of the dark night sky itself.  
This process in illustrated in
Figure~\ref{fig:twilights}.   The relevant IRAF tasks involved are {\it imcombine}, {\it illumination}, and
{\it ccdproc}

This concludes the basic CCD reductions, and one is now ready to move on to extracting and calibrating the spectrum itself.

\begin{figure} [htp]
\epsscale{0.30}

\plotone{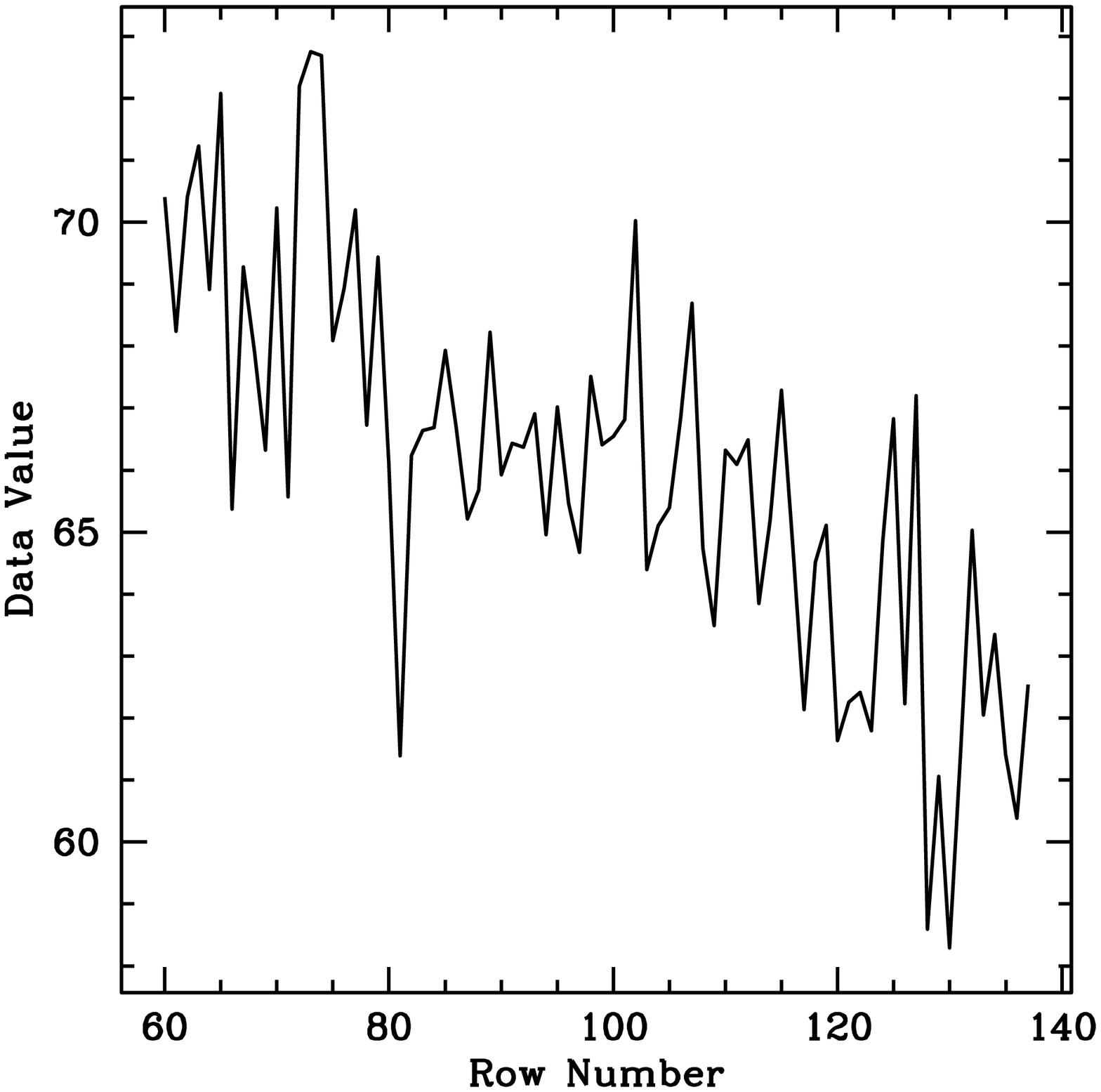}
\plotone{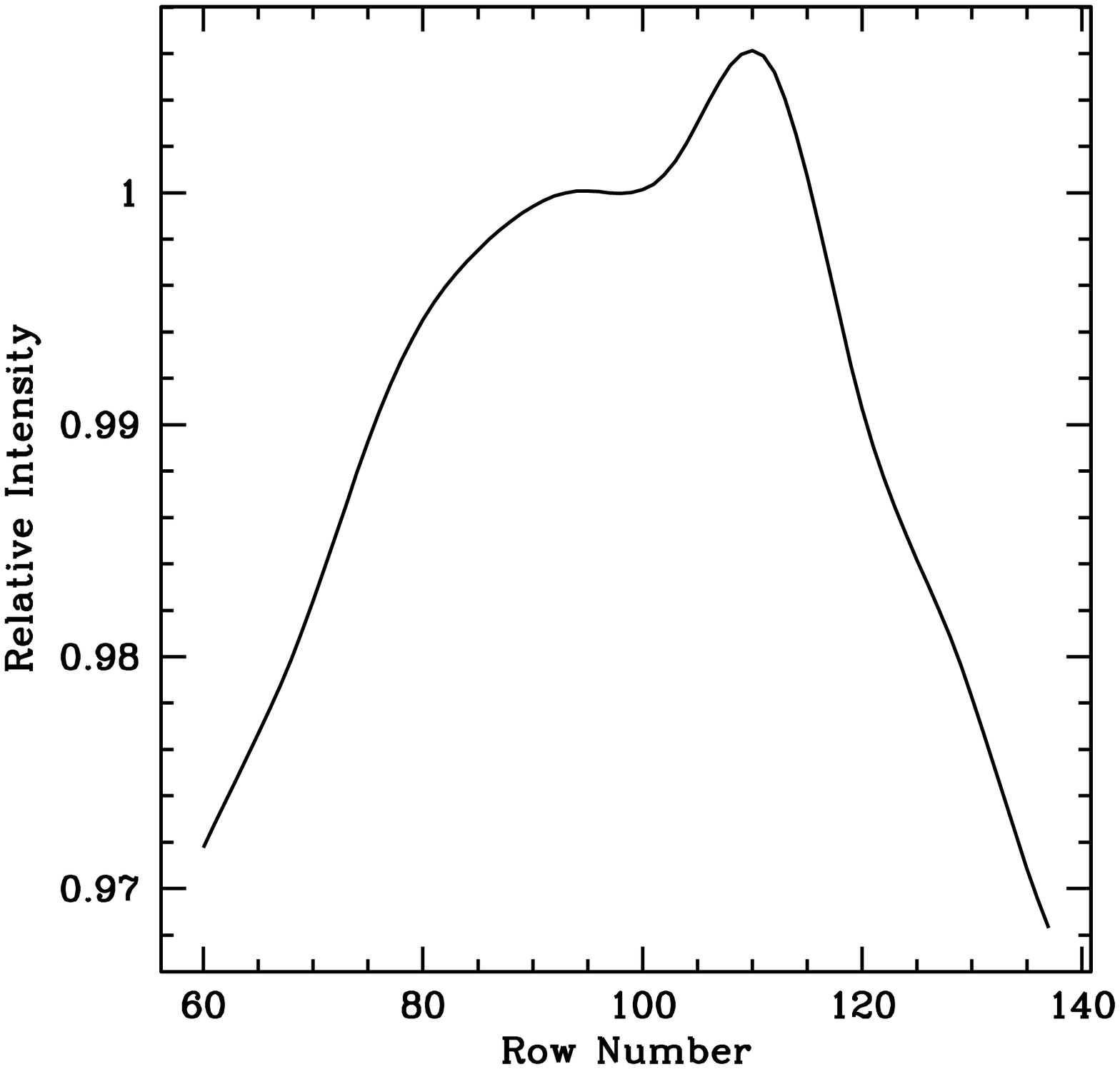}
\plotone{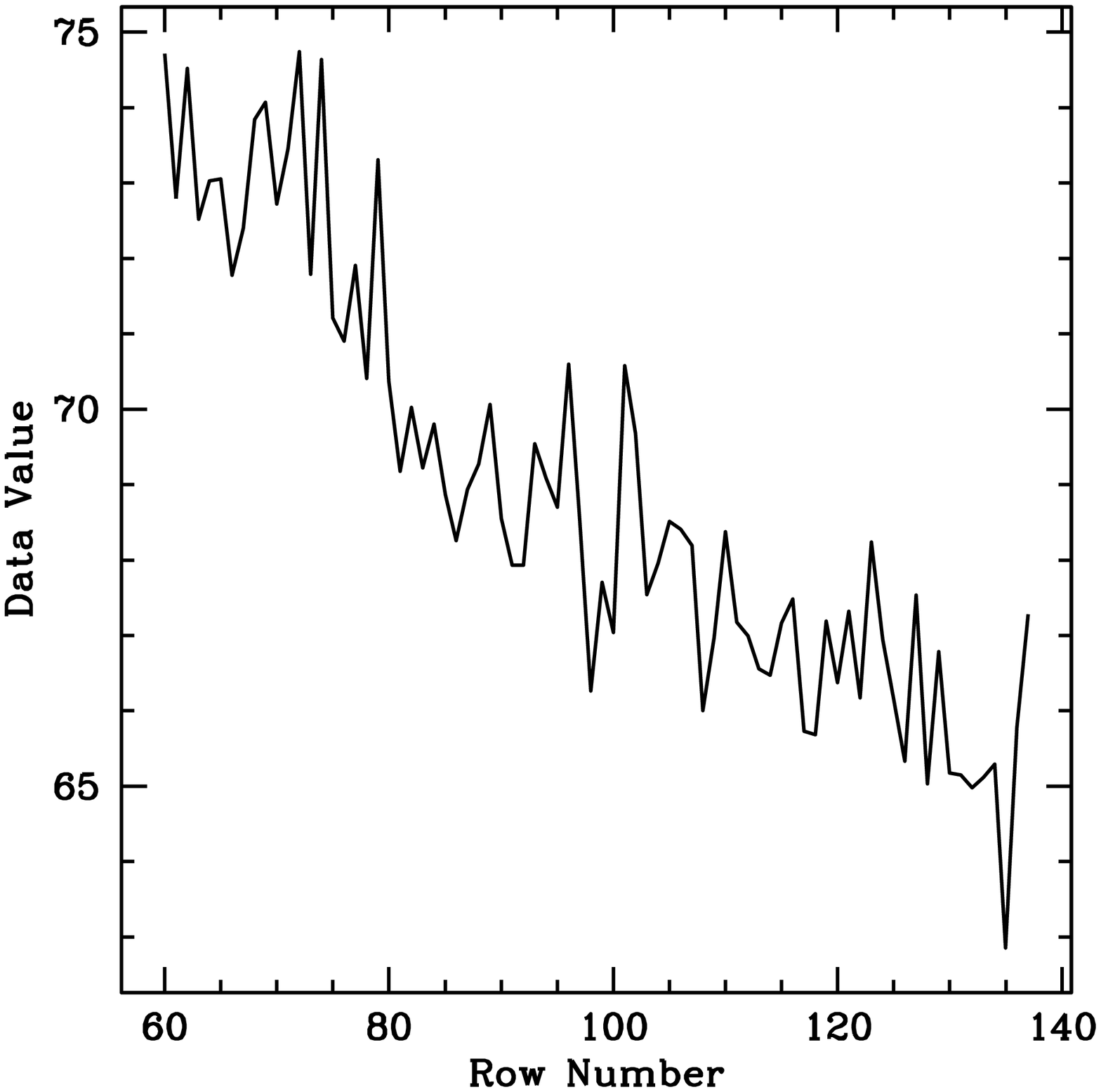}
\plotone{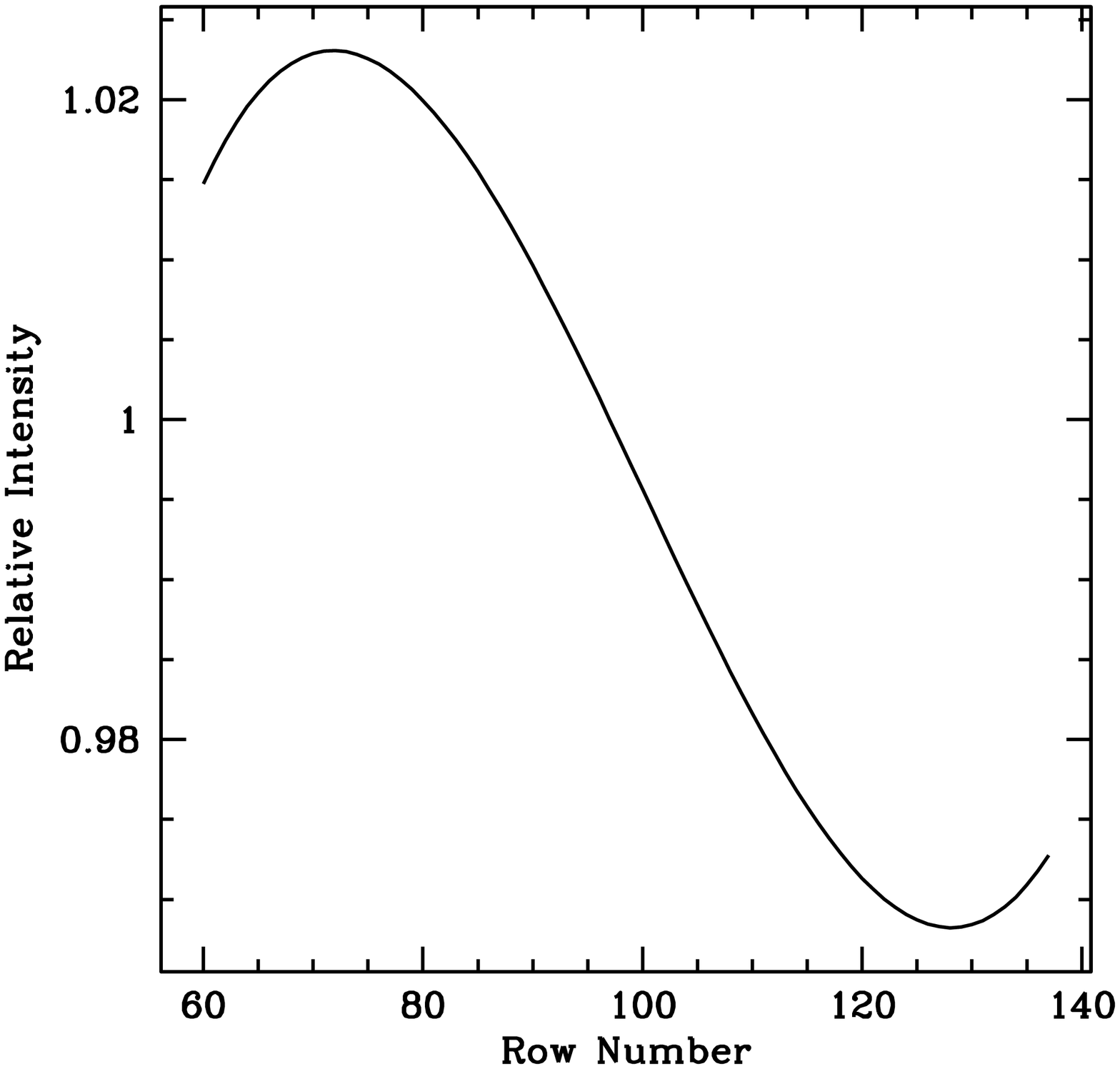}
\plotone{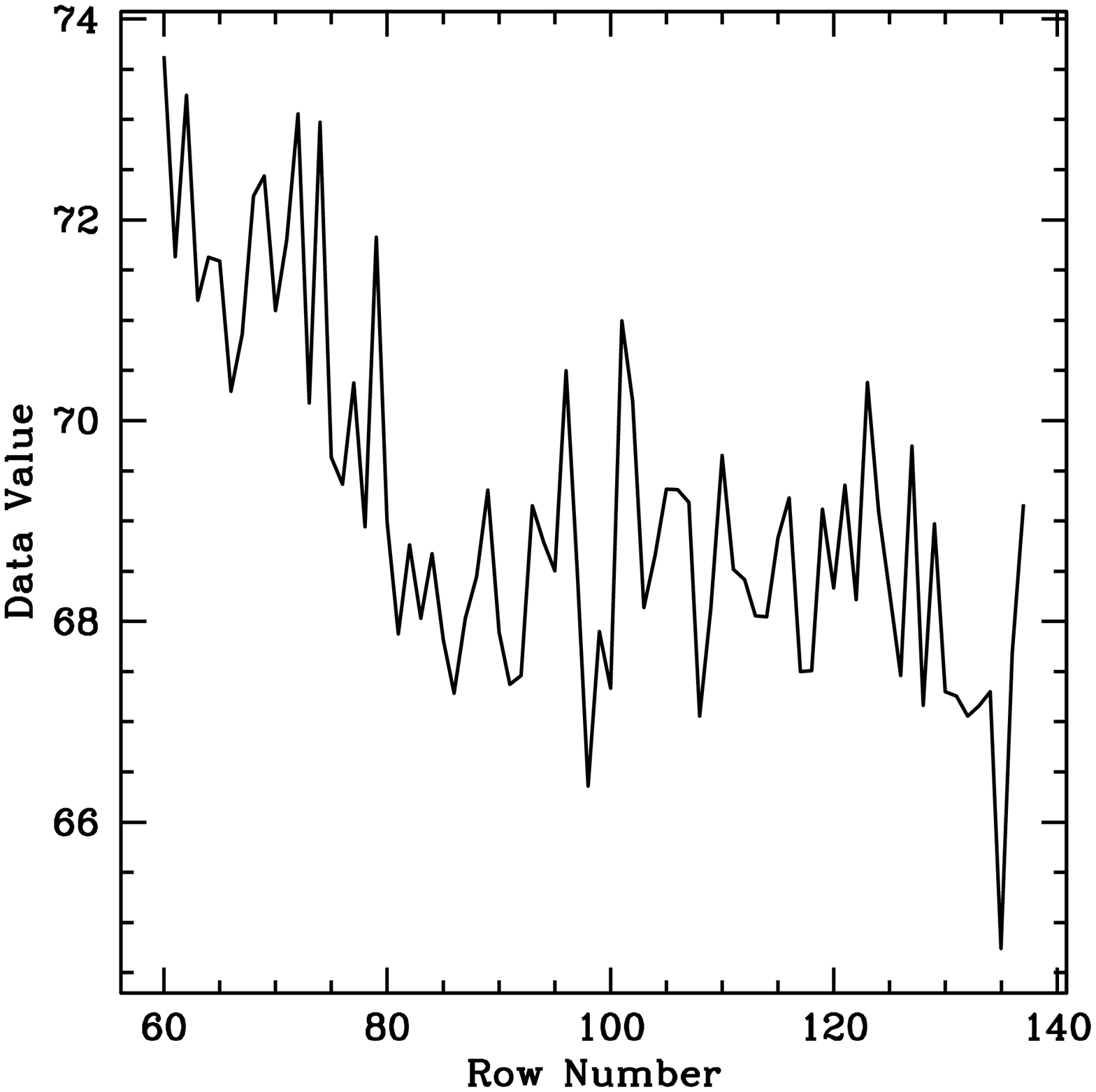}

\caption{\label{fig:twilights} The spatial profile after flattening by the dome flat (step~\ref{step-flat}) is shown
at the upper left.  Clearly there is a gradient from left to right.  The profile of the twilight sky is shown in the middle.  It's not a particularly
good match, and after dividing by it we see that the situation is not improved (top right).   The profile of all of the sky exposures (after
division by the twilight profile) is shown at lower left, and was used to re-correct all of the data.  After this, the spatial profile was
flatter, as shown at lower right, although there is now a modest kink.}
\end{figure}

\item{\bf Identification of Object and Sky.}  At this stage one needs to understand the location of the stellar 
spectrum on the detector, 
decide how wide an extraction window to use, and to select where the sky regions should
be relative to the stellar spectrum. 
The location of the star on the detector will doubtless be a mild function of wavelength, due
either to slight non-parallelism in how the grating is mounted or simply atmospheric differential refraction.   This map of the location of the stellar spectrum with wavelength is often
referred to as the ``trace".  Examples are shown in  Figure~\ref{fig:ap}.  In IRAF this is done
either as part of {\it apall} or via {\it doslit}.

\begin{figure}[htp]
\epsscale{0.4}
\plotone{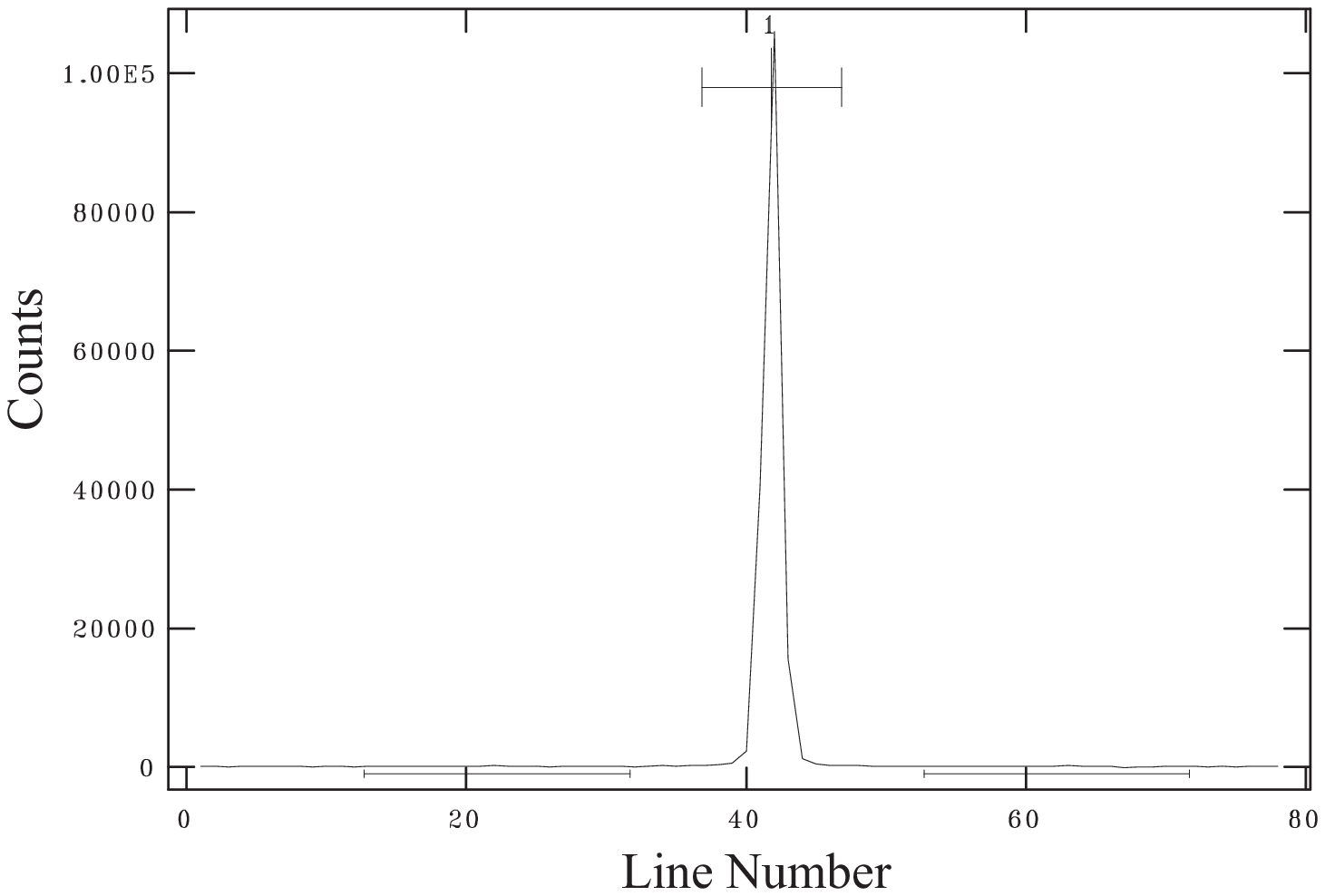}
\plotone{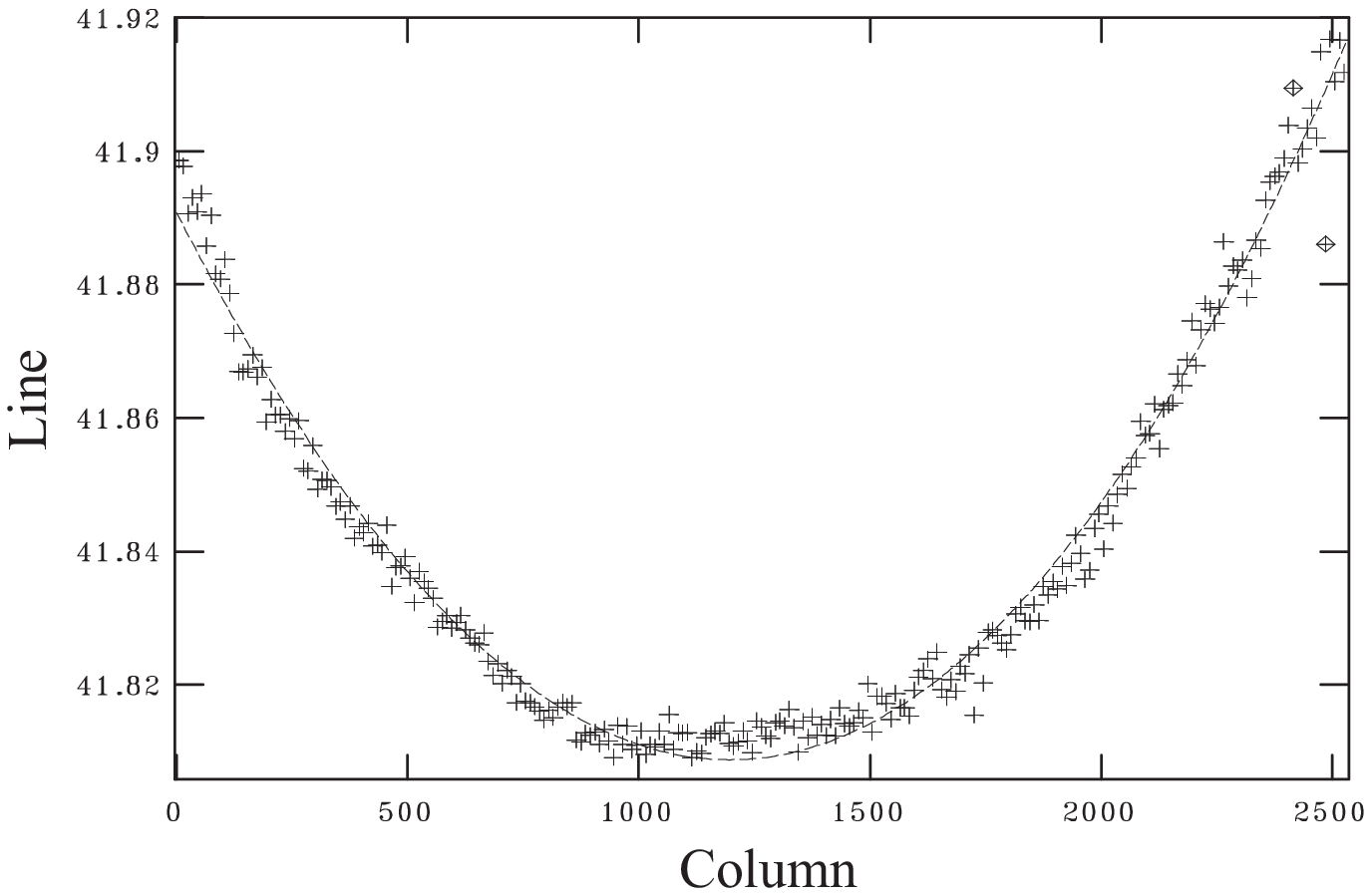}
\caption{\label{fig:ap} Locating the spectrum.  On the left is a cut along the spatial profile near the middle of the array.
A generously large extraction aperture (shown by the numbered bar at the top) has been defined, and
sky regions located nicely away from the star, as shown by the two bars near the bottom.  On the right  the trace of
the spectrum along the detector is shown.  Although the shape is well defined, the full variation is less than a tenth of a pixel. 
}
\end{figure}

\item \label{wave} {\bf Wavelength calibration.}  Having established the aperture and trace of the stellar spectrum, it behooves one 
to extract the
comparison spectrum in exactly the same manner 
(i.e., using the same trace), and then identify the various lines and perform a smooth fit to the wavelength
as a function of pixel number along the trace.  Figure~\ref{fig:comp} shows
a HeNeAr comparison lamp.  
In IRAF this is done either by {\it identify} or as part of {\it doslit}.

\begin{figure}[htp]
\epsscale{0.42}
\plotone{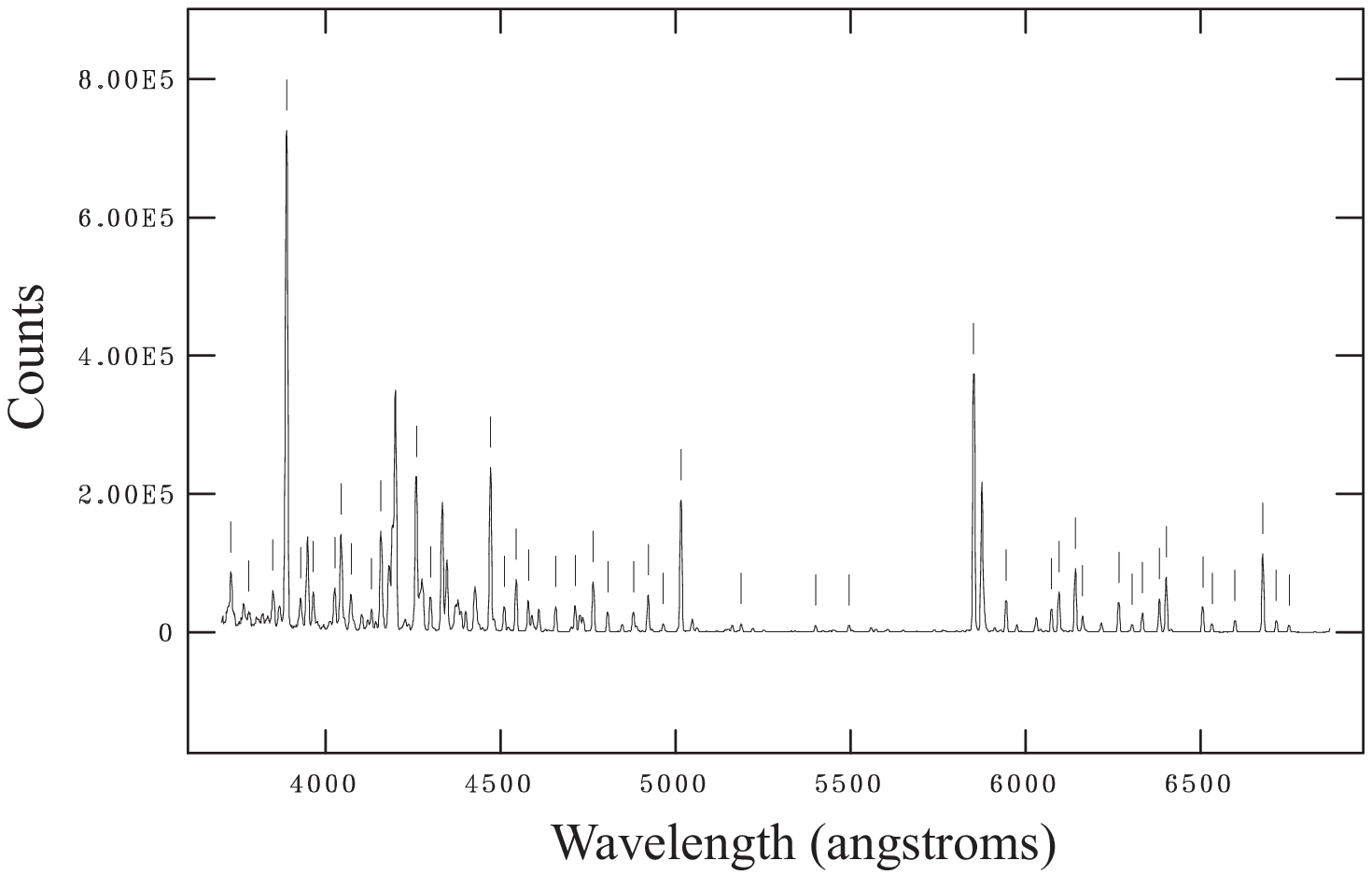}
\plotone{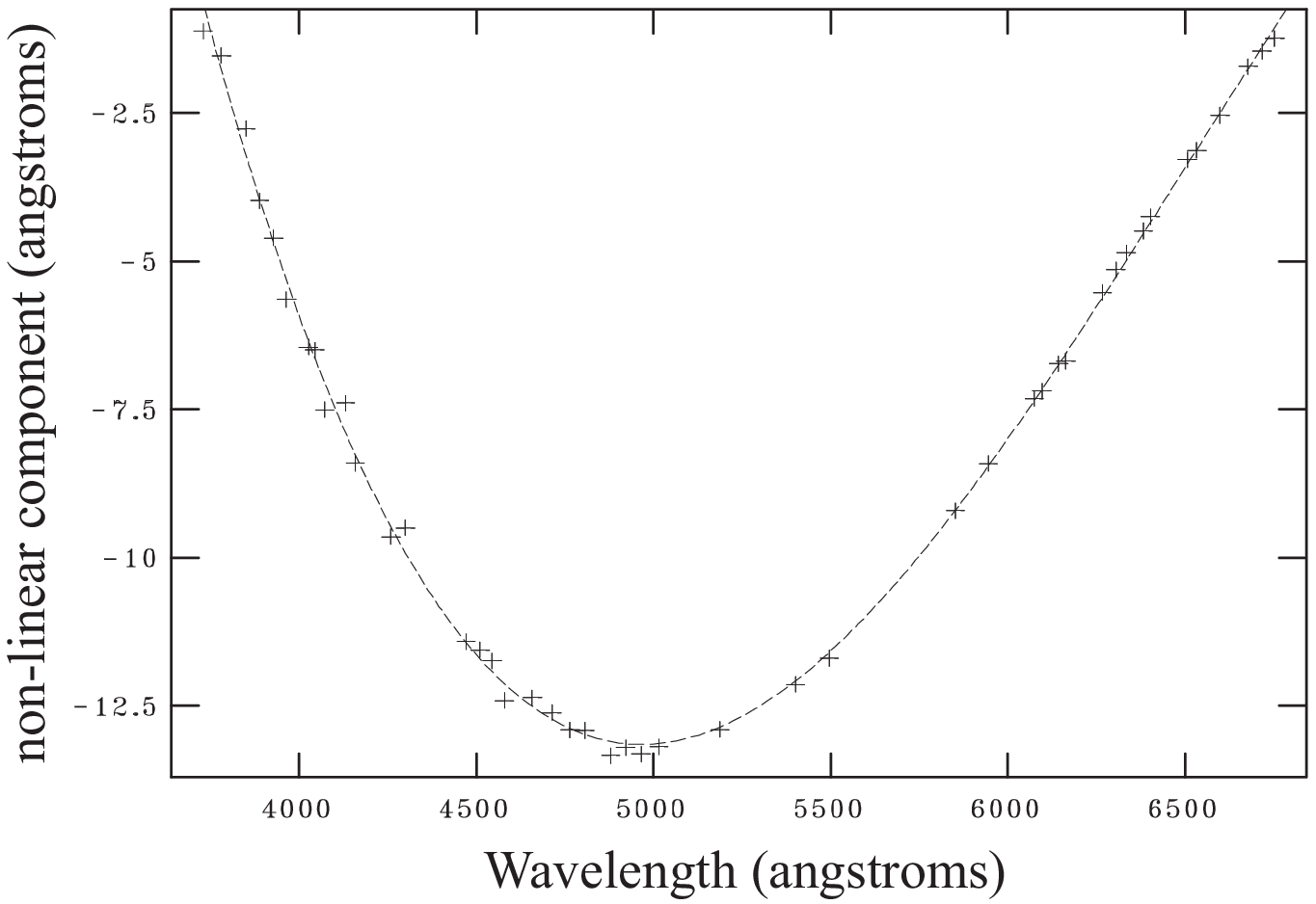}
\caption{\label{fig:comp}  Wavelength calibration. On the left is shown a HeNeAr spectrum, with prominent lines marked and identified
as to wavelength.  On the right is shown the non-linear portion of the fit of a first order cubic spline; the residuals from the fit were about 0.2\AA.}

\end{figure}

\item{\bf Extraction.} Given the location of the spectrum, and knowledge of the wavelength calibration, one needs to then extract the spectrum,
by which one means adding up all of the data along the spatial profile, subtracting sky, and then applying the wavelength solution.
Each of these steps requires some decision about how to best proceed. 

First, how does one establish an unbiased sky estimate?  To guard against faint, barely resolved stars in the sky aperture, one often
uses a very robust estimator of the sky, such as the median.  To avoid against any remaining gradient in the spatial response, one
might want to do this on either side of the profile an equal distance away.  The assumption in this is that the wavelength is very close
to falling along columns (or lines, depending upon the dispersion axis).  If that is not the case, then one needs to geometrically
transform the data.

Second, how should one sum the spectrum along the stellar profile?  One could simply subtract the sky values,  and then sum over the spatial profile, weighting each point equally,
but to do this would degrade the signal-to-noise ratio. Consider:
 the data near the sky add little to the
signal, as the difference is basically zero, and so adding those data in with the data of
higher signal will just add to the noise.  Instead, one usually uses a weighted sum, with
the statistics expected from the data (depending upon the gain and the read-noise).  
Even better would be to use the ``optimal extraction" algorithm described below (\S~\ref{Sec-optimal}), which would also allow one to
filter out cosmic rays.

Third, how should the wavelength solution be applied?  
From the efforts described in step \ref{wave} above,
one knows the wavelength as a function of pixel value in the extracted spectrum.  Generally one 
has used a cubic spline or something similarly messy to characterize it.  So, rather than try to
put a description of this in the image headers, it is usually easier to ``linearize" the spectrum, such that
every pixel covers the same number of Angstroms.  The interpolation needs to be carefully done in order
not to  harm to the data.

In IRAF these tasks are usually handled by {\it apsum} and {\it dispcor} or else as part of
{\it doslit}.

\item{\bf Finishing up: normalization or flux calibrating?}  In some instances, one is 
primarily interested in modeling line profiles, measuring equivalent
widths in order to perform abundance analysis, measuring radial velocities, or 
performing classical spectral type classification spectra.
In this case, one wants to normalize the spectrum.   In other cases one might want to model
the spectral energy distribution, determine reddening, and so on, which requires one
to ``flux the data"; i.e., determine the flux as a function of wavelength.
 Figure~\ref{fig:finishing} shows a raw extracted spectrum of Feige 34, along with a normalized version of the spectrum and a fluxed-calibrated version of the spectrum.

\begin{figure}[htp]
\epsscale{0.32}
\plotone{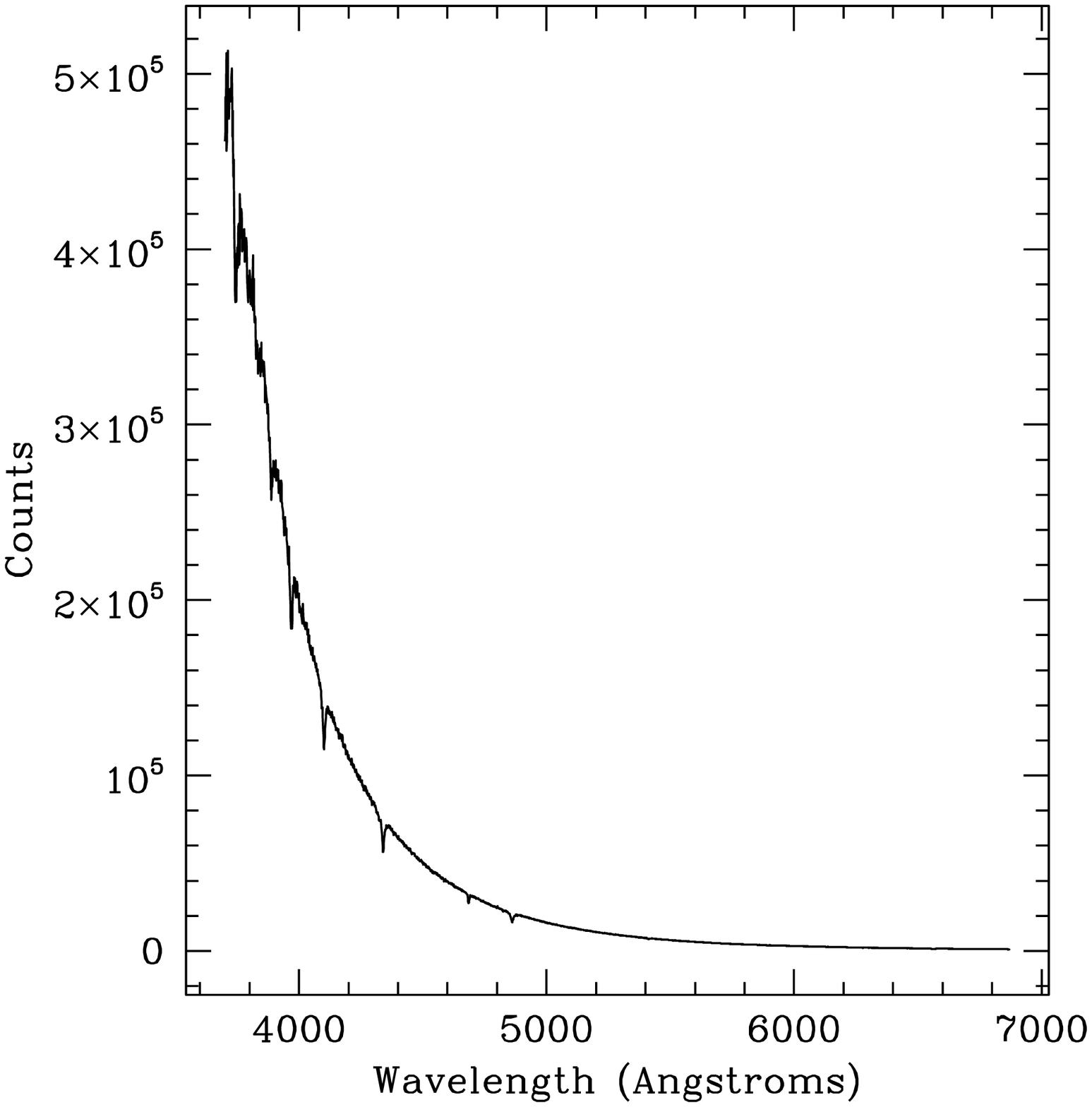}
\plotone{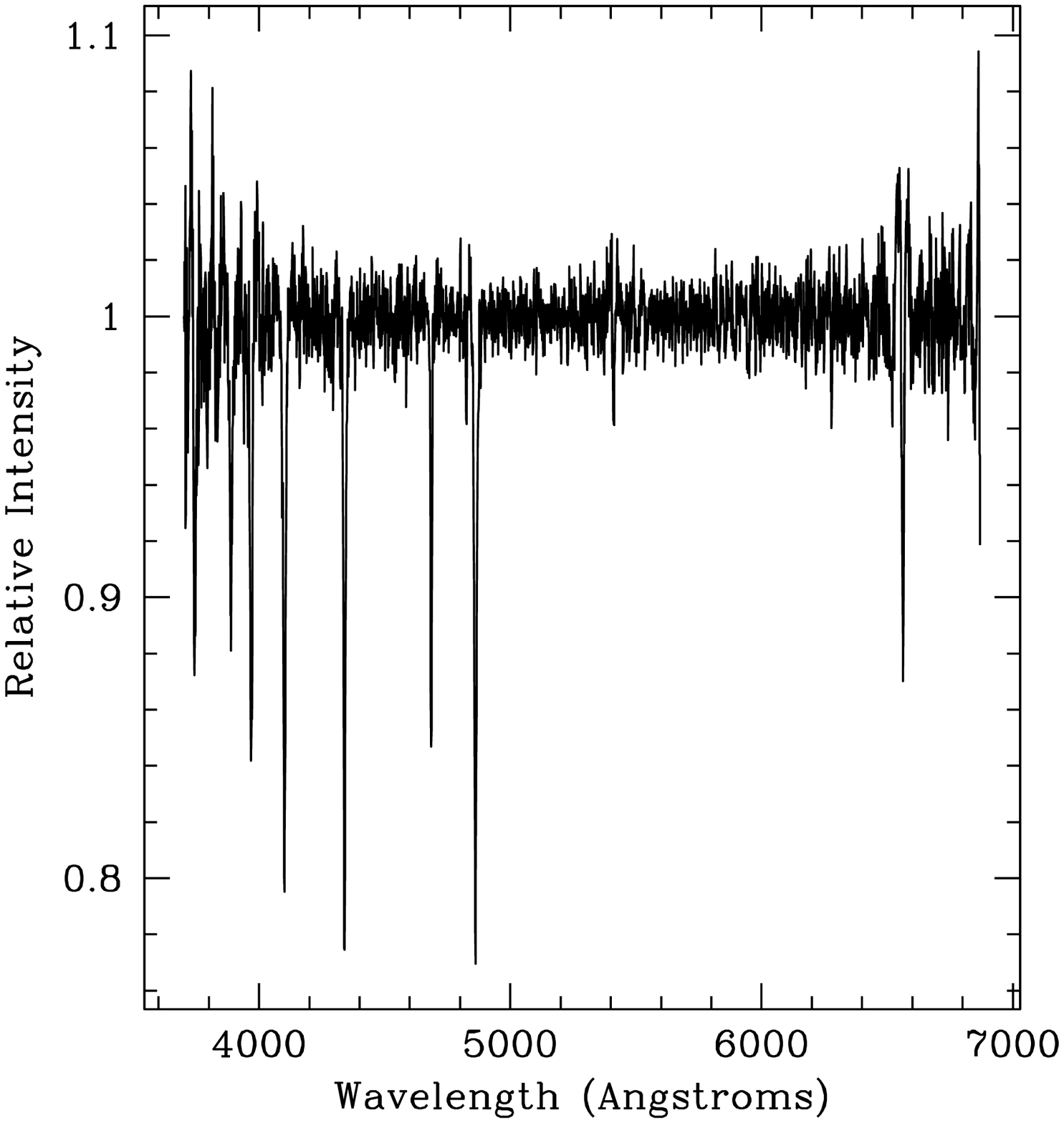}
\plotone{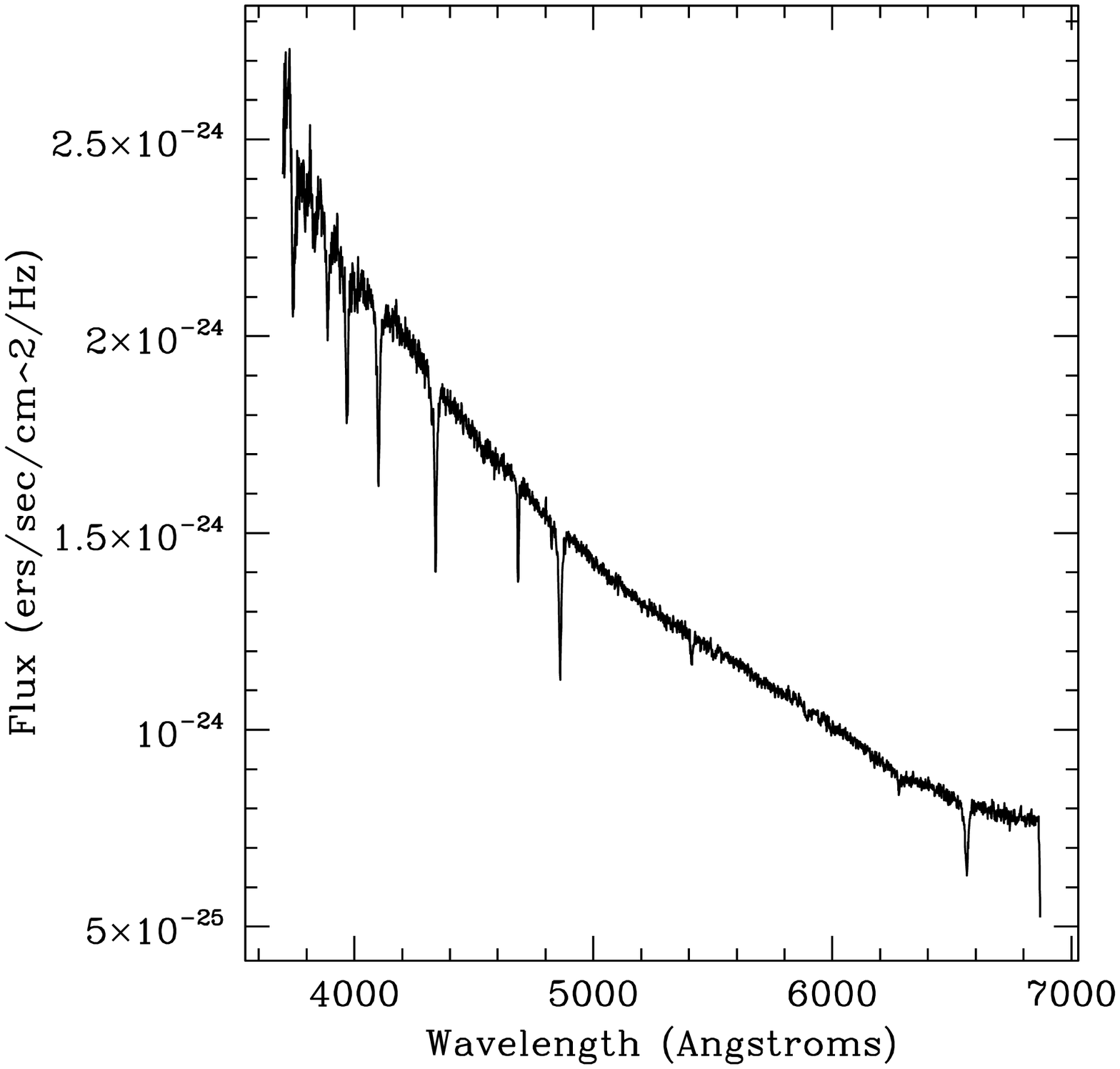}
\caption{\label{fig:finishing}  Finishing the reductions.  The left panel shows the 
extracted spectrum
of the star Feige 34.  There are far more counts in the blue than in the red owing to the
normalization of the flat field.  The middle panel shows the normalized version of the same
plot; a higher order function was fit to the extracted spectrum with low and high points being
rejected from the fit iteratively.  The right panel shows the fluxed version of the spectrum,
with $F_\nu$ vs. wavelength.}
\end{figure}

Normalization is usually achieved by fitting a low order function to the spectrum.  In order
to exclude absorption lines, one might wish to eliminate from the fit any point more than
(say) 2.5$\sigma$ below the fit and iterate the fit a few times.  At the same time, one might
want to avoid any emission lines and reject points 3$\sigma$ too high.  The relevant
IRAF task is {\it continuum}.

 Spectrophotometric
standards are stars whose fluxes have been measured in various wavelength ranges, some
of them at 50\AA\ intervals\footnote{The fluxes are often expressed in terms of
spectrophotometric
``AB magnitudes", which are equal to $-2.5\log f_\nu - 48.60,$ where $f_\nu$ is the flux
in ergs cm$^{-2}$ s$^{-1}$ Hz$^{-1}$.  The spectrophotometric magnitude is essentially
equal to $V$ for a wavelength of 5555\AA.}. By observing the standard stars and summing the counts over the same bandpasses used to calibrate the standards, one can associate a count rate (as a function of wavelength) with the published fluxes.   In practice, one uses a low order
fit to the observed counts per second and
the spectrophotometric magnitudes for all of the standard stars, applying a grey shift (if needed) to reduce the effects of slit losses.    Optionally one might try to derive a wavelength-dependent
extinction curve with such data, but experience suggests that the 3-5\% errors in the
calibration of the standards preclude much improvement over the mean extinction curve\footnote{Note that this is not equivalent to adopting a mean  extinction; instead, that 
mean extinction term is absorbed within the zero-point of the fit.  Rather, this statement
just means that changes in the extinction tend to be independent of wavelength (grey).}.
Finally one corrects the observed data by the mean extinction curve and applies the
flux calibration to the data.  The final spectrum is then in terms of either $f_\nu$ or $f_\lambda$ vs wavelength.

The relevant IRAF tasks are {\it standard}, {\it sensfunc}, and {\it calib}.

\end{enumerate}

The steps involved in this reduction are straightforward, and when observing
stars using a long slit it is hard to imagine why an astronomer would not have fully
reduced data at the end of the night.  Carrying out these reductions in real time allows
one to see that the signal-to-noise ratio is as expected,
and avoids there being any unpleasant surprises after the the observing run.
The IRAF task {\it doslit} was specifically designed with this in mind, to allow a quick-look
capability that was in fact done sufficiently correctly so that the data reduced at
the telescope could serve as the final reduction.

\subsubsection{Multi-object techniques}

Most of what has just been discussed will also apply to multi-object data, be it
fibers or slitlets.  There are a few differences which should be emphasized.

For fibers, sky subtraction is accomplished by having some fibers assigned to blank sky.
However, how well sky subtraction works is entirely dependent upon how well the flat-fielding
removes fiber-to-fiber sensitivity variations (which will be wavelength dependent) {\it as well as}
how well the flat fields match the vignetting of a particular fiber configuration.  Since the vignetting is
likely to depend upon the exact placement of a fiber within the field, the best solution is to
either model the vignetting (as is done in the Hectospec pipeline) or to make sure that
flat fields are observed with each configuration.  For some telescopes and instruments 
(such as CTIO/Hydra) it
is more time efficient to simply observe some blank sky, making a few dithered exposures
near the observed field.

For multi-slits one of the complications is that the wavelength coverage of each slitlet will
depend upon its placement in the field, as discussed above in \S~\ref{Sec-multi}. This results
in some very challenging reductions issues, particularly if one plans to flux-calibrate.   Consider
again the IMACS instrument (\S~\ref{Sec-IMACS}).  There are eight chips, each of which will have
its own sensitivity curve, and even worse, the wavelength coverage reaching each chip will depend upon
the location of each slitlet within the field.  The perfect
solution would be to observe a spectrophotometric standard down each of the slits, but that would hardly
be practical.  Flux calibrating multislit data is {\it hard}, and even obtaining normalized spectra can be a
bit of a challenge.

\subsubsection{NIR techniques}
\label{Sec-irred}

When reducing near-infrared spectra, the basic philosophy outlined in \S~\ref{Sec-Basicred} for the optical generally applies.  But there are two significant differences between the infrared and optical.  First the detectors are different,  as discussed in \S~\ref{Sec-irspectrographs}.  
This leads to some minor modifications of the standard optical reduction.  
But, more importantly, observing in the near-infrared presents significant challenges that do not occur in the optical, in the form of significant background radiation and strong absorption from the EarthÕs atmosphere. 
For sky, which is often much brighter than the science target, a sky frame is obtained equal in exposure time to the object. For point sources this is often accomplished by moving the object along the slit (dithering). Alternate dithers can be subtracted to remove the sky. Or, the entire dither set might be median combined to form a sky frame.

Although this process works relatively well for subtracting the night-sky emission, it does 
nothing for removing the telluric absorption features.  Instead, telluric standards (stars with nearly featureless intrinsic spectra) need to be observed at nearly {\it identical}  airmasses 
 as the
program stars: differences of less than 0.1 airmasses is advisable.   Why? The IR telluric absorption spectrum is due to a number of molecules (e.g., CO, CO$_2$, H$_2$O, CH$_4$, etc.),
which can vary in relative concentration, especially H$_2$O.  In addition,
some of the absorption lines are saturated and thus do not scale linearly with airmass.  Both of these factors can make it difficult or impossible to scale the telluric absorption with airmass.   
  
Additional calibration observations and reduction steps will be required to address these challenges.  Newcomers to spectroscopy might well
want to reduce optical spectra before attempting to tackle the infrared.

Infrared hybrid arrays do not have an overscan strip along their side like in a CCD.   However, bias structure is a very real problem and absolutely must be removed.  Direct bias images are not typically used to do this.  This is because in the infrared, one of the first stages in reduction is always subtracting one data frame from another, of the same integration time, but with differing sky locations or with differing sources on or off. Thus the bias  is automatically removed as a result of always subtracting ``background" frames, even from flat-field and comparison exposures.

During the day, one should take the following, although they may not always be needed in the reductions:

\begin{itemize}
\item{\bf Dark frames} are used to subtract the background from twilight flats and comparison lamps, 
so it is important that the darks have the same exposure times as the
twilight and comparison lamp exposures. Dark frames are also used to make a bad pixel mask.

\item{\bf A bad pixel mask} is constructed from (1) the dark frames, which identify hot pixels,  and from (2) the dome flats, which
are used to identify dead and flaky pixels. 

\item{\bf Dome flats}  are critical.    And, they are done differently in the near-infrared.
They must be obtained for 
{\it each} grating setting and filter combination you use during the night, the same as with optical data. Furthermore, they should be
obtained with exposure levels that push to the linearity limits in order to maximize the signal-to-noise ratio.  This linearity limit
can be as low as 10,000 e$^-$ for some IR detectors.  However, {\it unlike the case for the optical},
one must also obtain an equivalent set (in terms of exposure time and number) of dome flats with the dome illumination turned off, as the dome flat itself is radiating in the NIR.
One can demonstrate the necessity for this by a simple experiment: subtract an image obtained with the lamps turned off from an
image with the lamps turned on, and see how many counts there really are over the entire wavelength range.  One 
may be surprised how many counts were removed from the original exposure, particularly in the longer wavelength $K$ band.
 But it is this {\it subtracted} dome flat that will be used to flatten the data frames, so there needs to be a lot of counts in it in order 
 not to reduce the signal-to-noise!  For the $K$ band, where there is high background counts and because the dynamic range of the detector is
 small, one may need 20 or more flat-on, flat-off pairs to get sufficient real signal. 
\item{\bf Quartz flats} will be taken during the night, but daytime exposures will test the exposure levels and reveal any structural differences
between it and the dome and twilight flats.  Of course, dark frames with matching integration times will be needed.

\item{\bf Comparison lamps} should be taken throughout the night, but obtaining a few during the day may also prove to be useful.
Be sure to get dark frames to match the integration time.
\end{itemize}

\noindent
During the night, one will take:
\begin{itemize}
\item{\bf Telluric standard stars} are (nearly) featureless stars that will be used to derive the
spectrum of the 
 telluric absorption bands for removal
from the program data.   These are not spectrophotometric stars. They must be chosen to lie at similar (within 0.1) 
airmasses 
as the target stars for the reasons explained above, 
and observed within an hour or two of the target (more often for observations beyond 3$\mu$m).
\item{\bf Quartz Lamps} may be useful or not, depending upon the specific instrument.
\item{\bf Comparison lamp} exposures are a good idea.  In the infrared, one can almost always use the night
sky emission lines as the wavelength reference source.  But separate comparison lamp exposures may prove useful if there turns out
not to be enough counts in the sky lines, or if some regions are too void of sky lines, such as the long-K region.  It is also a good idea
to take these if you are moving the grating often,  the program integration times are short, or  if the dispersion is high.  
\item{\bf Twilight flats }may wind up not being used, but they could be handy.  Many NIR spectrographs will need some illumination correction
(as in the optical)  and the twilight flats are the easiest means to make such corrections.  Be sure to get dark frames of the same integration time to remove dark current and bias levels later. 
\item{\bf Stellar flats} are spectra of a very bright star which is moved by small amounts along the slit. The images can be co-added to create a very rough point-source flat, and used much like a twilight flat to check on illumination corrections or other irregularities which can uniquely appear in point-source observations. 
\end{itemize}

The basic two-dimensional reduction steps for NIR spectroscopic data are presented here; one should review \S~\ref{Sec-Basicred} to
compare and contrast these with what is involved in optical reductions.  The relevant IRAF tasks are mentioned for convenience,
although again the goal here is not to identify what buttons to press but rather to be clear about the steps needed.

\begin{enumerate}

\item \label{subtract} {\bf Subtract one slit position frame from another.}    This step removes the bias and all other uniform additive sources, such as
evenly illuminated sky. (Lumpy, nebular emission will subtract horribly.)  In long-slit spectroscopy one has to observe the star
in two (or more) positions along the slit.  This basic subtraction step needs to be done while one is observing, and it behooves the
observer to {\it always look at the result.}  There should be a zero-level background throughout, with two stars, one positive, one negative, running along the dispersion axis.   The primary thing to check is to see if the OH night-sky emission lines are canceling.  There may be  residual positive and/or negative OH lines running perpendicular to the dispersion axis.   This can happen just
as a consequence of the
temporal and spatial variations in the OH spectra, but certainly will happen if light clouds move through the field.   As there will be many frames, try all  combinations with unique slit positions to find the best
OH removal.  Nothing else should be there, and if there is, there are some serious problems.  The only IRAF tool needed is {\it imarith}.

\item {\bf Construct and apply a master, normalized, featureless flat.}  The ``lamp on" flats and the ``lamp off" flats are median averaged
using a rejection algorithm (IRAF: {\it imcombine}) and then the ``lamps off" average is subtracted from the ``lamps on" average to produce a
master flat.  As with the optical, one must use one's own judgement about how to normalize.  This flat is then divided
into each of the subtracted frames.  The process is repeated for each grating/filter combination. 

\item {\bf Construct and use an illumination correction.}  This step is identical to that of the optical, in that twilight flats can be used to 
correct any vignetting left over from the flat-field division. Alternately, telluric standards can be obtained at the same slit positions as science targets. Making a ratio of the object to telluric standard will also correct for vignetting along the slit.

\item \label{bad} {\bf Make the bad pixel mask.}  A good bad pixel mask can be constructed from a set of dark and lamps-on flat field images.  A histogram
from a dark image reveals high values from ``hot" pixels.   Decide where the cut-off is.  Copy the dark image, call it `hot', set all pixels below this level to zero, then set all pixels above zero to 1.   Display a histogram of your flat and decide what low values are unacceptable. Copy the flat image calling it dead and set all values below this low value to 1.  Set everything else to 0.   Finally, take many identically observed flat (on) exposures, average them and determine a sigma map (in IRAF, this is done using {\it imcomb} and entering an image name for ``sigma").   Display a histogram, select your upper limit for acceptable sigma, and set everything below that to zero in the sigma map.  Then set everything above that limit to 1.  Now average (no rejection) your three images: dead, hot and sigma.  All values above 0.25 get set to 1.0 and {\it voila!},
one has a good bad pixel mask.   One should then examine it to see if one was too harsh or too lax with your acceptable limits.

\item {\bf Trace and extract spectrum.}  This step is nearly identical to what is done for the optical: one has to identify the location of the
stellar spectrum on the array and map out its location as a function of position.  Although the
sky has already been subtracted to first order (Step~\ref{subtract}), one might want to 
do sky subtraction again during this stage for a couple of reasons, namely  if one needs 
to remove astrophysical background (nebular emission or background stellar light), or if the previous sky subtraction left strong residuals due to temporal changes, particularly in the sky lines. Be sure {\it not} to use optimal extraction, as the previous sky subtraction has altered the
noise characteristics.  If you do subtract sky at this stage, make sure it is the median of many values, else one will add noise.
(It may be worth reducing a sample spectrum with and without sky subtraction turned on in the extraction process to see which is better.)
Bad pixels can be flagged at this stage using the bad pixel mask constructed in Step~\ref{bad} and will disappear
(one hopes!) when all of the many extracted frames are averaged below.  In IRAF the relevant task is {\it apall}.

\item {\bf Determine the wavelength scale.}  One has a few choices for what to use for the wavelength calibration, and the right
choice depends upon the data and goals.  One can use the night-sky emission spectrum as a wavelength reference, and in fact at
high dispersion at some grating settings these may be the only choice.  (One needs to be at high dispersion though to do so as many of the
OH emission lines are hyperfine doubles; see Osterbrock et al.\ 1996, 1996.) In this case, one
needs to start with the raw frames and median average those with unique stellar positions with a rejection algorithm that will get rid of the
stellar spectra.  Alternatively, in other applications one may be able to use the comparison lamp exposures after a suitable dark has been removed.  Whichever is used,
one needs to extract a one-dimensional spectrum following the same location and trace as a particular stellar exposure.  One can
then mark the positions of the reference lines and fit a smooth function to determine the wavelength as a function of position along the
trace. One should keep fit order reasonably low.   A good calibration should result in a
wavelength scale whose 
uncertainty is smaller than one tenth of the resolution element.
In principle one can also use the telluric absorption lines for wavelength calibration, but this is only recommended if there is no other recourse, as it is very time consuming.  The comparisons will have to be extracted and calibrated for each stellar trace.  The resulting
fits can be applied to the stellar spectra as in the optical.  The relevant IRAF tasks are {\it identify}, {\it reidentify}, and {\it dispcor}.

\item {\bf Average the spectra.}  If one took the data properly, there should be 6-12 spectra that can be averaged to increase the signal-to-noise and remove bad pixels.  One needs to check that the spectra have the same shape.   The spectra
need to be averaged in wavelength, not pixel, space, with a rejection algorithm to remove the bad pixels and with appropriate scaling
and weighting by the number of counts.  In IRAF the relevant task is {\it scombine}.

\item {\bf Remove telluric absorption.}  How do the spectra look?  Not so good?  That's because there is still one last important step left, namely
the removal of telluric absorption features.   The  telluric standards need to be treated with the same care and effort used for the
target spectra.   There is a bit of black magic that must occur before one can divide the final target spectrum by the final telluric spectrum in (wavelength space) to remove the Earth absorption lines, namely that any stellar features intrinsic to the telluric standards have to be
removed. A-type stars are often used as telluric standards as they have hardly any lines other than hydrogen. But the hydrogen lines are huge, deep and complex: broad wings, deep cores and far from Gaussian to fit, particularly if $v\sin i$ is low.  Some prefer to use early G-type
 dwarfs, correcting the intrinsic lines in the G star by using a Solar spectrum, available from the National Solar Observatory.  For lower resolution work
 or in a low signal-to-noise regime, this is probably fine and the differences in metallicity, $v\sin i$, temperature and gravity between the telluric and the Sun may not be significant.   Probably the best plan is to think ahead and observe at least one telluric star during the night for which the intrinsic spectrum is already known to high precision (see the appendix in Hanson et al.\ 2005).   This gives a direct solution from which you can boot-strap additional telluric solutions throughout the night.  More about this is discussed below in \S~\ref{Sec-ObsNIR}.
 \end{enumerate}

\subsection{Further Details}

The description of the reduction techniques above were intended as a short introduction to
the subject.  There are some further issues that involve both observation and reduction
techniques that are worth discussing in some additional depth here; in addition, there are some
often-neglected topics that may provide the spectroscopist with some useful tools. 

\subsubsection{Differential Refraction}
\label{Sec-parallactic}

Differential refraction has two meanings to the spectroscopist, both of them important.  First, there is the issue of refraction as a function
of wavelength.  Light passing through the atmosphere is refracted, so that an object will appear to be higher in the sky than it
really is.  The amount of refraction is a function of the wavelength and zenith distance, with blue light being refracted the most,
and the effect being the greatest at low elevation.  
If one looks at a star near the horizon with sufficient magnification one will
notice that the atmosphere itself has dispersed the starlight,
with the red end of the spectrum nearest the horizon, and the blue part of the spectrum further from the horizon.  The second meaning
has to do with the fact that the amount of refraction (at any wavelength) depends upon the object's zenith distance, and hence
the amount of refraction will differ across the field in the direction towards or away from the zenith.

This wavelength dependence of refraction has important implications for the spectroscopist.  If one is observing a star at low elevation
with the slit oriented parallel to the horizon, the blue part of the star's light will be above the slit, and the red part of the star's
light below the slit if one has centered on the slit visually.  Thus much of the light is lost.  Were one to instead rotate the slit
so it was oriented perpendicular to the horizon then all of the star's light would enter the slit, albeit it at slightly different spatial locations
along the slit.  So, the spectrum would appear to be tilted on the detector, but the light would not have been selectively removed.

The position angle of the slit on the sky is called the {\it parallactic angle}, and so it is good practice to set the slit to this orientation
if one wishes to observe very far from the zenith.  How much does it matter?  Filippenko (1982) computed the amount of refraction expected relative to 5000\AA\ for a variety of airmasses making realistic assumptions. Even at a modest airmass of 1.5, the image at 4000\AA\ is displaced upwards (away
from the horizon) by 0.71 arcsec compared to the image at 5000\AA.  So, if one were observing with a 1-arcsec slit, the 4000\AA\ image would
be shifted out of the slit!   The degree of refraction scales as the tangent of the zenith distance, with the wavelength dependence
a function of the ambient temperature, pressure, and amount of water vapor in the atmosphere (see Filippenko 1982 and
references therein for details).  

The parallactic angle $\eta$ will be 0$^\circ$ or 180$^\circ$ for an object on the meridian (depending, respectively, if the object
is north or south of the zenith); at other hour angles $h$ and declinations $\delta$ the relationship is given by
$$\sin \eta = \sin h \cos \phi / [1-(\sin \phi \sin \delta+ \cos \phi \cos \delta \cos h)^2]^{0.5},$$
where $\phi$ is the observer's latitude.  Fortunately most telescope control systems compute the parallactic angle automatically
for the telescope's current position.

When using fibers or multi-slit masks there is not much one can do, as the fiber entrance aperture is circular, and the multi-slit masks
are designed to work at one particular orientation on the sky.  Thus these instruments almost invariably employ an atmospheric
dispersion corrector (ADC).  (There are several ways of constructing an ADC, but most involve
two counter-rotating prisms; for a more complete treatment, see Wynne 1993.) 
This is good for an additional reason having to do with the second meaning of ``differential refraction,"
namely the fact that objects on one side of the field will suffer  slightly different refraction than objects on the other side of the
field as the zenith distances are not quite the same.  Imagine that the instrument has a 1$^\circ$ field of view, and that one is
observing so that the ``top" of the field (the one furthest from the horizon) is at a zenith distance of 45$^\circ$.  The lower
part of the field will have a zenith distance of 46$^\circ$, and the difference in tangent between these two angles is 0.036.
At 5000\AA\ the typical amount of refraction at a zenith distance of 45$^\circ$ is roughly 1 arcmin, so the differential refraction
across the field is 0.036 arcmin, or about 2 arcsec!  Thus the separation between multi-slits or fibers would have to be adjusted by this
amount in the absence of an ADC.   Among the fiber positioners listed in Table~4, only Hydra on WIYN lacks an ADC.
There one must employ rather large fibers and adjust the position of the fibers depending upon the proposed wavelength
of observation. 
Nevertheless, ADCs have their drawbacks, and in particular their transmission in the near-UV may be very poor.

\subsubsection{Determining Isolation}
\label{Sec-isolate}

An interesting question arises when obtaining spectra of stars in crowded fields: how much light from neighboring objects
is spilling over into the slit or fiber?   In principle this can be answered given a complete catalog of sources.
For a star centered in a slit with a width of $2a$, the relative contamination from a star a separation $s$ away will depend
upon the seeing.  We can characterize the latter by a Gaussian with a $\sigma$ of $0.85f/2,$ where the 
seeing full-width-at-half-maximum is $f$.
Following equation (8) in Filippenko (1982), the relative contribution of a nearby star is
$$10^{(\Delta V/-2.5)} F (a,s,\sigma)/ F(a,0,\sigma)$$ where the definition of $F$ depends upon whether or not the second star is located
partially in the slit or not.  Let a1=a-s and a2=a+s if the star is in the slit ($s<a$).  Then
$F=0.5(G(a1,\sigma)  + G(a2, \sigma))$.  (Use this for the denominator as well, with a1=a2=a.)  If the star is located outside
of the slit ($s>a$) then $F=0.5(G(a1,\sigma) - G(a2, \sigma))$.  $G(z,\sigma)$ is the standard Gaussian integral,
$$G(z,\sigma) = \frac{1}{\sqrt{2\pi\sigma}} \int_{-z}^z e^{-x^2/2\sigma^2} dx$$
The simplifying assumption in all of this is that the slit has been oriented perpendicular to a line between the two stars.
But, this provides a mechanism in general for deciding in advance what stars in one's program may be too crowded to
observe.

\subsubsection{Assigning Fibers and Designing Multi-slit Masks}
\label{Sec-assign}

In order to design either a fiber configuration or a multi-slit mask, one invariably runs highly customized software which
takes the celestial coordinates (right ascension and declination) of the objects of interest and computes optimal centers,
rotation, etc. that allow the fibers to be assigned to the maximum number of objects, or the most slitlet masks to be machined
without the spectral overlapping.   However, a key point to remind the reader is that such instruments work only if there are
alignment stars that are on the same coordinate system as the program objects.  In other words, if one has produced coordinates
by using catalog ``X" to provide the reference frame, it would be good if the alignment stars were also drawn from the same
catalog.   This was much harder ten years ago than
today, thanks to the large number of stars in uniform astrometric catalogs such as the 2MASS survey or the various USNO
publications.  The most recent of the latter is the
CCD Astrograph Catalogue Part 3 (UCAC3).  One advantage of the proper motion catalogs such as the 
UCAC3 is one can then assure that the relative proper motion between the alignment stars and the program objects
are small.  

This point bears repeating: there is a danger to mixing and matching coordinates determined from one catalog
with coordinates from another.  The coordinates need to be on the same system, or there is significant risk of being quite
disappointed in the final throughput at the telescope.

\subsubsection{Placing Two Stars on a Long Slit}
\label{Sec-twofer}

If one is observing multiple objects whose separations are smaller than the slit length (such as stars within a nearby galaxy or
within a star cluster), and one does not need to be at the parallactic angle (either because the instrument has an ADC or because one is observing either near the zenith or over a small wavelength range) one may want to multiplex by 
placing two stars on the slit by rotating the
slit to the appropriate angle.  To plan such an
observation, one must first precess the coordinates of both stars to the current equinox, as precession itself introduces a rotation.
Then, one must compute the ``standard coordinates", i.e., de-project the spherical coordinates to the tangent plane.  

Assume
that one of the two stars is going to be centered in the slit, and that its precessed coordinates are $\alpha_1$ and $\delta_1$,
converted to radians. Assume that the other star's coordinates are $\alpha_2$ and $\delta_2$, where again these are expressed
in radians after precession.  Then the standard coordinates $\xi$ and $\eta$ of star 2 will be

$$\xi=\cos \delta_2  \sin (\alpha_2-\alpha_1) / F$$

$$\eta=(\sin \delta_2  \cos \delta_1 - \cos \delta_2   \sin \delta_1 \cos (\alpha_2-\alpha_1))/F,$$

\noindent
where 
$$F=\sin \delta_2  \sin \delta_1 + \cos \delta_2  \cos \delta_1 \cos (\alpha_2 - \alpha_1). $$

\noindent
The position angle from star 1 to star 2 will then simply be the arctangent of ($\xi/\eta$).  If $\eta$ is 0, then the
position angle should be 90$^\circ$ or 270$^\circ$ depending upon whether $\xi$ is positive or negative, respectively;  if $\xi$ is 0, then the position angle
should be 0$^\circ$ or 180$^\circ$, 
depending on whether $\eta$ is positive or negative, respectively.  The distance between the two objects will be $\sqrt{\xi^2+\eta^2}$.

\subsubsection{Optimal Extraction}
\label{Sec-optimal}

With CCDs coming into common use as detectors, Horne (1985) pointed out that simply summing the data over the spatial profile of a point
source did an injustice to the data, in that it degraded the signal-to-noise ratio.  Consider the case of a faint source with lots of sky background.
A large extraction aperture in which the data were summed in an unweighted manner would be far noisier than one in which
the extraction aperture was small and excluded more sky: the sky adds only noise, but no new information.  The mathematically
correct way to extract the data is to construct the sum by weighting each point by the inverse of the variance. (Recall that the variance is the
square of the expected standard deviation.)  

The assumption in the Horne (1985) algorithm is that the spatial profile P($\lambda$) varies slowly and smoothly with wavelength $\lambda$.
At each point along the spatial direction $i$ at a given wavelength $\lambda$, $C_i(\lambda)$ photons are measured.  This value is the sum of  the number of photons from the star $A_i(\lambda)$ and  of the sky background $B(\lambda)$, where the latter is assumed to be a constant
at a given $\lambda$.    In the absence of optimal-weighting, one would determine the total sum $T(\lambda)$ of the sky-subtracted
object  by:
$$T(\lambda)=\sum_i A_i(\lambda)=\sum_i (C_i(\lambda)-B(\lambda))$$
\noindent
as $A_i(\lambda)=C_i(\lambda)-B(\lambda)$.  In practice the summation would be performed over some ``sensible" range to include
most of the spatial profile.  For optimal weighting one would instead weight each point in the sum by $W_i (\lambda)$:
$$T(\lambda)=\sum_i W_i (\lambda) (C_i(\lambda)-B(\lambda))/ \sum_i W_i(\lambda)$$   

The weighting function $W_i(\lambda)$ is taken to be $P^2(\lambda)/\sigma_i^2(\lambda)$, where $P(\lambda)$ is the intrinsic
spatial profile of the spectrum at wavelength $\lambda$. $P$ is usually forced to vary smoothly and slowly with wavelength by
using a series of low order functions to represent the profile as a function of wavelength, and is normalized in such a way that the integral of $P(\lambda)$ at a particular $\lambda$ is always unity.  (For the more complicated situation that corresponds to echelle or other highly distorted spectra,
see Marsh 1989.)
The variance $\sigma_i^2(\lambda)$ is readily determined if one assumes that the statistical uncertainty of the data is described
by simple Poisson statistics plus read-noise.  The errors add in quadrature, and will include the read-noise $R$. The variance will also include
the photon-noise due to the object $\sqrt{A_i(\lambda)}$ and the photon-noise due to the sky $\sqrt{B (\lambda)}$.  Finally, the variance will also include a term that represents 
the uncertainty in the background determination.  Typically one determines the sky value by averaging over 
a number $N$ of pixels located far from the star, and hence the uncertainty in the sky determination is $\sqrt{B(\lambda)/(N-1)}$.
Thus $$\sigma_i^2(\lambda)=R^2+A_i (\lambda)+ B (\lambda)  + B (\lambda)/(N-1).$$ 
One can eliminate cosmic-rays by substituting  $T(\lambda) P_i (\lambda)$ for $A_i(\lambda)$:
$$\sigma_i^2(\lambda)=R^2+T(\lambda) P_i(\lambda) + B (\lambda) + B(\lambda)/(N-1),$$ and continuing to solve for $T(\lambda)$
iteratively. 

\begin{figure}[htp]
\epsscale{1.0}
\plotone{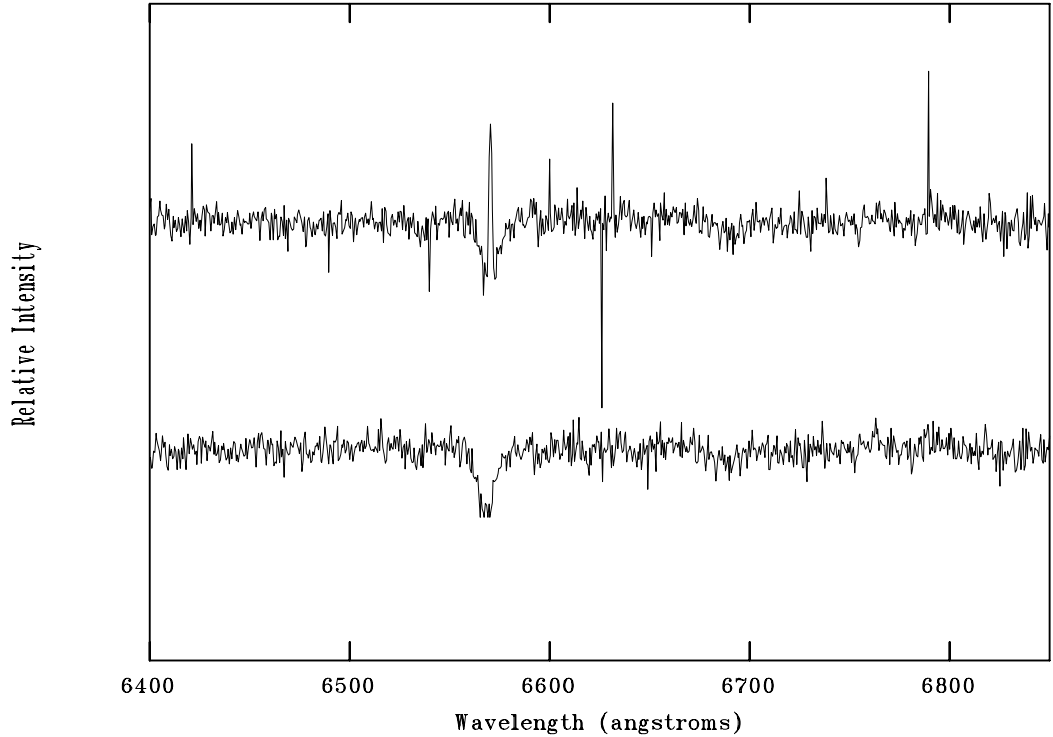}
\caption{\label{fig:optimal} Two reductions of a STIS spectrum obtained of an early O-type star in the LMC, LH101:W3-24.  The 
upper spectrum is the standard reduction produced by the {\it HST} STIS ``CALSTIS" pipeline, while the lower spectrum has been 
reduced using the  CCD spectral reduction package {\it doslit} 
in IRAF using optimal extraction.  The signal-to-noise ratio in the upper spectrum is
18 in a spike-free region, while that of the lower spectrum is 22.  The many cosmic-ray spikes in the upper further degrade the signal-to-noise ratio.
Note in particular the difference in the profile of the strong H$\alpha$ absorption line at 6562\AA.
From Massey et al.\ (2004).  Reproduced by permission of the AAS.}
\end{figure}

An example of the improvement obtained by optimal extraction and cleaning is shown in Fig.~\ref{fig:optimal}, where Massey et al.\ (2004)
compare a standard pipeline reduction version of an {\it HST} STIS optical spectrum (upper spectrum)
with one re-reduced using IRAF with the optimal
extraction algorithm (lower spectrum). 

\subsubsection{Long-Slit Flat-Fielding Issues}
\label{Sec-flats}

With conventional long-slit spectroscopy, generally two kinds of flats are needed: (1) a series of exposures of a featureless continuum
source illuminating the slit (such as a dome flat) and (2) something that removes any residual spatial mis-match between the flat-field
and the night-sky.  Typically  exposures of bright twilight are used for (2), although dithered exposures of the night sky itself would be
better, albeit costly in the sense of telescope time.

\paragraph{Featureless Flats.} 
\label{Sec-flat1}
The featureless flats are primarily useful to remove pixel-to-pixel gain variations within the CCD: some pixels are a little more or a little
less sensitive to light than their neighbors.  Good flat-fielding is vital in the high signal-to-noise ratio regime.  But, with modern chips it is a little
less needed than many realize at modest and low signal-to-noise ratios.  And poor flat-fielding is worse than no flat-fielding.  How one evaluates what is needed is briefly discussed here.

First,  one needs to consider the heretical question of what would be the effect of
 {\it not} flat-fielding one's data?
The answer to this depends upon the signal-to-noise ratio
regime one is attempting to achieve.  If CCDs were perfectly uniform and required
no flat-fielding, one would expect that the scatter $\sigma$ (root-mean-square variation, or rms)
in counts would be related to the number of counts $n$ as
$$\sigma=\sqrt{ng}/g$$ where 
$g$ is the gain in e$^-$ per count.  (For simplicity it is assumed here
that 
read-noise $R$ is inconsequential, i.e., $ng>>R^2$, and that the background counts are not
significant.)   
Consider chip``3" in the 8-chip
mosaic used for the f/4 IMACS camera.  The gain is 0.90 e/ADU.   
 According to the formula above,
for a flat field with 6055 counts, 
one would expect an rms of only 82 counts, but in reality the rms is 95 counts.
  The extra scatter is due to the
 ``graininess" (non-uniformity) of a CCD.  Let p be the average ratio of the gain of one pixel compared to another.  Then the noise
(in e$^-$) added to the signal will just be $npg$ and the total rms $\sigma_T$ (what one measures, in counts) is
\begin{equation}
\label{Equ-p}
\sigma_T^2=n/g + n^2p^2.
\end{equation}
Solving for $p$ in the above example, 
one finds that $p=0.008$,
i.e., about 0.8\%.  
The signal-to-noise $S$ will be $$S=ng/\sqrt{ng+n^2p^2g^2}.$$
Thus if one had a stellar spectrum exposed
to the same level, one would achieve a signal-to-noise of 63 rather than 73.  
An interesting implication of this is that if one is after a fairly small
signal-to-noise ratio, 10 per pixel, say (possibly 30 when one integrates over the spectral resolution elements
and spatially) then if one assumes perfect flat fielding $ng=100$.  Adding in the additional
noise term with p=0.008 leads to the fact that the signal-to-noise one would achieve
would be 9.97 rather than 10.  In other words, if one is in the low signal-to-noise regime
flat-fielding is hardly necessary with modern detectors.  

It should be further emphasized that were one to use 
bad flat-fields (counts comparable to those of the object) one actually can significantly
{\it reduce} the signal-to-noise of the final spectra.  
How many counts does one need for a ``good" flat field?  The rule of thumb is that one does
not want the data to be limited by the signal in the flat field.  Again let $n$ be the number of counts
per pixel in the spectrum.  Let $m$ be the number of counts per pixel in the flat-field.   Consider this
purely from a propagation of errors point of view.  If $\sigma_f$ is the error in the flattened data, and
$\sigma_n$ and $\sigma_m$ are the rms-scatter due to photon-statistics in the spectrum and flat-field,
respectively, then $$\frac{\sigma_f^2}{(ng)^2} = \frac{\sigma_n^2}{(ng)^2}+\frac{\sigma_m^2}{(mg)^2}.$$
(The assumption here is that the flat has been normalized to unity.)  But each of those quantities is really just
the inverse of the signal-to-noise squared of the respective quantities; i.e., the inverse of the
final signal-to-noise squared will be the sum of the inverses of the signal-to-noise squared of the raw spectrum and
the inverse of the signal-to-noise squared of the flat-field.  To put another way:
$$1/S^2 = 1/ng  + 1/mg,$$ where $S$ is the final signal-to-noise.  The quantity $1/ng$ needs to be much greater than $1/mg$.
If one has just the same number of counts in one's flat-field as one has in one's spectrum, one has
degraded the signal-to-noise by $1/\sqrt{2}$.

Consider that one would like to achieve a signal-to-noise ratio per spectral resolution element of 500, say.
That basically requires a signal-to-noise ratio per pixel of 200 if there are seven pixels in the product of the
spatial profile with the number of  pixels in a spatial resolution element.  To achieve this 
one expects to need something like $200^2$=40,000 e$^-$ in the stellar spectrum.  
If one were not to flat-field {\it that}, one would
find that the signal-to-noise ratio was limited to about 100 per pixel, not 200. If one were to obtain only 40,000
counts in the  flat field, one would achieve a signal-to-noise ratio of 140.  To achieve a signal-to-noise ratio of 190 (pretty close to what was
wanted) one would need about 400,000 e$^-$ (accumulated) in all of the flat-field exposures---in other words, about
10 times what was in the  program exposure.    In general, obtaining 10 times more counts in the flat than
needed for the program objects will mean that the signal-to-noise ratio will only be degraded by 
5\% over the ``perfect" 
flat-fielding case.  (This admittedly applies only to the peak counts in the stellar profile,
so is perhaps conservative by a factor of two or so.)

What if one cannot obtain enough counts in the flat field to achieve this?  For instance, if one is trying to
achieve a high signal-to-noise ratio in the far blue (4000\AA, say) it may be very hard to obtain enough counts with standard
calibration lamps and dome spots.  One solution is to simply ``dither" along the slit: move the star to several
different slit positions, reducing the effect of pixel-to-pixel variations.

A remaining issue is that there is always some question about exactly how to normalize
the featureless dome flat.  If the illumination from a dome spot (say) had the same color
as that of the objects or (in the case of background-limited work) the night sky, then
simply normalizing the flat by a single number would probably work best, as it would
remove much of the grating and detector response.  But, in most instances the featureless
flat is much redder than the actual object (or the sky), and one does better by fitting the
flat with a function in the wavelength direction, and normalizing the flat by that function.
In general, the only way to tell which will give a flatter response in the end is to 
try it both ways on a source expected to have a smooth spectrum, such as a spectrophotometric
standard star.  This is demonstrated above  in \S~\ref{Sec-Basicred}, Step~\ref{step-flat}.

\paragraph{Illumination Correction Flats.}
\label{Sec-flat2}

The second kind of flat-fielding that is useful is one that improves the slit illumination
function.  As described above,  one usually uses a featureless flat from observations of the
dome spot to
remove the pixel to pixel variations. To a first approximation, this also usually corrects for
the slit illumination function, i.e., vignetting along the slit, and correction for minor irregularities
in the slit jaws.  Still, for accurate sky subtraction, or in the cases where one has observed an
extended source and wants accurate surface brightness estimates, one may wish to correct
the illumination function to a better degree than that.

A popular, albeit it not always successful, way to do this is to use exposures of the bright
twilight sky.  Shortly after the sun has set, one takes exposures of the sky offsetting the
telescope slightly between exposures to guard against any bright stars that might fall
on the slit.  The spectrum appears to be that of a G2~V.  How then can this be used to 
flatten the data?  In this case one is  interested only in correcting the data
along the spatial axis.  So,  the data can be collapsed along the wavelength axis and 
examined to see if there is any residual non-uniformity   in the spatial direction after the
featureless flat has been applied.  If so, one might want to fit the spatial profile with a low
order function and divide that function into the data to correct it. 

The twilight sky exposures come for free, as the sky is too bright
to do anything else.  But, in some critical applications, the twilights may not match the illumination of the
dark night sky well enough.  The only solution then is to use some telescope time to observe
 relatively blank night sky, dithering between exposures
in order to remove any faint resolved stars.  Since these illumination-correction data are going to be collapsed
in the wavelength direction, one does not need 
very many counts per pixel in order to achieve high signal-to-noise ratio in the spatial profile, but excellent bias correction
is needed in this case.
 
\paragraph{Summary.} When using a CCD for spectroscopy for the first time, it is worthwhile to understand how
``grainy" the pixel-to-pixel variations are in the first place in order to better understand one's flat-fielding needs.
One can do this by obtaining a flat-field exposure and comparing the rms of a fairly uniform section with that
expected from photon statistics.  If $n$ are the number of counts, $g$ is the gain, and $\sigma_T$ is the total
rms (also in counts), then the intrinsic pixel-to-pixel variation $p$ can be found by solving Equation~\ref{Equ-p}:
$$p=\sqrt{\sigma_T^2/n^2-1/ng}.$$  It is only in the case that the precision needed is comparable to this quantity $p$
that one really has to worry about flat-fielding.  In other words, if $p=0.01$ (1\%) as is typical,
and one wants a signal-to-noise ratio
of 100 (1\%), flat-fielding is needed.  If one wants a signal-to-noise ratio of 20 (5\%), flat-fielding is not going to improve
the data.  In addition, one needs to obtain a final flat field that has enough counts in order not to degrade one's data.
In general one wants the flat field to have about ten times the number of counts (per pixel) than 
the most heavily exposed pixel in the
object spectrum in order not to damage the signal-to-noise significantly. 
A good rule of thumb is to take the desired signal-to-noise per spectral resolution element, square it, and aim for that
number of counts (per pixel) in the flat field.  This assumes that the number of pixels in the spatial profile times the number of
pixels in the resolution element is of order 10. 

Slit illumination corrections can be obtained (poorly but cheaply) from bright twilight flats, or (well but expensive) by
observing blank sky at night.  However, one needs only enough counts integrated over the entire wavelength range
to achieve 0.5\% precision (40,000 e$^-$ per spatial pixel integrated over all wavelengths); i.e., if there are 2000 pixels in the wavelength direction one doesn't
need more than about 20e$^-$ per pixel.  At these low count levels though accurate bias removal is essential.

\subsubsection{Radial velocities and Velocity Dispersions}
\label{Sec-RVSTDS}

Often the goal of the observations is to obtain radial velocities of the observed object.
For this, one needs to obtained sufficient calibration to make sure that any flexure in the
instrument is removed.  Even bench-mounted instruments may ``flex" as the liquid N$_2$ in
the CCD dewar evaporates during the night.  The safe approach is to make sure that the
wavelength calibration spectra are observed both before and after a series of integrations,
with the wavelength scale then interpolated.

To obtain radial velocities themselves, 
the usual technique is to observe several radial velocity standard stars of spectral type
similar to the object for which one wants the velocity, and then cross-correlate the spectrum
of each standard with the spectrum of the object itself, averaging the result. 

Spectra are best prepared for this by normalizing and then subtracting 1 so that the 
continuum is zero.  This way the continuum provides zero ``signal" and the lines provide
the greatest contrast in the cross-correlation.  IRAF routines such as {\it fxcor} and {\it rvsao}
will do this, as well as (in principle) preparing the spectra by normalizing and subtracting the
continuum.

One way of thinking of cross-correlation
is to imagine the spectrum of the standard and the program object each contain a single
spectral line.  One then starts with an arbitrary offset in wavelength and sums the two spectra.  If the 
lines don't match up, then the sum is going to be zero.  One then
shifts the velocity of one star slightly relative to the other and recomputes the sum.  When the lines 
begin to line up, the cross-correlation will be non-zero, and when they are best aligned the cross-correlation is at a maximum.  In practice such 
cross-correlation is done using Fourier transforms.   The definitive reference to this
technique can be found in Tonry \& Davis (1979). 

The Earth is both rotating and revolving around the sun.  Thus the Doppler shift of an
object has not only the object's motion relative to the sun, but also whatever the radial
component is of those two motions.  This motion is known as the heliocentric correction.
The rotation component is at most $\pm$ 0.5 km s$^{-1},$ while the orbital motion is
at most $\pm$29.8 km s$^{-1}$.  Clearly the heliocentric correction will depend both on
latitude, date, time of day, and the coordinates of the object.  When one cross-correlates
an observation of a radial velocity standard star against the spectrum of a program object,
one has to subtract the standard star's heliocentric correction from its cataloged value,
and then add the program object's heliocentric correction to the obtained relative velocity.

It should be noted that cross-correlation is not always the most accurate method for measuring
radial velocities.  Very early-type stars (such as O-type stars) have so few lines that it is often
better to simply measure the centers of individual spectral lines.  Broad-lined objects, such as
Wolf-Rayet stars, also require some thought and care as to how to best obtain radial velocities.

The velocity dispersion of a galaxy is measured using a similar method
in order to obtain the width of the cross-correlation function.  The intrinsic line widths
of the radial velocity standards must be removed in quadrature.  Thus sharp-lined radial
velocity standards work better than those with wide lines.

\paragraph {\bf Precision Radial Velocities.}  The desire to detect exoplanets has changed the meaning
of ``precision" radial velocities from the 1-2 km per second regime to the 1-100 {\it meters} per second regime.  Several
methods have been developed to achieve this.  The traditional method of wavelength calibration does not achieve
the needed precision as the comparison arcs and the starlight are not collimated in exactly the same
method.  Early work by Campbell \& Walker (1979) achieved a precision of 10 m s$^{-1}$ using a hydrogen fluoride cell inserted
into the beam that provided an evenly spaced absorption spectrum.  More modern techniques achieve 1 m s$^{-1}$ precision.
One example is the High-Accuracy Radial Velocity Planetary Searcher (HARPS) deployed on on ESO's 3.6-m telescope
on La Silla.  It uses a 
conventional ThAr discharge lamp for wavelength reference.  The ThAr source and the star's light each enter its own fiber,
which then feeds a bench-mounted, extremely stable spectrograph.  Both the ThAr reference source and the star are
observed simultaneously. (See Pepe et al.\ 2004.)

The recent emphasis on planet detection among late-type (cool) stars has driven some of this work into the NIR,
where ammonia gas cells provide a stable reference (Valdivielso et al.\ 2010, Bean et al.\ 2010).

\paragraph  {\bf Laboratory Wavelengths.}  \label{linelists}
The traditional source of wavelengths for astronomers has been 
Moore (1972), a reprint of her original 1945 table.   More up-to-date line lists can be found on the Web, allowing one
to search the National Institute of Standards and Technology Atomic Spectra Database.  The
official site is http://www.nist.gov/physlab/data/asd.cfm, but a very useful search interface
can be found at
http://www.pa.uky.edu/$\sim$peter/atomic/.  One must make sure that one is using the ``air" wavelengths rather than
``vacuum" wavelengths when observing from the ground.

One problem is that many lines in stellar (and reference sources!)
are actually blends of lines.  So, tables of ``effective wavelengths" can be found for stellar lines for stars of different spectral
types. These are generally scattered throughout the literature; see, for example, Conti et al.\ (1977) for a line list for O-type stars.  

A good general line list (identifying what lines may be found in what type of stars) is the revised version of
the ``Identification List of Lines in Stellar Spectra (ILLSS) Catalog) of Coluzzi (1993), which
can be obtained from http://cdsarc.u-strasbg.fr/cgi-bin/Cat?VI/71.  Also useful is the Meinel et al.\ (1968) {\it Catalog of Emission Lines in Astrophysical Objects} and the Tokunaga (2000) list
of spectral features in the NIR region.  Lists of wavelengths of the night sky OH lines can be found in both  Osterbrock et al.\ (1996, 1997) and Oliva \& Origlia (1992).

\subsubsection{Some Useful Spectral Atlases}

The spectroscopist is often confronted by the ``What Is It?" question.  Of course the answer may be a quasar or other extragalactic object, but if the answer is some sort of star, the following resources may be useful in identifying what kind.

\begin{itemize}
\item Jacoby, Hunter, \& Christian (1984) {\it A Library of Stellar Spectra} provides moderate
resolution (4.5\AA) de-reddened spectrophotometry from 3500-7400\AA\ for normal stars of various spectral types and luminosity classes.  Intended primarily for population synthesis,  one deficiency of this work is the lack of identification of any
spectral features.  The digital version of these spectra may be found through VizieR.

\item Turnshek et al.\ (1985) {\it An Atlas of Digital Spectra of Cool Stars} provides spectra
of mid-to-late type stars (G-M, S and C) along with line identifications.  

\item Walborn \& Fitzpatrick (1990) {\it Contemporary Optical Spectral Classification of
the OB Stars---A Digital Atlas} provides moderate resolution normalized spectra from 3800-5000\AA\ of early-type stars, along with line identification.  The spectral atlas only
extends to the early B-type stars for dwarfs, and to B8 for supergiants.  

\item Hanson et al.\ (2005) {\it A Medium Resolution Near-Infrared Spectral Atlas of O and Early-B Stars} provides moderately high resolution ($R\sim$8000-12,000) spectra of O and early B-type
stars in the H- and K-bands.   The lower resolution spectra shown in Hanson et al.\ (1996)
also remains very useful for spectroscopists working at $R\leq 1000$.

\item Hinkle et al.\  (2003) describe in details various {\it High Resolution Infrared, Visible,
and Ultraviolet Spectral Atlases of the Sun and Arcturus.}

\end{itemize}

\subsection{Observing Techniques: What Happens at Night}

 One of the goals of this chapter has been to provide observing tips, and possibly the best
way of doing this is provide some examples of what  some typical nights are actually like.
Included here are examples of observing with a long-slit
spectrograph,  observing with a fiber spectrograph,  and some advice on what to do when 
observing with a NIR spectrometer.

A common theme that emerges from these (mostly true) stories is that the observers spend a lot
of time thinking through the calibration needs of their programs.  For the optical this is mainly
an issue of getting the flat-fields ``right" (or at least good enough), while there are more subtle issues
involved in NIR spectroscopy.  Throughout these the same philosophy holds: obtaining useful
spectra involves a lot more than just gathering photons at the right wavelength.

\subsubsection{Observing with a long-slit spectrograph}
\label{Sec-Obslong}

The GoldCam spectrometer on the Kitt Peak 2.1-meter provides an interesting example of a
classical long-slit instrument.  The observing program described here was aimed at obtaining
good ($<$5\%) spectrophotometry of a sample of roughly 50 northern Galactic red supergiant stars 
whose $V$-band magnitude ranged from 6 to 11.  The goal was to match both the spectral
features and continuum shapes to a set of stellar atmosphere models in order to determine basic 
physical properties of the stars, especially the effective temperatures.   Broad wavelength coverage
was needed (4000-9000\AA), but only modest spectral resolution (4-6\AA) as the spectra of these
late-type stars are dominated by broad molecular bands.  An observing program
was carried out for southern stars using a similar spectrograph on the Cerro Tololo 1.5-m telescope.

At the stage of writing the observing proposal, the choice of the telescopes was pretty well set by the fact that these
stars were bright and the exposure times on even 2-meter class telescopes would be short.  The number of nights
needed would be dominated by the observing overhead rather than integration time.  
The gratings and blocking filters that would meet the requirements were
identified, based upon the needed spectral resolution and wavelength coverage (\S~\ref{Sec:Basics}).  It was not possible to obtain the needed spectral resolution with a single grating setting over the entire
wavelength range, and so the northern (2.1-m) sample would have to be observed twice, once with a 600 line/mm grating
that would provide wavelength coverage from 3200\AA\ to 6000\AA, and a 400 line/mm grating that would provide
coverage from 5000-9000\AA.  Both gratings would be used in first order.  The blue grating would thus require no blocking filter,
as inspection of Figure~\ref{fig:orders} reveals: at 6000\AA\ the only overlap would be with 3000\AA\ from the second
order.  Given the poor atmospheric transparency at that wavelength, and the extreme red color of such stars, it is safe to assume that
there would be negligible contamination.  For the red setting (5000-9000\AA) there would be overlap with second order light $<$4500\AA, and so a GG495 blocking filter was chosen.  As shown by 
Figure~\ref{fig:filters}, this filter would begin to have good transmission for wavelengths greater than 5000\AA,
 and no transmission below 4750\AA, a good match.
 
 The 2.1-m GoldCam does not have any sort of ADC, and so the major challenge would be to observe in such a way that good
 spectrophotometry was maintained: i.e. that differential refraction (\S~\ref{Sec-parallactic}) 
 would not reduce the accuracy of the 
 fluxed spectra. 
 The GoldCam spectrograph is mounted on a rotator with a mechanical encoder, and would have to be turned by hand (at zenith) to the
 anticipated parallactic angle for each observation.  How much ``slop" in the parallactic angle could be tolerated would be a function of time
 during the night for any given star, and in advance of the observing these values
 were computed. It was clear that the slit would have to be kept open as 
  wide as possible without
 sacrificing the needed spectral resolution, and this was one of the factors that entered into the grating choice.  With the higher dispersion
 blue grating one could have a slit width of 3 arcsec (250$\mu$m) and still have 4\AA\ spectral resolution, while with the lower dispersion red grating
 one needed a skinnier 2 arcsec slit (170$\mu$m) in order to maintain adequate resolution (6\AA), but this was made possible by the fact that
 differential refraction was less significant in the red.  Without doing the math (using the equations given in \S~\ref{Sec-parallactic} and adapting the equations in \S~\ref{Sec-isolate}), all of this would have been a guess.

The first afternoon of the run was spent performing the basic setup for both gratings.  First the grating was inserted, the blocking filter (if any)
was put in the beam.  The grating tilt was set to some nominal value based on the instrument manual and the desired central wavelength
setting.  The HeNeAr comparison source was turned on, and an exposure made to see if the grating tilt was a little too red or
blue.  This required some finagling, as bad columns dominate the first 300 columns of the 3000 pixel detector,
and the focus gets soft in the final 500 columns.  Thus 2200 columns are useful, but the center of these is around column 1600 and not 1500.
Several exposures of the comparison arc were needed, with small tweaks (0.02-0.1 degrees) of the grating tilt used to get things just so.

Next, the spectrograph was focused.  The HeNeAr comparison source was left on, and the slit set to a skinny 100$\mu$m.
Typically the collimator should be at a certain distance
(the ``auto-collimate" position) from the slit so that it is just filled by the diverging beam from the slit (Figure~\ref{fig:spect}).  But this luxury is
achieved only in spectrographs in which the camera lens can be moved relative to the detector to obtain a good focus of the spectrum.
In many instances---and in fact, one might have to say {\it most} instances---that is not the case, and the camera focus is fixed or at least
difficult to adjust.  Instead, the collimator is moved slightly in order to achieve a good focus.  Each of the two gratings would require a different
collimator setting as one had a blocking filter and the other did not; inserting such a filter after the slit changes the optical
path length slightly.

The observers decided to start with the red grating, since this was their first night and the
red observations should be a little less demanding.  It was not practical to change gratings during the night
and so there would be specific nights devoted to either the red or blue observations.  The grating tilt and focus were adjusted
to the values found earlier.  The slit was opened to the needed amount, and the optical path was carefully checked. Was the
appropriate blocking filter in place?  Was the collimator set to the right focus?  

The detector covers 512 (spatial) rows but each star would cover only a few of them.  Good sky subtraction was important particularly in the blue
(the observations would be made with considerable moonlight) but even so, this  was unnecessarily excessive.
The chip was reformatted to read out only the central 250 rows.
The observers could have spatially binned the data by 3 (say) and reduced the read-out time, 
but to do so would slightly compromise the ability to reject cosmic rays by having a nicely sampled spatial
profile (\S~\ref{Sec-optimal}).  Besides, it was clear that the observing overhead would be dominated by the need to move the telescope to
zenith and manually rotate to the parallactic angle, not the read-out time of the chip.  The telescope could 
 be slewed during the readouts.  (Telescopes are usually tracking during read-out.)

With the CCD and spectrograph set, the observers next proceeded to take some calibration data. 
Since there was still plenty of time before dinner, 
they decided to make a bad pixel mask.   The calibration source housed two lamps: the HeNeAr source and a quartz lamp.  The
quartz lamp could provide a ``featureless flat" but its illumination of the slit was sufficiently different that it provides a very poor
match to the night sky compared to the dome flat. It does, however, have the advantage that it can be run in place.
For just hitting the detector with enough light to identify bad pixels it would be plenty
good enough.  A five second exposure had 30,000 counts, plenty of counts, and well below the expected saturation of the detector.
Two more were obtained just to protect against a single exposure having been hit by a strong cosmic ray.  In order to obtain frames
with a scant number of counts, a 5-magnitude (100$\times$ attenuation) ``neutral density" filter was placed in front of the slit.
This resulted in a five
second exposure having about 300 counts.  A series of 50 of these would take about an hour to run, given the read-out time, but 
the average would then have good statistics. During the break, 
the observers attempted to get the music system connected to their iPods, and discussed some unanticipated
flat-fielding issues.

Long experience at the 2.1-m had taught the observers
 that the dome flat exposures do a far better job at matching the illumination of the night
sky in the spatial direction than do the internal quartz exposures.  But, even superficial inspection of the bad pixel mask data revealed
that there were significant, 10-20\% fringes in the red ($>$7000\AA) region.  How to remove these?  If the instrument were absolutely stable (no
flexure) then the fringes should divide out in the flat-fielding process.  The internal lamp offered an additional option: 
it could be used in
place without moving to the dome spot.  Thus for safety the observers decided to take the standard dome flat exposures but also planned
to take some internal quartz lamp exposures during the night at various positions and see how much (if any) the fringes moved.
The blue data would be straightforward, and just require long exposures as the dome spot has poor reflectivity in the far blue.

During the course of the run the observers discovered that the fringes moved significantly during the first half hour after the nightly fill of the
CCD dewar but were quite stable after that.  So, in the end they wound up combining the quartz lamp exposures taken throughout the
night and using that as the featureless flat in the red, and using dome flats as the featureless flat for the blue. 

The mirror cover was next opened and the telescope moved to the dome flat position.  The illumination lamps were turned on, and the
comparison optics (HeNeAr/quartz) were removed from the beam.  A short test exposure was run.  Much surprise and consternation 
was expressed when a nearly blank exposure read out.  What was the problem?   Generations of astronomers have answered this
in the same way: think about the light path from the one end to the other and at each point
consider what could be blocking the light.
The lamps were on.  The telescope was pointed in the right position, as confirmed by visual inspection.  The mirror covers were open.
The comparison optics were out, at least according to the control unit.  Wait! The filter wheel above the slit was still set to the 5-magnitude
neutral density filter.  Setting this back in the clear position solved the problem.  A series of 5 dome flats were quickly obtained, and the
telescope was slewed back to zenith and the mirror cover closed.  The observers went to dinner, leaving a series of 15 biases running.

Shortly before the sun set, the observers filled the CCD dewar, 
opened the telescope dome, and brought the telescope fully on line with tracking turned on.  After watching the sunset, they hurried
back inside, where they took a series of exposures of the twilight sky.  These would be used to correct for the mismatch (a few percent)
between the projector flats and the night sky illumination along the slit, improving sky subtraction.  They slewed to a bright star nearly
overhead, and checked that the pointing was good.  The slit was visible on the TV camera, with the reflective metal to either side showing the
sky.  

Next they moved the telescope back to zenith so they could manually adjust the rotator to the parallactic angle planned
for the first observation, which
would be of a spectrophotometric standard.
The star would be relatively near the zenith, and so knowing exactly when they could get started was not critical, 
as the allowed tolerance on the
rotator angle is very large for good spectrophotometry.  
They moved the platform out of the way, and slewed the telescope to the star.  When the sky was judged
to be sufficiently dark, they carefully centered it in the slit and begin a 5 minute exposure.  Since all of the exposures would be short,
they decided not to bother with the considerable overhead of setting up the guider, but would hand guide for all of the exposures,
using the hand paddle to tweak the star's position on the slit if it seemed to be slightly off-center.  They observed two more
spectrophotometric standards, each time first moving the telescope to zenith, unstowing the platform, rotating to the parallactic angle,
restowing the platform, and slewing to the next target.  There was no need to measure radial velocities (a difficult undertaking with
broad-lined stars) and so a single HeNeAr comparison would be used to reduce all of the data during the night.

 After each exposure read down, the data were examined by running IRAF's {\it splot} plotting routine in a somewhat unconventional manner.
 The dispersion axis runs along rows, and normally one would plan to first extract the spectrum before using {\it splot}.  Instead, the
 astronomers used this as a quick method for measuring the integrated counts across the spatial profile subtracting off the bias and sky level, by specifying {\it splot image[1600,*]}
 to make a cut across the middle of the spectrum.  The ``e" keystroke which is usually used to determine an equivalent width is then run
 on the stellar profile to determine the number of counts integrated under the profile, listed as the ``flux".  This neatly removes any
 bias level from the counts and integrates across the spatial profile. This could be checked at several
 different columns (500, 1600, 2200) to make sure there are good counts everywhere.  After things
 settled down for the night, the
 the observers were in a routine, and used {\it ccdproc} to trim the data and remove the overscan. Flat-fielding would  be left until they had
 thought more about the fringes.  Nevertheless, 
  {\it doslit} could be used to extract the spectrum with a wavelength calibration.  
 
 The observers began observing their red supergiant sample.  The {\it splot} trick proved essential to make sure they were obtaining
 adequate counts on the blue side, given the extreme cool temperatures of these stars.  Every few hours they would take a break from the 
 red supergiants to observe spectrophotometric standards, two or three in a row.  By the end of the first night they had observed
 28 of their program objects, and 11 spectrophometric standards, not bad considering the gymnastics involved in going to the parallactic
 angle.  Some older telescopes have rotators that are accessible remotely (CTIO and KPNO 4-meters) while all alt-az telescopes have
 rotators that can be controlled remotely by necessity.
 
The analogous observations at CTIO were obtained similarly.  Since the detector there was smaller, three gratings were needed to obtain full
wavelength coverage with similar dispersion.  Going to the parallactic angle was even less convenient since the
control room was located downstairs from the telescope.
 Fortunately the slit width at the CTIO telescope could be controlled
remotely, and therefore the observations were all made with two slit settings, a narrow one for good resolution, and a really wide one to
define the continuum shape.

\begin{figure}[htp]
 \epsscale{0.9}
 \plotone{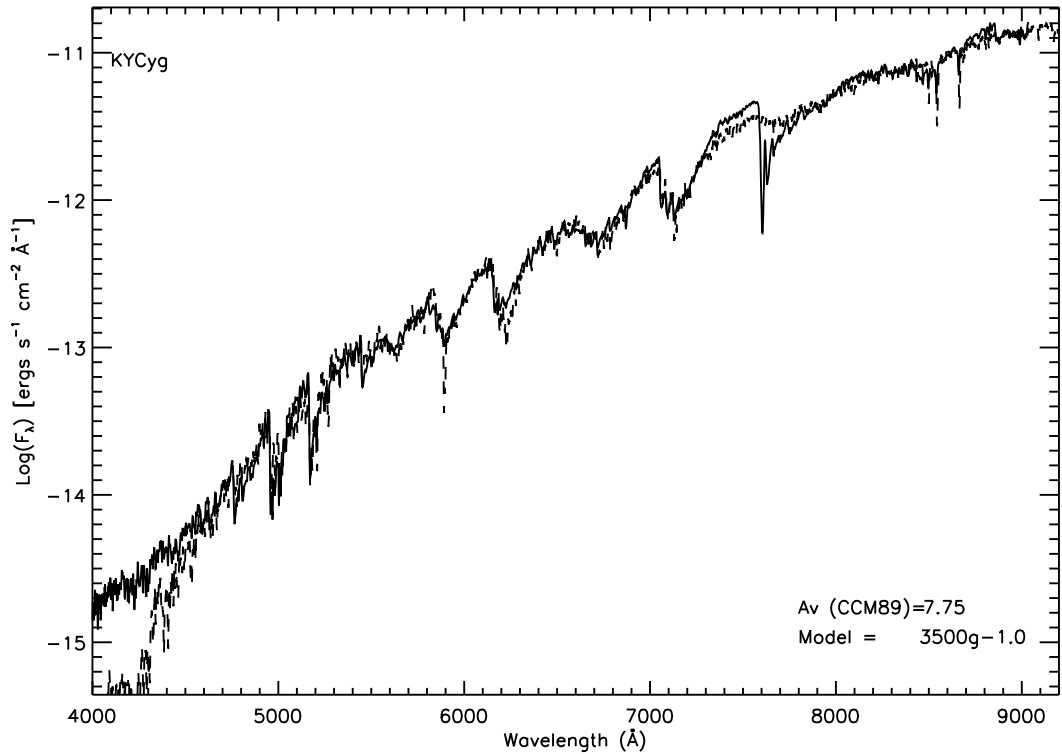}
 \caption{\label{fig:emily} Model fitting KY Cyg. The solid black shows the fluxed spectrum
 obtained as part of the 2.1-m GoldCam observing described here.  The dotted line shows
 the best fit model, with an effective temperature of 3500 K and a log g=-1.0 [cgs].   The
 study by Levesque et al. (2005) showed that this was one of the largest stars known.  The extra
 flux in the star in the far blue is due to scattering by circumstellar dust.}
 \end{figure}

In the end the data were all fluxed and combined after several weeks of work.  A few stars had been observed both from CTIO and
KPNO and their fluxed spectra agreed very well.  The  comparison of these spectra with model atmospheres began.
 A sample spectrum, and model fit, are shown in Figure~\ref{fig:emily}.  The work (Levesque et al.\ 2005) established the first modern
 effective temperature scale for red supergiants, removing the discrepancy between evolutionary theory and the ``observed" locations
 of red supergiants in the H-R diagram, discovered circumstellar reddening due to these stars dust production, and identified the 
 three ``largest stars known," not bad for a few nights of hard labor hand guiding!
 
\subsubsection{Observing with a Multi-Fiber Spectrometer}
\label{Sec-Obsfibers}
The CTIO 4-m Hydra fiber spectrometer was commissioned in late 1998, and was the
third version of this instrument, with earlier versions having
been deployed on the KPNO 4-m and  at the WIYN 3.5-m telescope (Barden \& Ingerson 1998).
It consists of 138 fibers each with 2.0-arcsec (300$\mu$m) diameter, and is located at the Ritchey-Chretien focus covering a 40 arcmin diameter field.  It is used with an ADC, removing the need to worry about differential refraction when placing the fibers.
The observing program described here was carried out in order to identify 
yellow supergiants in the Small Magellanic Cloud based on their radial velocities measured
using the strong Ca II triplet lines ($\lambda \lambda$ 8498, 8542, 8662) in the far red.
In addition, the observations would test the use of the OI $\lambda 7774$ line as a
luminosity indicator at the relatively low metallicity of the SMC. 

The astronomers needed to observe about 700 stars, only a small fraction
of which were expected to be {\it bona fide} supergiants.  The rest would be foreground stars in
the Milky Way.  This list of 700 stars had been selected on the basis of color and magnitude,
and had already been culled from a much larger sample based on having negligible proper motions.  The stars were relatively bright ($V<14$)
 and the major limitation would be the
overhead associated with configuring the Hydra instrument, which requires about 8 seconds
per fiber, or  20 minutes in total.   Because the SMC is large compared even to Hydra's field of
view, several dozen fields would be needed to cover most of the targets.

The situation was further complicated by the desire to not only have spectra in the far red
(including at least 7770-8700\AA) but also in the blue to obtain some of the classical
luminosity indicators used for Galactic yellow supergiants.
To observe each fiber configuration twice, on ``blue nights" and ``red
nights" would require twice as many observing nights, given the large overhead in 
each fiber configuration. 

The fibers feed a bench spectrograph mounted in a dark room a floor below the telescope.
Changing the grating tilt might require refocusing the spectrograph (a manual operation,
impractical at night) but simply changing blocking filters could be done remotely.  If the
filters were of similar thickness, and if the camera's focus was fairly achromatic (which
would be expected of a Schmidt camera) then one could configure the fibers, observe in
the red, and simply by changing blocking filters, observe in the blue.   No one was quite
sure if this would work, as no one could remember the spectrograph having been
used this way, but it would be easy enough to check on the first afternoon of the run.
A 790 line/mm grating blazed at 8500\AA\ in first order was available, and would yield
2.6\AA\ resolution in the red in first order, and 1.3\AA\ resolution in the blue in second order,
providing wavelength coverage of 7300-9050\AA\ in the red and 3650-4525\AA\ in the blue.
Obviously the red observations would require a blocking filter that removed light $<$4525\AA,
while the blue observations needed a red cut-off filter that removed any light $>$7300\AA.
Among the available filters, an OG515 did an excellent job in the first case, and a BG39 did
a good job in the second case while still transmitting well over the region of interest (see Figure~ \ref{fig:filters}).  
(The same argument was presented above in \S~\ref{Sec-RCSpec}.)

Prior to the observing run, thirty fiber configuration fields had been designed in order to obtain as many
of the target stars as possible.  Since bad weather is always a possibility (even at Cerro Tololo) the
fields were designed in a particular order, with field centers chosen to include the maximum number
of stars that had not been previously assigned.  Although the fiber configuration program is flexible in providing
various weighting schemes for targets, it was found necessary to slightly rewrite the code to allow for stars
that had been previously assigned to be added ``for free", i.e., without displacing any not-yet assigned star.
 (It helped that the first author had written the
original version of the code some years back.)   The process took a week or more to refine the code, but
the assignments themselves then were straightforward.

The first afternoon at the telescope, the astronomers arrived to find that everything appeared to be in good
shape.   Instrument support personnel had inserted the grating and blocking filter, checked the grating tilt,
and had focused the spectrograph, substantiating the fact that the focus was unchanged between the red
and the blue setups.   A comparison arc had been used to focus the spectrograph, and examination confirmed
the expectation that at the best focus the spectral resolution covered about seven pixels.  The observers decided thus
to bin by a factor of 2 in the wavelength direction.  Even though radial velocities were desired, there was no advantage
in having that many pixels in a spectral resolution element: 3 would be plenty, and 3.5 generous,
according to the  Nyquist-Shannon criterion (\S~\ref{Sec:Basics}).  No binning was applied
to the spatial direction as clean separation of one fiber from another is desirable.

The fibers could be positioned only with the telescope at zenith.  This is quite typical for fiber instruments.  Fibers
are not allowed to ``cross" or get so close to another fiber to disturb it, and thus in order to have reliable operations the
fibers are configured only at a certain location.   In mid afternoon then the astronomers had the telescope moved to the
zenith and configured the fibers into a circle for observing the dome spot.

With fiber instruments, sky subtraction is never ``local", as it is with a long slit. Some fibers are pointing at objects, while
other fibers have been assigned to clean sky positions.  In order to subtract the sky spectrum from the object spectrum,
flat-fielding must remove the fiber-to-fiber sensitivity, which itself is wavelength dependent. In addition, it must compensate
for the different illumination that a fiber will receive from the sky when it is placed in the middle as opposed to somewhere near
the edge.  In other words, under a perfectly clear sky, the same fiber would have somewhat different counts looking at blank
sky depending upon its location.

In addition, the pixel-to-pixel gain variations need to be removed, just as in long slit observations.  But, the profiles of the fibers output
are quite peaked, and they may shift slightly during the night as the liquid nitrogen in the dewar turns to gas and the weight
changes. 

On Hydra CTIO there are four possible flats one can take: an instrument support person can place a diffuser glass in back of the
fibers, providing a somewhat uniform illumination of the CCD when the fibers are pointed at a bright light source.  This is called
a ``milk flat", and would be suitable for removing the pixel-to-pixel gain variations.  A second flat is the dome flat, with the fibers
configured to some standard (large circle) configuration.  This would also work for pixel-to-pixel gain variations  as long as the
output location of the fibers were stable on the detector.  A third flat involves putting in a calibration screen and illuminating it 
with a lamp.  This can be done in place with the fibers in the same position for the actual observations (unlike the dome flats)
but the illumination by the lamp is very non-uniform, and thus has little advantage over the dome flat in terms of removing the
vignetting. The fourth possibility is to observe blank sky with the same configuration.  Since the SMC F/G supergiants were bright,
and sky subtraction not critical, it was easy to eliminate the fourth possibility.

Exposures of the dome flat quickly revealed that although there was plenty of light for the
red setting, obtaining a proper flat (one that
would not degrade the observations; see \S~\ref{Sec-flats}) in the blue would take on the order of days, not minutes. 

Given this, an arguable decision was reached, namely that the observing would (provisionally) rely upon the calibration screen
flats obtained at each field during the night.  There were several arguments in favor of using the calibration screen flats.
 First, by having the rest of the fibers stowed,
and only the fibers in use deployed, the flat-field would be useful in unambiguously identifying which fibers mapped to which
slit positions on the detector.  Second, and more importantly, it provided a real-time mapping of the trace of each fiber on the
array.  Third, it would cost little in overhead, as the radial velocities already required observing the HeNeAr calibration lamps with
the screen in place, and that most of the overhead in the calibration itself would be moving the calibration screen in and out of the beam.
The definitive argument, however, was that the stars were very bright compared to the sky, and so even if there was no sky subtraction,
the science data would not be much compromised.  Had the objects been comparable to the sky values, the best alternative
would have been to do blank sky exposures, despite the use of extra telescope time.
In any event,  dome flats in the red were run each afternoon as it provided a good chance to exercise the instrument during the afternoon
and ascertain that everything remained copacetic.

Prior to dinner, the observers configured the instrument to their first field.  A problem was immediately revealed: one of the assigned
fibers did not deploy.  Why? Although there are 138 fibers, several fibers have become broken over time or have very low through-put
and they are ``locked" into the park position.  A ``concentricities" file is provided with the software used to assign fibers and test the
configurations, and after a little probing it became clear that the concentricities file used in the assignments had been out of date.  Therefore,
assignments for the entire thirty fields would have to be recomputed.  Fortunately most of the preparation work was simply in getting the
software system set up, and before dinner the observers had managed to get the first few fields recomputed, enough to get them
going, and the remainder were easily recomputed during the night. 
The new configuration files were transferred from the observer's laptop to the instrument computer, and the
first configuration was again configured, this time without incident.  The observers began a series of biases running and left for
dinner. 

Shortly before sunset the instrument assistant opened the dome to allow any heat in the dome to escape.  The first actual target would
be a bright radial velocity standard star (\S~\ref{Sec-RVSTDS}).  Without disturbing the other fibers, the astronomers moved an unused
fiber to the center of the field, and deployed an unused alignment fiber to the location of another bright star near the radial velocity
standard.  As discussed in \S~\ref{Sec-assign} there has to be some way to align (and guide!) fiber instruments.  In the case of
Hydra, these functions are accomplished by means of any of 12 ``field orientation probes" (FOPs).  These are each bundles of 
5 fibers around a central sixth fiber.  These are deployed like regular fibers, but the other ends of these fiber bundles are connected
to a TV rather than feeding the spectrograph.  Thus an image of six dots of light are seen for each FOP.  When the telescope
is in good focus, the centering is good, and the seeing is excellent, all of the light may be concentrated in the central fiber of the
six.   The telescope is guided by trying to maximize the amount of light in each of the central FOP fibers.  In principle, a single FOP should
be sufficient for alignment and guiding, since the only degrees of freedom are motions in right ascension and declination, and not rotation.  
But in practice a
minimum of 3 is recommended.  The assigned SMC fields had 3-5 each, but for the bright radial velocity standard a single FOP was
judged sufficient as the exposure would be a few seconds long at most and no guiding would be needed.

Once the two new fibers were in place, the focal plane plate was ``warped"; i.e., bent into the curved focal surface using a vacuum.
 (The fibers had to be deployed onto the plate when it was flat.)  The telescope was slewed to the position of the radial velocity standard,
 and the ``gripper"---the part of the instrument which moves the fibers, was inserted into the field.  The gripper has a TV camera mounted
 on it in such a way that it can view the reflection of the sky.  Thus by positioning the gripper over a deployed fiber (such as a FOP) one
 can also see superimposed on the image any stars near that position.  In this case, the gripper was placed in the center of the field
 of view, and the bright radial velocity standard carefully centered.  The gripper was then moved to the single ``extra" FOP and the
 presence of a bright star near that position was also confirmed.  As the gripper was removed from the field of view, the light from the
 single FOP was visible.  The telescope was next focused trying to maximize the light in the central fiber of the bundle. 
 
 While this was going on, the observers carefully checked the spectrograph configuration using the spectrograph GUI.
  Was the correct blocking filter in place for the red?  Were the grating tilt and other parameters still set to what they were in the afternoon?
  When the operator announced that the telescope was focused, the observers then took a 5 second exposure.  At the end, the CCD
  read down, and light from a single fiber was obvious.  A cut across the spectrum showed that there were plenty of counts.  The  voltage
  on the TV was then turned down to protect the sensitive photocathode, the calibration screen was moved into place, and both a projector
  flat and HeNeAr comparison arc exposure were made.  The first observation was complete!
  
  Rather than waste time removing the two extra fibers (which would have required going back to the zenith), the telescope was 
  slewed to the SMC field for which the fibers had been configured.  The gripper was moved into the field, and sent to one of the
  deployed FOPs.  A bright star was seen just to the upper left.  The gripper was then moved to a second FOP. Again a bright star was
  seen just to the upper left. These must be the alignment stars.  The operator then moved the telescope to center the reflection of the
  star image on the FOP.  Going to a third FOP confirmed that there was now a bright star superimposed on that FOP.  The gripper
  was moved out of the field, and the images of 5 illuminated FOPs appeared on the guider TV.  The guider was activated, and after a
  short struggle the telescope motion seemed to be stable.  ``Okay," the operator announced.  The astronomers took three exposures
  of five minutes each.  Guiding was stopped, the voltage was turned down on the TV, the calibration screen reinserted into the beam,
  and a short projector flat and HeNeAr exposure were made.  Then the blocking filter was changed from the red (OG515) to the
  blue (BG39), and new projector flats and HeNeAr exposures were made.  The calibration screen was removed.  Examination of the
  FOP guide TV showed that the telescope had drifted only slightly, and guiding was again initiated.  The observers took three ten minute
  exposures for blue spectra of the same stars.  Then the telescope was moved to the zenith, the plate flattened, and the next field was
  configured.  The process was repeated throughout the night, interrupted from time to time to observe new radial velocity standards
  in the red.
  
\begin{figure}[htp]
  \epsscale{0.6}
  \plotone{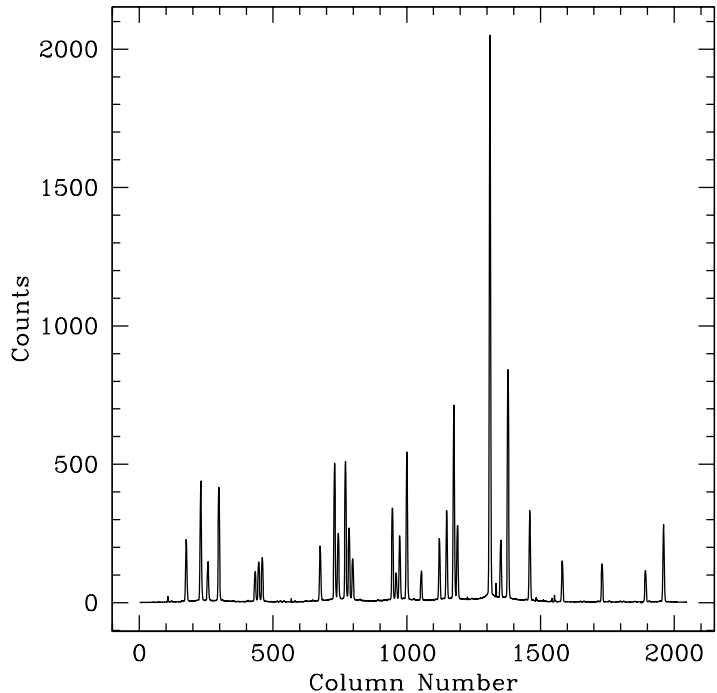}
  \caption{\label{fig:map} Spatial cut across Hydra data.  The spectra of many stars have been obtained in a single exposure,
  but which star is which?}
  \end{figure}

  Throughout the night the observers made cuts through the spectra, but the first efforts to reduce the data to ``final" spectra
  failed as there was an ambiguity in how the slit positions were numbered.  The assignment files assigned fiber 103 to a specific star.
  But, where did fiber 103 map to on the detector?  The concentricities file was supposed to provide the mapping between fiber
  number and slit position, but the image headers also contained a mapping.  These agreed for the first couple of dozen fibers but
  after that there was an offset of one.  After a few dozen more fibers they differed by several.  The problem appeared to be that
  there were gaps in the output slit.  The concentricities file numbered the slit positions consecutively in providing the mapping, while
  the header information was derived apparently assuming there was more or less even spacing.  The problem this introduced was
  not just being sure which object was which, but which spectra were that of sky in order to sky subtract. Figure~\ref{fig:map} shows
  the problem.

Fortunately (?) the following night was cloudy, and the astronomers spent the evening with the telescope pointed at zenith,
  creating a mapping between fiber number, slit position in header, and pixel number on the detector.  In the case of any
  uncertainty, a fiber could be moved from the outer region to the central region and exposed to the calibration screen, and the
  position on the detector measured unambiguously.   The third night was clear, and by then data could be reduced correctly in
  real time using the mapping and the IRAF task {\it dohydra}.
  
\begin{figure}[htp]
  \epsscale{0.65}
  \plotone{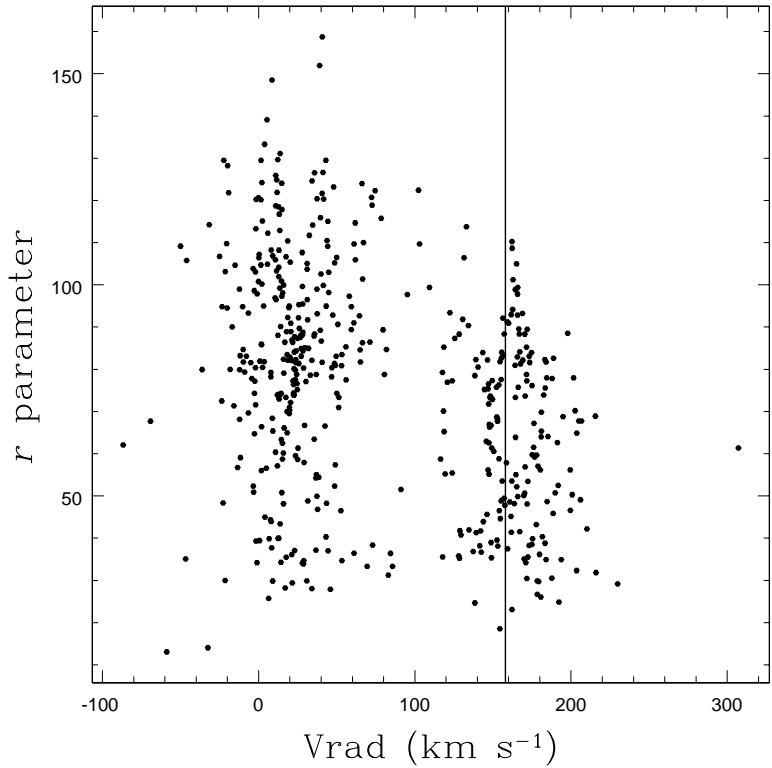}
  \caption{\label{fig:rvs} The radial velocities of SMC F/G supergiant candidates.  The Tonry \& Davis (1979) {\it r} parameter is plotted
  against the radial velocity of approximately 500 stars observed with Hydra at the CTIO 4-m telescope.  The stars with radial velocities
  $<$100 km s$^{-1}$ are Milky Way foreground stars.  The line at 158 km s$^{-1}$ denotes the systematic radial velocity of the SMC. Based upon Neugent et al.\ (2010).}
  \end{figure}

  How well did the project succeed? Radial velocities were obtained for approximately 500 stars.  Figure~\ref{fig:rvs} shows the Tonry \& Davis (1979) {\it r} parameter (a measure of how well the
  cross-correlation worked) versus the radial velocity of each star.  There are clearly two distributions, one centered around a velocity of zero
  (expected for foreground stars) and one centered around 160 km s$^{-1}$, the radial velocity of the SMC.  All together 176 certain
  and 16 possible SMC supergiants were found, and their numbers in the H-R diagram were used to show that the current generation
  of stellar evolutionary models greatly over estimate the duration of this evolutionary phase (Neugent et al.\ 2010).  Since a similar
  finding had been made in the higher metallicity galaxy M31 (using radial velocities from Hectospec; Drout et al.\ 2009) the
  SMC study established that uncertainties in the mass-loss rates on the main-sequence must not be to blame.  The OI $\lambda 7774$
  line proved to be a useful luminosity indicator even at relatively low metallicities.
  
\subsubsection{Observing with a NIR Spectrometer}
\label{Sec-ObsNIR}

With near-infrared observations, virtually all the steps taken in preparing and undertaking a run in the optical are included and will not be repeated here.  What will be done here is a review of the few additional steps required for near-infrared spectroscopy.  These steps center around the need to remove OH sky emission lines (numerous and strong, particularly in the $H$-band) and to correct for the absorption lines from the Earth's atmosphere.  It will not be possible to do a very thorough reduction during the night like one can with the optical, but one can still perform various checks to ensure the data will have sufficient signal-to-noise and check if most of the sky and thermal emission is being removed.

\begin{figure*}[htp] 
\centering
\includegraphics[angle=270,width=16.0cm]{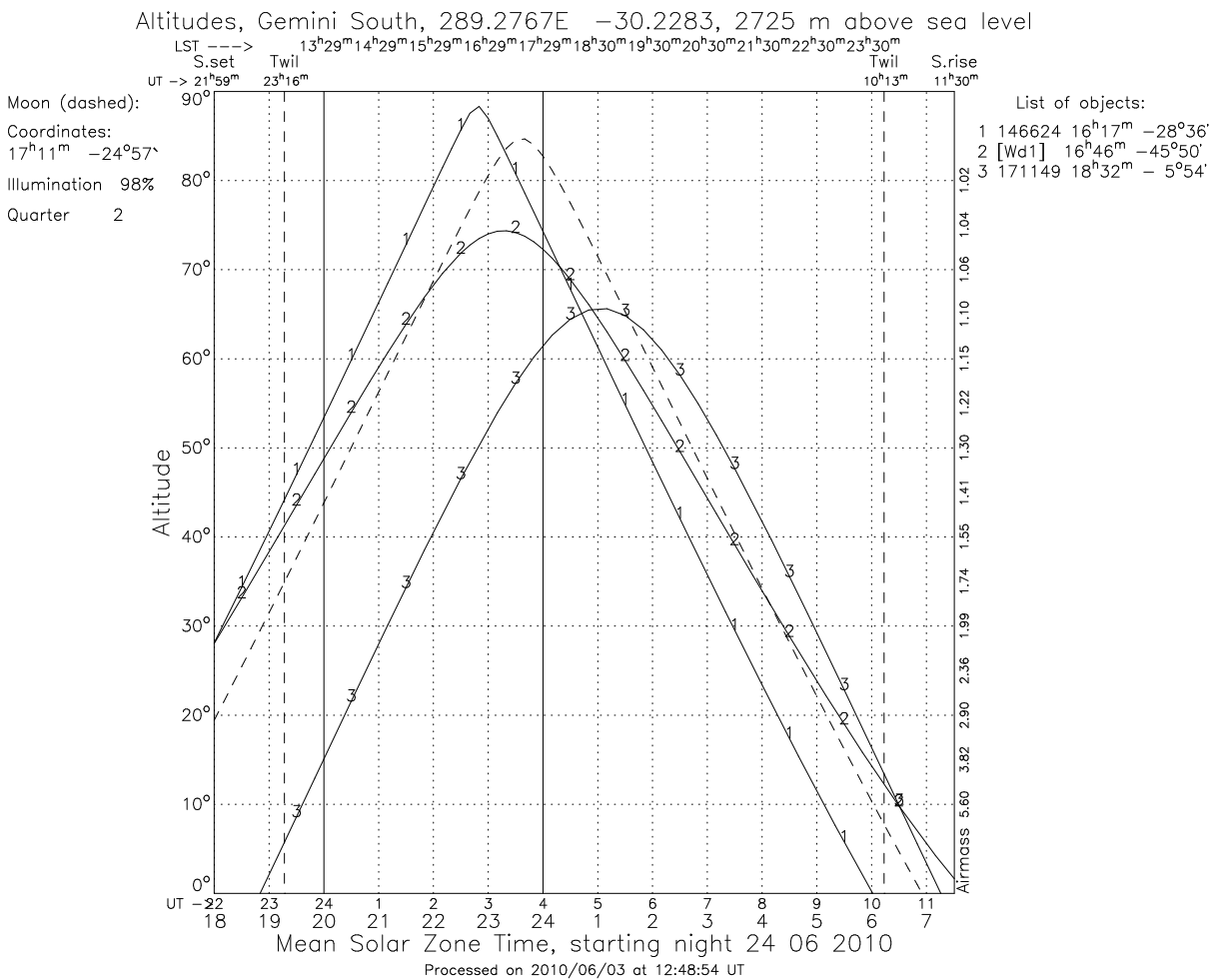}
\caption{\label{fig:airmass}Output from the Staralt program.  Here the date, observatory location and objects for the run have been entered uniquely.  Object 2 is the science target, Westerlund 1.  Object 1 and Object 3 are HD numbers for telluric standards, taken from Hanson et al.\  (2005).  }
\end{figure*}

The near-infrared spectroscopist is far more obsessed with airmass than the optical spectroscopist.  It is advised that the observer plot out the airmass of all of the objects (targets and standards) well in advance for a run.   This can be done using online software, 
http://catserver.ing.iac.es/staralt/index.php. The output of this program is given in Figure~\ref{fig:airmass}.  For this run, the authors were extraordinarily lucky that two of the telluric standards from the Hanson et al.\ (2005) catalog were well suited to be observed during a recent SOAR run: HD 146624 during the first half, and HD 171149 during the second half.   Looking at this diagram, one can make the best choices about when to observe the telluric standard relative to any observations made of a target object.  If the target observations take about 30 minutes (this includes total real time, such as acquisition and integration) then observing Telluric Object 1 just before the program target during the early part of the night will mean the target star will pass through the exact same airmass during its observations, optimizing a telluric match.  Later in the night, Object 3 can also be used, though observed after the target.   As hour angle increases, airmass increases quickly.  Note the non-linear values of the ordinate on the right of Figure~\ref{fig:airmass}. One must be ready to move quickly to a telluric standard or the final spectra may be quite disappointing.   This observer has been known to trace in red pen in real time on such a diagram, the sources being observed as time progresses, to know when it is time to move between object and telluric and vice versa.  The goal should be to observe the telluric
standard when its airmass is within 0.1 of that of the observation of the program object.

How to select telluric standards?  As was mentioned in \S~\ref{Sec-irspectrographs}, early-A dwarfs or solar analogues are typically used.  Ideally, one should seek telluric standards which are bright (for shorter integration times), have normal spectral types (no anomalies), and are not binaries (visible or spectroscopic).  But also, location in the sky is important.  Referring back to Figure~\ref{fig:airmass}  again, stars passing close to zenith at meridian have a different functional form to their airmass curves than do stars that remain low even during transit.  This can make it hard to catch both target and telluric at the same airmass if their curves are very different.  So, attempt to select a telluric standard that has a similar declination to the target object, but that transits 30-60 minutes before or after the target object.  

Background emission is the second serious concern for the infrared spectroscopist.  Even optical astronomers are aware of the increase in night sky brightness with increasing wavelength, with U and B brightness of typically $>$22 mag arcsec$^{-2}$, V around 21.5, R $\sim$ 21 and I $\sim$ 20 at the best sites.  But this is nothing compared to what the infrared observer must endure.  A very nice review of infrared astronomy is given by Tokunaga (2000) that all new (and seasoned!) infrared astronomers should read.  He lists the sky brightness in mag arcsec$^{-2}$ as 15.9, 13.4 and 14.1 at $J$, $H$, and $K_s$.  For the $L$ and $M$ bands, the sky is 4.9 and around 0 mag arcsec$^{-2}$, respectively!  In the latter bands, this is dominated by thermal emission, while in the $J$, $H$ and $K_s$, it is dominated by OH airglow.  This background emission will dominate one's spectrum if not removed. 

Removal of background emission is done by stepping the object along the slit between exposures or periodically offsetting the telescope to a blank field.  Since the background emission is ubiquitous, offsetting a compact target along the slit allows one to measure the background spectrum at the first target position in the second exposure, and vice versa.  In the near-infrared, where the sky background intensity is modest, one may step the target to several positions in the slit, observing the target all of the time, while simultaneously observing the sky background in the rest of the slit.  However, if the field is densely populated with stars, or the target itself is extended or surrounded by nebulosity, it is necessary to offset to a nearby patch of blank sky periodically to obtain the sky spectrum for background subtraction.  At mid-infrared wavelengths, the sky background is significantly larger and even small temporal variations can overwhelm the signal from the science target, so it is necessary to carry out sky subtraction on a much shorter time scale.  This is often done by taking very short exposures (to avoid saturation) and chopping the target on and off the slit at a few Hz, typically using a square-wave tip/tilt motion of the telescope secondary mirror.  The chopping and data taking sequences are synchronized, so that the on- and off-source data can be stored in separate data buffers and the sky subtraction carried out in real time.

How often to step along the slit?  This depends on a few things.  One always wants to maximize the counts for any single step integration, letting the exposure time be determined by the limits of the detector.  Remember that you must stay within the linear regime {\it while including} background emission in any single frame!   If the integration is too long in the $H$-band and the OH airglow lines are saturated, they won't properly subtract.  Always check that the counts are not too high before any subtraction is done.  The number of steps should be at least four, to remove bad pixels and six is a more typical minimum number.   Build up signal-to-noise through multiple sets of optimized offsets, returning to a telluric standard as needed between sets.   Finally, always check, as the data is coming in, that when you do subtract one slit (or sky offset) position from another that you do get zero counts outside of the star.  What can go wrong here, as was mentioned in \S~\ref{Sec-irspectrographs},  is that the strengths of the OH bands vary with
time.  This is particularly true if clouds move in or the seeing changes between slit positions,
with the result that the OH lines will not cancel entirely and reduction will be much more difficult.  
 If this occurs, one is forced to use a shorter integration per step to find a time frame over which the sky emission is sufficiently stable.   
 
 Note that even if one does have ``perfect" sky subtraction of the OH spectra, the large signal in the OH lines invariably adds noise to the final reduced spectra.  This is unavoidable, and makes the entire concept of defining the signal-to-noise-ratio in the NIR tricky to define, as it is bound to
vary depending upon the OH spectra within a particular region. 

What could possibly go wrong?  A lot.  The second author has never worked on an infrared spectrometer that didn't offer additional challenges which fell outside the standard operation as listed above.   To keep this brief, four of the most common problems that occur will now be discussed.  They include: flexure, fringing, wavelength calibration problems, and poor telluric matching.  Each of these can lead to greatly reduced signal-to-noise in the spectra, far below what would be predicted by counts alone.  

\begin{itemize}
\item{\bf Flexure.} No spectrograph is absolutely rigid!
Flexure can be a larger issue for infrared spectrographs because the optics must be cooled to reduce the thermal background, and it is challenging to minimize the thermal conduction to the internal spectrograph bench while also minimizing the mechanical flexure between the bench and outer structure of the instrument.
Depending on how the instrument is mounted, when observing to the east, versus looking to the west, for instance, or as the telescope passes meridian, the internal light path will shift due to structure shifts.  The amount of flexure depends on the instrument and telescope set up.  Previous users or the support astronomer for the instrument can help the new user decide if this needs to be considered and how to mitigate the effects.  This is best addressed during the observing, keeping telluric standards, lamps and objects on the same side of the meridian for instance.  

\item{\bf Fringing.} Optical CCDs fringe in the red wavelength regions due to interference within the surface layers of the detector; for IR spectrographs, fringing can occur due to parallel
optical surfaces, such as blocking filters, in the optical path.
While in principle, this should cancel out with the dome flats, due to flexure in the system and light path differences, they typically do not cancel well. Common remedies include obtaining quartz lamps at the exact same sky
location as the observations.   This might work and should be included in
the observing plan.
However, the fringes are often not similar between the quartz lamp and point sources.  The second author has instead turned to Fourier filtering methods to simply remove fringes outright.   Software exists within IRAF in the STSDAS package to lock in on the fringe and remove it.  This is easily done with flat fields, and virtually impossible for point source images.  Stellar flats (\S~\ref{Sec-irred}) can be used to create a very crude two-dimensional illumination of a point source.  If the fringe pattern is relatively strong, the Fourier filtering packages should be able to lock in on the overarching pattern and create a fringe correction which can be applied to all your frames before extraction.

\item{\bf Wavelength Calibration.}  For many long slit spectrometers, the wavelength solution is a function of position on the slit.   It was already suggested that wavelength calibration should be applied using comparison lamp solutions which were extracted at the exact same location as the star was extracted.   If this is not done, then there will be slight variations in solution with slit position and the resolution will be inadvertently reduced by co-adding such data.  Moreover, this can lead to even more serious problems later on when applying telluric corrections.

\item{\bf Telluric Correction.}  The telluric standards need to have been observed at a very similar airmass and hopefully fairly close in time to the
program object.    When observing, one needs to be sure that the telluric standard observations extend to the same range of airmasses
as the program objects (both minimum and maximum).  If the match isn't great, one can interpolate as needed by using hybrid spectra of two telluric stars to get a better airmass correction.  Also, IRAF has very useful software which actually uses the telluric lines with a statistical minimization routine, to make the best match.  Finally, if one is working in an area of very strong telluric lines, there may be a good match
with the features {\it only} when the target and standard were observed in the {\it exact same location in the slit}.  This requires keeping all spectra separate until telluric removal, then combining the final set of spectra as the last step.

\end{itemize}

Possibly the strongest recommendation is that one needs to talk to a previous user of the instrument, preferably one who actually knows
what they are doing. Make sure that the answers make sense, though. (Better still would be to talk to several such previous users, and average
their responses, possibly using some strong rejection algorithm.)  Maybe they will even be willing to share some of their data before the
observing run, so that one can really get a sense of what things will look like. 

\section*{Wrap-Up and Acknowledgements
}

The authors hope that the reader will have gained something useful from this chapter.  They had a lot of fun in writing it.  If there was
any simple conclusion to offer it would be that astronomical spectroscopy is  a stool that must sit on three legs for it to be a useful:
simply taking data with sufficient counts and resolution is not enough.  One has to carefully think through the calibration requirements 
of the data as needed by one's program, and one must perform the reductions in such a way that honors the data.    Quantitative quick-look
at the telescope is essential for the process. 

The authors' knowledge of spectroscopy is what
it is due to contact with many individuals over the years, and in particular the authors want to acknowledge the influence of
Peter S. Conti, Bruce Bohannan, James DeVeny,
Dianne Harmer, Nidia Morrell, Virpi Niemela, Vera Rubin, and Daryl Willmarth.   Constructive comments were made on parts or the whole
of a draft of this chapter by
Travis Barman, Bob Blum, Howard Bond, Dianne Harmer, Padraig Houlahan,  Deidre Hunter, Dick Joyce,
Emily Levesque, Stephen Levine, 
Nidia Morrell, Kathryn Neugent, George Rieke, Sumner Starrfield, and
Mark Wagner, and Daryl Willmarth.  Neugent  was responsible for constructing many of the figures used here.
The authors gratefully acknowledge support from the National Science Foundation, most recently through AST-1008020 and AST-11009550.

\end{document}